\documentclass{article}
\usepackage[utf8]{inputenc}

\newcommand{\Gardenfors}{G\"ardenfors}

\newcommand{\St}{\mathsf{St}} 

\usepackage{amsmath}
\usepackage{amsthm}
\usepackage{amssymb}
\usepackage{stmaryrd}  

\newcommand{\kto}{\to} 

\theoremstyle{definition}
\newtheorem{Def}{Definition}
\newtheorem{definition}[Def]{Definition}

\newtheorem{lemma}[Def]{Lemma}

\newtheorem{remark}{Remark}

\newtheorem{example}{Example}

\usepackage{tikz-cd}

\usepackage{stackengine}

\newcommand{\cat}[1]{\ensuremath{\mathbf{#1}}}
\newcommand{\catC}{\cat{C}}

\newcommand{\id}[1]{\ensuremath{\mathrm{id}_{#1}}}

\newcommand{\Quant}{\cat{Quant}}

\newcommand{\hilbH}{\mathcal{H}} 
\newcommand{\hilbK}{\mathcal{K}} 

\newcommand{\discard}[1]{\ensuremath{\tinygroundnew_{#1}}}

\usepackage{tikz,xypic}
\usetikzlibrary{decorations.pathreplacing,decorations.markings,arrows.meta,backgrounds}
\pgfdeclarelayer{edgelayer}
\pgfdeclarelayer{nodelayer}
\pgfsetlayers{background,edgelayer,nodelayer,main}
\tikzstyle{whitedot}=[circle, draw=black, fill=white, inner sep=.4ex]
\tikzstyle{none}=[inner sep=0mm]

\usepackage{tikz,xypic}

\usetikzlibrary{decorations.pathreplacing,decorations.markings,arrows.meta,backgrounds}
\pgfdeclarelayer{edgelayer}
\pgfdeclarelayer{nodelayer}
\pgfsetlayers{background,edgelayer,nodelayer,main}

\tikzstyle{cdot}=[circle, draw=black, fill=black!25, inner sep=.4ex] 
\tikzstyle{bigdot}=[dot, inner sep=0pt]
\tikzstyle{whitedot}=[circle, draw=black, fill=white, inner sep=.4ex]
\tikzstyle{greydot}=[circle, draw=black, fill=black!25, inner sep=.4ex] 
\tikzstyle{blackdot}=[circle, draw=black, fill=black, inner sep=.4ex]
\tikzset{arrow/.style={decoration={
    markings,
    mark=at position #1 with \arrow{>[length=2pt, width=3pt]}},
    postaction=decorate},
    reverse arrow/.style={decoration={
    markings,
    mark=at position #1 with {{\arrow{<[length=2pt, width=3pt]}}}},
    postaction=decorate},
    thick/.style = {line width=0.15em}
}

\newcommand{\tinycomult}[1][cdot]{
\smash{\raisebox{-2pt}{\hspace{-5pt}\ensuremath{\begin{pic}[scale=0.4,yscale=1]
    \node (0) at (0,0) {};
    \node[#1, inner sep=1.5pt] (1) at (0,0.55) {};
    \node (2) at (-0.5,1) {};
    \node (3) at (0.5,1) {};
    \draw (0.center) to (1.center);
    \draw (1.center) to [out=left, in=down, out looseness=1.5] (2.center);
    \draw (1.center) to [out=right, in=down, out looseness=1.5] (3.center);
    \node[#1, inner sep=1.5pt] (1) at (0,0.55) {};
\end{pic}
}\hspace{-3pt}}}}

\newcommand{\tinycap}{\smash{\raisebox{-3pt}{\hspace{-2pt}\ensuremath{\begin{pic}[scale=0.2, yscale=-1]
   \pgftransformscale{1.5} \draw[arrow=.6, scale = 1] (0,0) to[out=-90,in=-90,looseness=1.5] (1.5,0);
\end{pic}}}}}

\newenvironment{pic}[1][] {\begin{aligned}\begin{tikzpicture}[scale=2.0, font=\tiny,#1]}{\end{tikzpicture}\end{aligned}} 

\newif\ifvflip\pgfkeys{/tikz/vflip/.is if=vflip}
\newif\ifhflip\pgfkeys{/tikz/hflip/.is if=hflip}
\newif\ifhvflip\pgfkeys{/tikz/hvflip/.is if=hvflip}

\newenvironment{picc}[1][]
{\begin{aligned}\begin{tikzpicture}[font=\tiny,#1]}
{\end{tikzpicture}\end{aligned}}

\newlength\minimummorphismwidth
\newlength\stateheight
\newlength\minimumstatewidth
\newlength\connectheight
\tikzset{colour/.initial=white}

\tikzstyle{pure}=[line width=.7pt]

\makeatletter

\pgfdeclareshape{groundd}
{
    \savedanchor\centerpoint
    {
        \pgf@x=0pt
        \pgf@y=0pt
    }
    \anchor{center}{\centerpoint}
    \anchorborder{\centerpoint}

    \anchor{north}
    {
        \pgf@x=0pt
        \pgf@y=0.16\stateheight
    }
    \anchor{south}
    {
        \pgf@x=0pt
        \pgf@y=0pt
    }
    \saveddimen\overallwidth
    {
        \pgfkeysgetvalue{/pgf/minimum width}{\minwidth}
        \pgf@x=\minimumstatewidth
        \ifdim\pgf@x<\minwidth
            \pgf@x=\minwidth
        \fi
    }
    \backgroundpath
    {
        \begin{pgfonlayer}{main} 
        \pgfsetstrokecolor{black}
        \pgfsetlinewidth{1.25pt}
        \ifhflip
            \pgftransformyscale{-1}
        \fi
        \pgftransformscale{0.5}
        \pgfpathmoveto{\pgfpoint{-0.5*\overallwidth}{0pt}}
        \pgfpathlineto{\pgfpoint{0.5*\overallwidth}{0pt}}
        \pgfpathmoveto{\pgfpoint{-0.33*\overallwidth}{0.33*\stateheight}}
        \pgfpathlineto{\pgfpoint{0.33*\overallwidth}{0.33*\stateheight}}
        \pgfpathmoveto{\pgfpoint{-0.16*\overallwidth}{0.66*\stateheight}}
        \pgfpathlineto{\pgfpoint{0.16*\overallwidth}{0.66*\stateheight}}
        \pgfpathmoveto{\pgfpoint{-0.02*\overallwidth}{\stateheight}}
        \pgfpathlineto{\pgfpoint{0.02*\overallwidth}{\stateheight}}
        \pgfusepath{stroke}
        \end{pgfonlayer}
    }
}

\usepackage{tikz,xypic}
\usetikzlibrary{decorations.pathreplacing,decorations.markings,arrows.meta,backgrounds,shapes}
\usetikzlibrary{circuits.ee.IEC}
\pgfdeclarelayer{edgelayer}
\pgfdeclarelayer{nodelayer}
\pgfsetlayers{background,edgelayer,nodelayer,main}
\tikzstyle{none}=[inner sep=0mm]
\tikzstyle{every loop}=[]
\tikzstyle{mark coordinate}=[inner sep=0pt,outer sep=0pt,minimum size=3pt,fill=black,circle]

\tikzset{arrow/.style={decoration={
    markings,
    mark=at position #1 with \arrow{>[length=2pt, width=3pt]}},
    postaction=decorate},
    reverse arrow/.style={decoration={
    markings,
    mark=at position #1 with {{\arrow{<[length=2pt, width=3pt]}}}},
    postaction=decorate}
}

\tikzstyle{upground}=[circuit ee IEC,thick,ground,rotate=90,scale=1.5]
\tikzstyle{upgroundwhite}=[circuit ee IEC,thick,ground,rotate=90,scale=1.5, fill=white]
\tikzstyle{downground}=[circuit ee IEC,thick,ground,rotate=-90,scale=1.5]
\tikzstyle{downgroundnorm}=[circuit ee IEC,thick,ground,rotate=-90,scale=1.5, fill=white]

\newcommand{\mapminh}{5mm} 

\newcommand{\maplw}{0.7pt} 

\tikzstyle{box}=[map]
\tikzstyle{medium box}=[medium map]
\tikzstyle{dot}=[inner sep=0mm,minimum width=2mm,minimum height=2mm,draw,shape=circle]  
\tikzstyle{black dot}=[dot,fill=black]
\tikzstyle{white dot}=[dot,fill=white,,text depth=-0.2mm]
\tikzstyle{grey dot}=[dot,fill=black!25] 

\tikzstyle{corner1}=[box,fill=white, font=\footnotesize] %
\tikzstyle{corner2}=[dot,fill=white, font=\footnotesize] %
\tikzstyle{corner3}=[dot,fill=black!25, font=\footnotesize] %
\tikzstyle{corner4}=[dot,fill=black, font=\footnotesize] %

\tikzstyle{scalar}=[circle,draw,inner sep=2pt, line width=\maplw] 

\usetikzlibrary{shapes.misc, positioning}

\tikzset{stateshape/.style={append after command={
   \pgfextra
        \draw[sharp corners, fill=white, line width = \maplw]%
    (\tikzlastnode.west)%
    [rounded corners=0pt] |- (\tikzlastnode.north)%
    [rounded corners=0pt] -| (\tikzlastnode.east)%
    [rounded corners=5pt] |- (\tikzlastnode.south)%
    [rounded corners=5pt] -| (\tikzlastnode.west);
   \endpgfextra}}}

\tikzset{effectshape/.style={append after command={
   \pgfextra
        \draw[sharp corners, fill=white, line width = \maplw]%
    (\tikzlastnode.west)%
    [rounded corners=0pt] |- (\tikzlastnode.south)%
    [rounded corners=0pt] -| (\tikzlastnode.east)%
    [rounded corners=5pt] |- (\tikzlastnode.north)%
    [rounded corners=5pt] -| (\tikzlastnode.west);
   \endpgfextra}}}

 \tikzstyle{map}=[draw,shape=rectangle, inner sep=2pt,minimum height=\mapminh, minimum width=5mm,fill=white]

\tikzstyle{point}=[fill=white,draw,shape=isosceles triangle,shape border rotate=-90,isosceles triangle stretches=true,inner sep=0.2pt,minimum width=0.5cm,minimum height=0.8mm,yshift=-0.0mm]
\tikzstyle{copoint}=[fill=white,draw,shape=isosceles triangle,shape border rotate=90,isosceles triangle stretches=true,inner sep=0.2pt,minimum width=0.5cm,minimum height=0.8mm,yshift=-0.0mm]
\tikzstyle{wide point}=[point, minimum width=12mm]
\tikzstyle{wide copoint}=[copoint, minimum width=12mm]

\tikzstyle{decomp}=[fill=white,draw,shape=isosceles triangle,shape border rotate=-90,isosceles triangle stretches=true,inner sep=0pt,minimum width=0.75cm,minimum height=4mm,yshift=-0.0mm]

\tikzstyle{decompwide}=[fill=white,draw,shape=isosceles triangle,shape border rotate=-90,isosceles triangle stretches=true,inner sep=0pt,minimum width=1.4cm,minimum height=5mm,yshift=-0.0mm]

\tikzstyle{decompflip}=[fill=white,draw,shape=isosceles triangle,shape border rotate=90,isosceles triangle stretches=true,inner sep=0pt,minimum width=0.75cm,minimum height=4mm,yshift=-0.0mm]

\tikzstyle{decompwideflip}=[fill=white,draw,shape=isosceles triangle,shape border rotate=90,isosceles triangle stretches=true,inner sep=0pt,minimum width=1.4cm,minimum height=5mm,yshift=-0.0mm]

\tikzstyle{medium map} = [map, minimum width = 12mm] 
\tikzstyle{semilarge map} = [map, minimum width = 15mm] 
\tikzstyle{large map} = [map, minimum width = 18mm] 

\tikzstyle{kpoint} =[point]
\tikzstyle{kpointadj} =[copoint]
\tikzstyle{kpointconj}=[dagpointconj] 

\makeatletter
\newcommand{\boxshape}[3]{%
\pgfdeclareshape{#1}{
\inheritsavedanchors[from=rectangle] 
\inheritanchorborder[from=rectangle]
\inheritanchor[from=rectangle]{center}
\inheritanchor[from=rectangle]{north}
\inheritanchor[from=rectangle]{south}
\inheritanchor[from=rectangle]{west}
\inheritanchor[from=rectangle]{east}
\backgroundpath{
\southwest \pgf@xa=\pgf@x \pgf@ya=\pgf@y
\northeast \pgf@xb=\pgf@x \pgf@yb=\pgf@y

\@tempdima=#2
\@tempdimb=#3

\pgfpathmoveto{\pgfpoint{\pgf@xa - 5pt + \@tempdima}{\pgf@ya}}
\pgfpathlineto{\pgfpoint{\pgf@xa - 5pt - \@tempdima}{\pgf@yb}}
\pgfpathlineto{\pgfpoint{\pgf@xb + 5pt + \@tempdimb}{\pgf@yb}}
\pgfpathlineto{\pgfpoint{\pgf@xb + 5pt - \@tempdimb}{\pgf@ya}}
\pgfpathlineto{\pgfpoint{\pgf@xa - 5pt + \@tempdima}{\pgf@ya}}
\pgfpathclose
}
}}

\boxshape{NEbox}{0pt}{3pt} 
\boxshape{SEbox}{0pt}{-3pt}
\boxshape{NWbox}{3pt}{0pt}
\boxshape{SWbox}{-3pt}{0pt}
\boxshape{rec-box}{0pt}{0pt}
\makeatother

\tikzstyle{cloud}=[shape=cloud,draw,minimum width=1.5cm,minimum height=1.5cm]

\tikzstyle{dagmap}=[draw,shape=NEbox,inner sep=2pt,minimum height=\mapminh,fill=white, line width = \maplw] %
\tikzstyle{dashedmap}=[draw,dashed,shape=NEbox,inner sep=2pt,minimum height=\mapminh,fill=white, line width = \maplw]
\tikzstyle{mapdag}=[draw,shape=SEbox,inner sep=2pt,minimum height=\mapminh,fill=white, line width = \maplw]
\tikzstyle{mapadj}=[draw,shape=SEbox,inner sep=2pt,minimum height=\mapminh,fill=white, line width = \maplw]
\tikzstyle{maptrans}=[draw,shape=SWbox,inner sep=2pt,minimum height=\mapminh,fill=white, line width = \maplw]
\tikzstyle{mapconj}=[draw,shape=NWbox,inner sep=2pt,minimum height=\mapminh,fill=white, line width = \maplw]

\tikzstyle{medium dagmap}=[draw,shape=NEbox,inner sep=2pt,minimum height=\mapminh,fill=white,minimum width=7mm, line width = \maplw]
\tikzstyle{semilarge dagmap}=[draw,shape=NEbox,inner sep=2pt,minimum height=\mapminh,fill=white,minimum width=9.5mm, line width = \maplw]
\tikzstyle{large dagmap}=[draw,shape=NEbox,inner sep=2pt,minimum height=\mapminh,fill=white,minimum width=12mm, line width = \maplw]

\makeatletter

\pgfdeclareshape{cornerpoint}{
\inheritsavedanchors[from=rectangle] 
\inheritanchorborder[from=rectangle]
\inheritanchor[from=rectangle]{center}
\inheritanchor[from=rectangle]{north}
\inheritanchor[from=rectangle]{south}
\inheritanchor[from=rectangle]{west}
\inheritanchor[from=rectangle]{east}
\backgroundpath{
\southwest \pgf@xa=\pgf@x \pgf@ya=\pgf@y
\northeast \pgf@xb=\pgf@x \pgf@yb=\pgf@y

\pgfmathsetmacro{\pgf@shorten@left}{\pgfkeysvalueof{/tikz/shorten left}}
\pgfmathsetmacro{\pgf@shorten@right}{\pgfkeysvalueof{/tikz/shorten right}}

\pgfpathmoveto{\pgfpoint{0.5 * (\pgf@xa + \pgf@xb)}{\pgf@ya - 5pt}}
\pgfpathlineto{\pgfpoint{\pgf@xa - 8pt + \pgf@shorten@left}{\pgf@yb - 1.5 * \pgf@shorten@left}}
\pgfpathlineto{\pgfpoint{\pgf@xa - 8pt + \pgf@shorten@left}{\pgf@yb}}
\pgfpathlineto{\pgfpoint{\pgf@xb + 8pt - \pgf@shorten@right}{\pgf@yb}}
\pgfpathlineto{\pgfpoint{\pgf@xb + 8pt - \pgf@shorten@right}{\pgf@yb - 1.5 * \pgf@shorten@right}}
\pgfpathclose
}
}

\pgfdeclareshape{cornercopoint}{
\inheritsavedanchors[from=rectangle] 
\inheritanchorborder[from=rectangle]
\inheritanchor[from=rectangle]{center}
\inheritanchor[from=rectangle]{north}
\inheritanchor[from=rectangle]{south}
\inheritanchor[from=rectangle]{west}
\inheritanchor[from=rectangle]{east}
\backgroundpath{
\southwest \pgf@xa=\pgf@x \pgf@ya=\pgf@y
\northeast \pgf@xb=\pgf@x \pgf@yb=\pgf@y

\pgfmathsetmacro{\pgf@shorten@left}{\pgfkeysvalueof{/tikz/shorten left}}
\pgfmathsetmacro{\pgf@shorten@right}{\pgfkeysvalueof{/tikz/shorten right}}

\pgfpathmoveto{\pgfpoint{0.5 * (\pgf@xa + \pgf@xb)}{\pgf@yb + 5pt}}
\pgfpathlineto{\pgfpoint{\pgf@xa - 8pt + \pgf@shorten@left}{\pgf@ya + 1.5 * \pgf@shorten@left}}
\pgfpathlineto{\pgfpoint{\pgf@xa - 8pt + \pgf@shorten@left}{\pgf@ya}}
\pgfpathlineto{\pgfpoint{\pgf@xb + 8pt - \pgf@shorten@right}{\pgf@ya}}
\pgfpathlineto{\pgfpoint{\pgf@xb + 8pt - \pgf@shorten@right}{\pgf@ya + 1.5 * \pgf@shorten@right}}
\pgfpathclose
}
}

\makeatother

\pgfkeyssetvalue{/tikz/shorten left}{0pt}
\pgfkeyssetvalue{/tikz/shorten right}{0pt}

\tikzstyle{dagpoint common}=[draw,fill=white,inner sep=1pt, line width = \maplw, minimum height = 4mm, yshift=1.2pt] 
\tikzstyle{dagpoint sc}=[shape=cornerpoint,dagpoint common]
\tikzstyle{dagpoint adjoint sc}=[shape=cornercopoint,dagpoint common]
\tikzstyle{dagpoint}=[shape=cornerpoint,shorten left=4pt,dagpoint common]
\tikzstyle{dagpointadj}=[shape=cornercopoint,shorten left=5pt,dagpoint common]
\tikzstyle{dagpointconj}=[shape=cornerpoint,shorten right=5pt,dagpoint common]
\tikzstyle{dagpointtrans}=[shape=cornercopoint,shorten right=5pt,dagpoint common]
\tikzstyle{dagpointsymm}=[shape=cornerpoint,shorten left=5pt,shorten right=5pt,dagpoint common]

\tikzstyle{widedagpoint}=[dagpoint, minimum width=1 cm, inner sep=2pt]
\tikzstyle{widedagpointadj}=[dagpointadj, minimum width=1 cm, inner sep=2pt]

\tikzstyle{every picture}=[baseline=-0.25em,scale=0.5]
\tikzstyle{label}=[font=\footnotesize,text height=1ex, text depth=0.15ex]

\usetikzlibrary[shapes]

\tikzset{
sidetriangle/.style = {regular polygon, regular polygon sides = 3, aspect = 1, shape border rotate = 90, draw, inner sep = 0, minimum width = 1.2cm}
}

\tikzset{
isoc/.style = {shape=isosceles triangle, shape border rotate = 180, isosceles triangle stretches = true, minimum width = 1.2cm, minimum height= 1.5cm, inner sep = 0.3}}

\tikzset{
coarse/.style = {shape = circle, fill = white, draw, inner sep = 0, minimum width =0.125cm}
}
\tikzset{
coarsesymbol/.style = {shape = circle, fill = white, inner sep = -0.7, minimum width = 0.125cm}
}

\tikzstyle{sidetriangle2}=[sidetriangle, minimum width = 2cm, fill=white]
\tikzstyle{sideisocsmall}]=[style=isoc, minimum width = 1cm, minimum height = 0.8cm, draw, fill=white, font=\Large]
\tikzstyle{sideisoc}]=[style=isoc, minimum width = 2cm, draw, fill=white, font=\Large]
\tikzstyle{sideisocmid}]=[style=isoc, minimum width = 2.5cm, draw, fill=white, font=\Large]
\tikzstyle{sideisocmedium}]=[style=isoc, minimum width = 3cm, draw, fill=white, font=\Large]

\newcommand{\tinygroundnew}{
\smash{
{\hspace{-3pt}
\ensuremath{
\begin{picc}[scale=1.0] 
    \node[upground, xscale=0.8, yscale=0.7] (1) at (0,0.16) {};
    \draw (0,0.03) to (0,-0.25);
\end{picc}
}\hspace{-1pt}}}}

\tikzstyle{label}=[font=\footnotesize,text height=1ex, text depth=0.15ex]

\tikzstyle{box}=[map]
\tikzstyle{medium box}=[medium map]
\tikzstyle{dot}=[inner sep=0mm,minimum width=2mm,minimum height=2mm,draw,shape=circle]  
\tikzstyle{black dot}=[dot,fill=black]
\tikzstyle{white dot}=[dot,fill=white,,text depth=-0.2mm]
\tikzstyle{grey dot}=[dot,fill=black!25] 

\tikzstyle{corner1}=[box,fill=white, font=\footnotesize] %
\tikzstyle{corner2}=[dot,fill=white, font=\footnotesize] %
\tikzstyle{corner3}=[dot,fill=black!25, font=\footnotesize] %
\tikzstyle{corner4}=[dot,fill=black, font=\footnotesize] %

\tikzstyle{scalar}=[circle,draw,inner sep=2pt, line width=\maplw] 

\tikzstyle{sharpstate}=[fill=white,draw,shape=isosceles triangle,shape border rotate=-90,isosceles triangle stretches=true,inner sep=0.2pt,minimum width=0.5cm,minimum height=0.8mm,yshift=-0.0mm]
\tikzstyle{sharpeffect}=[fill=white,draw,shape=isosceles triangle,shape border rotate=90,isosceles triangle stretches=true,inner sep=0.2pt,minimum width=0.5cm,minimum height=0.8mm,yshift=-0.0mm]
\tikzstyle{wide sharpstate}=[point, minimum width=12mm]
\tikzstyle{wide sharpeffect}=[copoint, minimum width=12mm]

\tikzstyle{point}=[sharpstate]

\usepackage{graphicx}%
\usepackage{multirow}%
\usepackage{amsmath,amssymb,amsfonts}%
\usepackage{amsthm}%
\usepackage{mathrsfs}%
\usepackage[title]{appendix}%
\usepackage{xcolor}%
\usepackage{textcomp}%
\usepackage{manyfoot}%
\usepackage{booktabs}%
\usepackage{algorithm}%
\usepackage{algorithmicx}%
\usepackage{algpseudocode}%
\usepackage{listings}%

\usepackage{tikzit}

\usepackage{apacite}

\usepackage{pifont}
\tikzstyle{seen}=[fill=lightgray, draw=black, shape=circle, minimum size=0.4cm, inner sep=0.1cm]
\tikzstyle{unseen}=[fill=white, draw=black, shape=circle, minimum size=0.4cm, inner sep=0.1cm]
\tikzstyle{none}=[]
\tikzstyle{full}=[->, draw=black]
\tikzstyle{dashed}=[->, draw=black]

\newcommand{\cc}{C} 
\newcommand{\crc}{C} 

\newcommand{\x}{\mathbf{x}}

\newcommand{\Class}{\cat{Class}}
\newcommand{\Prob}{\cat{Prob}}
\newcommand{\ConSp}{\cat{ConSp}}
\newcommand{\Tr}{\mathsf{Tr}}

\newcommand{\RXYZ}{R}
\newcommand{\RZYX}{R'}

\newcommand{\RX}[1]{R^X_{\theta_{#1}}}
\newcommand{\RY}[1]{R^Y_{\theta_{#1}}}
\newcommand{\RZ}[1]{R^Z_{\theta_{#1}}}

\newcommand{\RXC}[1]{R^X_{\thetac_{#1}}}
\newcommand{\RYC}[1]{R^Y_{\thetac_{#1}}}
\newcommand{\RZC}[1]{R^Z_{\thetac_{#1}}}

\newcommand{\Hdom}[1]{\hilbH_{\text{#1}}}

\newcommand{\thetac}{\phi^C} 

\newcommand{\Ha}{\hilbH_1} 
\newcommand{\Hb}{\hilbH_2} 
\newcommand{\Hc}{\hilbH_3} 
\newcommand{\Hd}{\hilbH_4} 

\newcommand{\z}{\mathbf{z}}
\newcommand{\Z}{\mathbf{Z}}

\newcommand{\con}{\mathbf{c}}

\usepackage{braket}

\title{From Conceptual Spaces to Quantum Concepts: Formalising and Learning Structured Conceptual Models}

\date{6 November 2023}

\author{Sean Tull, Razin A. Shaikh, Sara Sabrina Zemlji\v{c} and Stephen Clark\\
Quantinuum\\
17 Beaumont Street, Oxford, UK\\
\texttt{\normalsize \{sean.tull,razin.shaikh,sara.zemljic,steve.clark\}@quantinuum.com}}

\begin{document}

\maketitle

\begin{abstract}
In this article we present a new modelling framework for structured concepts using a category-theoretic generalisation of conceptual spaces, and
show how the conceptual representations can be learned
automatically from data, using two very different instantiations: one classical and one quantum. A contribution of the work is a thorough category-theoretic formalisation of our framework. We claim that the use of category theory, and in particular the use of string diagrams to describe quantum processes, helps elucidate some of the most important features of our approach. We build upon \Gardenfors' classical framework of \emph{conceptual spaces}, in which cognition is modelled geometrically through the use of convex spaces, which in turn factorise in terms of simpler spaces called \emph{domains}. We show how concepts from the domains of \textsc{shape}, \textsc{colour}, \textsc{size} and \textsc{position} can be learned from images of simple shapes, where concepts are represented as Gaussians in the classical implementation, and quantum effects in the quantum one. In the classical case we develop a new model which is inspired by the $\beta$-VAE model of concepts, but is designed to be more closely connected with language, so that the names of concepts form part of the graphical model. In the quantum case, concepts are learned by a hybrid classical-quantum network trained to perform concept classification, where the classical image processing is carried out by a convolutional neural network and the quantum representations are produced by a parameterised quantum circuit. Finally, we consider the question of whether our quantum models of concepts can be considered conceptual spaces in the \Gardenfors\ sense.
\end{abstract}

\section{Introduction}

The study of concepts has a long history in a number of related fields, including philosophy, linguistics, psychology and cognitive science \shortcite{murphy_concepts,conceptual_mind}. More recently, researchers have begun to consider how mathematical tools from quantum theory can be used to model cognitive phenomena, including conceptual structure. The general use of quantum formalism in psychology and cognitive science has led to an emerging area called quantum cognition \shortcite{aerts2009,pothos2013bbs}. The idea is that some of the features of quantum theory, such as entanglement, can be used to account for psychological data which can be hard to model classically. Examples include ordering effects in how subjects answer questions \shortcite{Trueblood2011} and concept combination \shortcite{aerts_gabora2005,tomas2015}.

Another recent development in the study of concepts has been the application of machine learning to the problem of how artificial agents can automatically learn concepts from raw perceptual data \shortcite{beta-vae,SCAN}. The motivation for endowing an agent with conceptual representations, and learning those representations automatically from the agent's environment, is that this will enable it to reason and act more effectively in that environment, similar to how humans use concepts \shortcite{lake_thinking_machines}. One hope is that the explicit use of concepts will ameliorate some of the negative consequences of the ``black-box" nature of neural architectures currently being used in AI.

In this article we present a new modelling framework for concepts based on the mathematical formalism used in quantum theory, and demonstrate how the conceptual representations can be learned automatically from data, using both classical and quantum-inspired models. A contribution of the work is a thorough category-theoretic formalisation of our framework, following \shortciteA{bolt2019interacting} and \shortciteA{tull2021categorical}. Formalisation of conceptual models is not new \shortcite{ganter:2016}, but we claim that the use of category theory \shortcite{applied_cat_theory}, and in particular the use of string diagrams to describe quantum processes \shortcite{coecke_kissinger_2017}, helps elucidate some of the most important features of our approach to concept modelling. This aspect of our work also fits with the recent push to introduce category theory into machine learning and AI more broadly. The motivation is to make deep learning less ad-hoc and less driven by heuristics, by viewing deep learning models through the compositional lens of category theory \shortcite{cat_ml}.

\shortciteA[p.1]{murphy_concepts} describes concepts as ``the glue that holds our mental world together". But how should concepts be modelled and represented mathematically? There are many modelling frameworks in the literature, including the \emph{classical theory} \shortcite{margolis_stanford}, the \emph{prototype theory} \shortcite{rosch1973natural}, and the \emph{theory theory} \shortcite{gopnik}.
Here we build upon \Gardenfors' framework of \emph{conceptual spaces} \shortcite{gardenfors2004conceptual,gardenfors2014}, in which cognition is modelled geometrically through the use of convex spaces, which in turn factorise in terms of simpler spaces called \emph{domains}.

Our category-theoretic formalisation of conceptual spaces allows flexibility in how the framework is instantiated and then implemented, with the particular instantiation determined by the choice of category.
First we show how the framework can be instantiated and implemented classically, by using the formalisation of ``fuzzy" conceptual spaces from \shortciteA{tull2021categorical}, and developing a probabilistic model based on Variational Autoencoders (VAEs) \shortcite{rezende14,kingma14}. Having ``fuzzy" probabilistic representations not only extends \Gardenfors' framework in a useful way, it also provides a natural mechanism for dealing with the vagueness inherent in the human conceptual system, and allows us to draw on the toolkit from machine learning to provide effective learning mechanisms. Our new model---which we call the \emph{Conceptual VAE}---is an extension of the $\beta$-VAE from \shortciteA{beta-vae}, with the concepts having explicit labels and represented as multivariate Gaussians in a factored conceptual space.

We use the Spriteworld software \shortcite{spriteworld19} to generate simple images consisting of coloured shapes of certain sizes in certain positions, meaning our conceptual spaces contain four domains: \textsc{colour}, \textsc{size}, \textsc{shape} and \textsc{position}. The main question we investigate for the classical model is a representational
learning one: can the Conceptual VAE induce factored representations in a
latent conceptual space which neatly separates the individual concepts, and under what conditions? Here we demonstrate that, if the system is provided with
supervision regarding the domains, and provided with the corresponding four
labels for each training instance (e.g. (\emph{blue}, \emph{small}, \emph{circle}, \emph{top})), then the VAE can learn Gaussians which faithfully represent the colour spectrum, for example. We also show
how the Conceptual VAE naturally provides a concept classifier, in the form of
the encoder, which predicts a Gaussian for an image that can be compared with
the induced conceptual representations using the KL divergence. 



Our second instantiation of the abstract framework uses a category for describing quantum processes \shortcite{coecke_kissinger_2017}. In this case, the images of shapes are represented as \emph{quantum states} in an underlying Hilbert space and concepts are \emph{quantum effects}. Applying a concept effect to an instance state yields a scalar, which we interpret as specifying how well the instance fits the concept.
The factoring of the conceptual space is represented naturally in our models through the use of the tensor product as the monoidal product. 
We choose to implement our quantum model using a hybrid quantum-classical network trained to perform concept classification, where the classical
image processing is carried out by a convolutional neural network \shortcite[Ch.9]{deep_learning} and the quantum representations are produced by a parameterised quantum circuit \shortcite{Benedetti2019}. Even though the framework
is instantiated at a level of abstraction independent of any particular implementation, the use-case we have in mind is one in which the models are (eventually)
run on a quantum computer, exploiting the potential advantages such computers may bring. Here the implementation is a classical simulation of a quantum
computation.\footnote{Note that we are not making any claims of ``quantum supremacy" \shortcite{preskill} for the particular set of quantum models that we implement in this article. However, we do anticipate the possibility of quantum models of concepts satisfying our framework which require quantum hardware for their efficient training and deployment, especially as we scale to more realistic datasets and larger quantum circuits.} 

We demonstrate how the training of the hybrid network produces conceptual representations in the Hilbert space which are neatly separated within the domains. We also show how discarding---which produces mixed
effects---can be used when the concept to be learned only applies to a subset of
the domains, and how entanglement (together with discarding) can be used to
capture interesting correlations across domains.

What are some of the main reasons for applying the formalism of quantum
theory to the modelling of concepts? First, it provides an alternative, and interesting, mathematical structure to the convex structure of conceptual spaces (see Section~\ref{sec:Is-Quantum-Conceptual}). Second, this structure comes with features which are well-suited to modelling concepts, such as entanglement for capturing correlations, and partial orders for capturing conceptual hierarchies.\footnote{Section~\ref{sec:combinations} describes entanglement; we leave the use of partial orders in experiments for future work.} Third, the use of the tensor product for combining domains leads to machine learning models with different characteristics to those typically employed in concept learning, such as the Conceptual VAE (i.e. neural
networks which use direct sum as the monoidal product plus non-linearities to
capture interactions between features) \shortcite{quantum_ml_features,PhysRevLett.122.040504}. The advantages this may bring, especially with the advent of larger, fault-tolerant quantum computers in the future, is still being worked out by the quantum machine learning community, but the possibilities are intriguing at worst and transformational at best.

Note that, in this article, our goal is to set out a novel framework for concept modelling, and demonstrate empirically---with two very different implementations---how concepts can be learned in practice. Further work is required to demonstrate that the framework can be applied fruitfully to data from a psychology lab---which is one of the goals of quantum cognition \shortcite{pothos2013bbs}---and also to agents acting in (virtual) environments---one of the goals of agent-based AI \shortcite{interactive_agents}. Note also that no claims are being made here regarding the existence of quantum processes in the brain, only that some cognitive processes can be effectively modelled at an abstract level using the quantum formalism.

The rest of the article is structured as follows. Section~\ref{sec:conspaces} provides a thorough category-theoretic formalisation of our modelling framework, using the language of string diagrams to describe the structured models. Section~\ref{sec:classical_impl} then describes our first instantiation of the framework, which is a novel adapation of the variational autoencoder. This section also contains experiments showing how Gaussian concept representations can be learned from images of coloured shapes. Section~\ref{sec:quantum-models} then describes our quantum instantiation, as well as a hybrid implementation applied to the same image data. The hybrid network uses a CNN for the classical image processing and a parameterised quantum circuit for inducing the concept representations (as quantum effects). Finally, Sections~\ref{sec:related_work} and \ref{sec:further_work} describe related and future work.

\section{Formalising Conceptual Spaces} \label{sec:conspaces}

\Gardenfors' framework of \emph{conceptual spaces} \shortcite{gardenfors2004conceptual,gardenfors2014} models conceptual reasoning in both human and artificial cognition. The approach models cognition geometrically, using convex spaces factorised in terms of ``elementary" spaces called \emph{domains}. Examples include the domains of \textsc{colour}, \textsc{taste}, \textsc{sound}, and \textsc{time}. Concepts are represented as convex regions, or more generally as ``fuzzy" functions defined over the space. We begin with a brief formalisation of this framework. While many have been presented \shortcite{aisbett2001general,rickard2007reformulation,lewis2016hierarchical,bechberger2017thorough}, we draw on the categorical approaches \shortcite{bolt2019interacting,tull2021categorical} and the latter's treatment of fuzzy concepts.

\begin{definition} \label{def:convex-space}
A \emph{convex space} is a set $Z$ coming with operations which allow us to take convex combinations, in that for all $z_1,\dots,z_n \in Z$ and $p_1, \dots, p_n \in [0,1]$ with $\sum^n_{i=1} p_i = 1$, there is an element:
\[
\sum^n_{i=1} p_i \cdot z_i 
\]
These combinations satisfy natural axioms; for example iterated combinations are given by multiplying weights as the notation suggests (see \shortciteA{bolt2019interacting} for details). Additionally we here require that $Z$ forms a \emph{measurable space}, meaning it comes with a $\sigma$-algebra of measurable subsets $\Sigma_Z \subseteq \mathbb{P}(Z)$.  
\end{definition} 

\begin{definition} \label{def:consp}
A \emph{conceptual space} is a convex space $Z$ given as a subset of a product of convex spaces:
\[
Z \subseteq Z_1 \times \dots \times Z_n
\]
where the product is equipped with element-wise convex operations.  We call an element $z = (z_1,\dots, z_n) \in Z$ an \emph{instance} of the conceptual space, following \shortciteA{clark_concepts}.
\end{definition}

Any factor $Z_i$ can be considered a conceptual space itself, with each $z_i$ an instance. A conceptual space is often written as a product of domains, such as \textsc{colour} or \textsc{sound}. Each domain itself factorises as a (subset of a) product of \emph{dimensions}. For example, the \textsc{sound} domain has the dimensions of \textsc{pitch} and \textsc{volume}. Here we simply use the neutral term ``factor" to treat either dimensions or domains.

\begin{definition} \label{def:crisp-concept}
A \emph{crisp concept} in a conceptual space $Z$ is a measurable  subset $\crc \subseteq Z$ which is \emph{convex}, meaning it is closed under convex combinations. When $z \in \crc$ we say $z$ is an \emph{instance of \crc}.
\end{definition} 

Convexity means that any point lying ``in-between" two instances of a concept will again form an instance of the concept, and is justified by \Gardenfors{} using experimental evidence in the division of colour space, and the ease of learning convex regions \shortcite{gardenfors2004conceptual}.

More generally, it is natural to consider concepts $\cc$ which are graded or ``fuzzy", so that the degree $\cc(z)$ to which $z$ is an instance of a concept can take any value from $0$ (``not at all") to $1$ (``fully satisfied"). In \shortciteA{tull2021categorical} it is shown that to be well-behaved compositionally and also satisfy a natural generalisation of convexity known as ``quasi-concavity", the membership function should satisfy the following. 

\begin{definition} \label{def:fuzzy-concept}
A \emph{fuzzy concept} of $Z$ is a measurable function $\cc \colon Z \to [0,1]$ which is \emph{log-concave}: 
\begin{equation} \label{eq:lc}
\cc(p  z + (1-p) z') \geq \cc(z)^p\cc(z')^{1-p}
\end{equation}
for all $z,z' \in Z$ and $p \in [0,1]$. A \emph{prototypical instance} of $C$ is an instance $z$ with $\cc(z) = \max_{w \in Z} \cc(w)$.
\end{definition} 

The prototypical instances of a fuzzy concept always form a crisp concept, and conversely any crisp concept $P$ forms a fuzzy concept via its indicator function $\cc = 1_P$.

\begin{example} \label{ex:RnCS}
Any convex subset $Z \subseteq \mathbb{R}^d$ forms a conceptual space, taking $\Sigma_Z$ to be the Lebesgue measurable subsets. Thus any product $Z = Z_1 \times \dots \times Z_n$ of convex subsets $Z_i \subseteq \mathbb{R}^{d_i}$ forms a conceptual space. In general any product of fuzzy concepts yields a new one on any convex subset $Z \subseteq Z_1 \times \dots \times Z_n$ via:
\begin{equation} \label{eq:prod-concept-space}
\cc(z) = \prod^n_{i=1} \cc_i(z_i) 
\end{equation}
\end{example} 


The following example of a fuzzy concept will form the basis of our classical implementation of the framework in Section \ref{sec:classical_impl}. 

\begin{example} \label{ex:gaussian-fuzzy-concept} 
We may define a fuzzy concept on $Z=\mathbb{R}^n$ from any multivariate Gaussian 
with mean $\mu$ and covariance matrix $\Sigma$:
\begin{align} \label{eq:Gaussian} 
\cc(z;\mu,\Sigma) &= e^{-\frac{1}{2}(z - \mu)^{\mathsf{T}}\Sigma^{-1}(z -\mu)} \\ 
& = e^{\sum^n_{i=1}-\frac{1}{2 \sigma_i^2}(z_i - \mu_i)^2}
\end{align}  
In the second line we restrict to the case where $\Sigma$ is diagonal, with $i$-th diagonal entry $\sigma_i^2$. In this case $\cc$ is given as a product of one-dimensional Gaussians $\cc_i(z_i;\mu_i,\sigma_i^2)$ as in \eqref{eq:prod-concept-space}.
\end{example}


\begin{example} A simple \textsc{taste} domain from \shortciteA{bolt2019interacting}, left-hand below, is given as a convex subset of $\mathbb{R}^3$ generated by the points \emph{sweet, bitter, salt} and \emph{sour}. Highlighted in red is a crisp concept for \emph{sweet}. Right-hand below shows a fuzzy concept on $\mathbb{R}^2$ from \shortciteA{tull2021categorical}.  From a set of exemplars (white crosses) the convex closure is formed, yielding the crisp concept $P$ given by the inner triangle. A fuzzy concept is then defined by $C(x) = e^{-\frac{1}{2\sigma^2}d(x,P)^2}$ where $d_H(x,P) = \inf_{p \in P} d(x,p)$, where each point in $P$ is prototypical. 
\[
\includegraphics[scale=0.17]{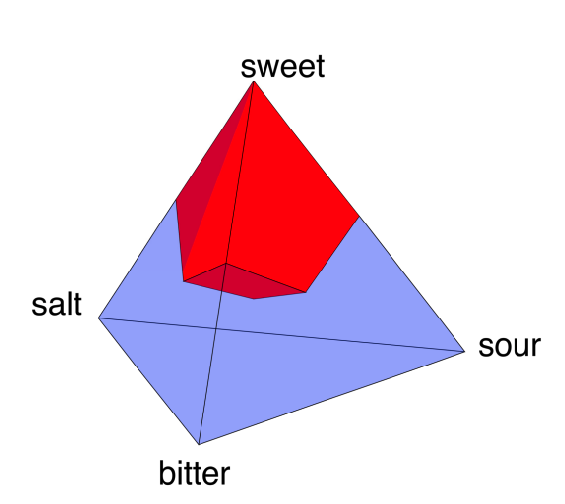}
\qquad 
\includegraphics[scale=0.3]{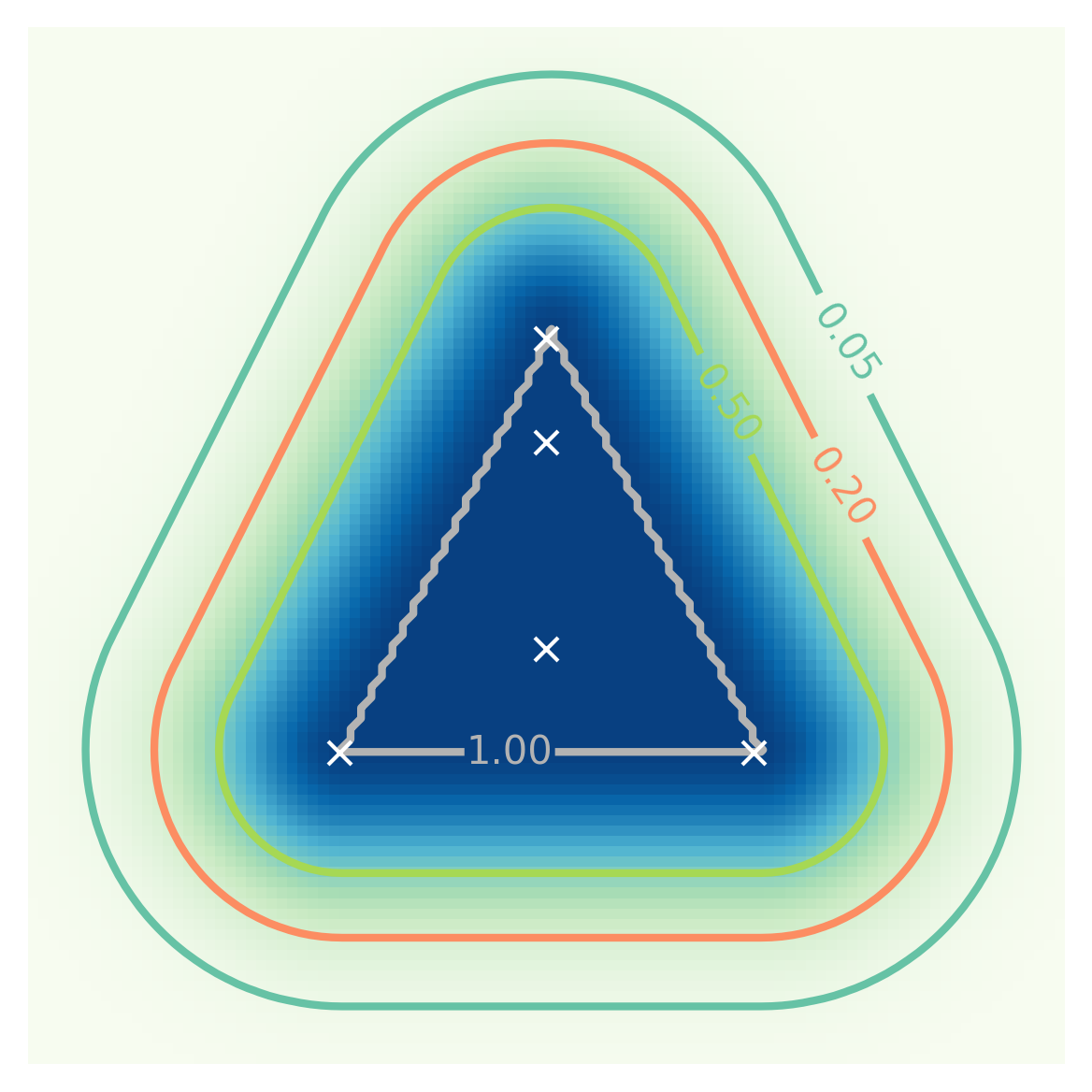}
\]
\end{example} 


\subsection{Categorical Setup} \label{sec:cats}

Our aim will now be to lift these basic notions from conceptual space theory into a general categorical framework, allowing us to pass them from the classical to the quantum setting in a principled manner. Here we introduce the categorical preliminaries. 

We will work in a \emph{symmetric monoidal category} $(\catC, \otimes ,I)$, making use of the graphical calculus \shortcite{selinger2010survey} in which objects are depicted as labelled wires, and morphisms $f \colon A \to B$ as boxes with input wire $A$ and output wire $B$, read here from bottom to top. 
Identities and sequential composition are depicted as follows. 
\[
\begin{tikzpicture}[tikzfig]
	\begin{pgfonlayer}{nodelayer}
		\node [style=none] (0) at (0.5, -1) {};
		\node [style=none] (1) at (0.5, 1) {};
		\node [style=map] (2) at (0.5, 0) {$\id{A}$};
		\node [style=none] (3) at (2, 0) {$=$};
		\node [style=none] (4) at (3.25, -1) {};
		\node [style=none] (5) at (3.25, 1) {};
		\node [style=label] (6) at (0.5, -1.5) {$A$};
		\node [style=label] (7) at (3.25, -1.5) {$A$};
		\node [style=label] (8) at (3.25, 1.25) {$A$};
		\node [style=label] (9) at (0.5, 1.25) {$A$};
	\end{pgfonlayer}
	\begin{pgfonlayer}{edgelayer}
		\draw (1.center) to (0.center);
		\draw (5.center) to (4.center);
	\end{pgfonlayer}
\end{tikzpicture} \qquad \qquad  \begin{tikzpicture}[tikzfig]
	\begin{pgfonlayer}{nodelayer}
		\node [style=none] (10) at (7, -1) {};
		\node [style=none] (11) at (7, 1) {};
		\node [style=map] (12) at (7, 0) {$g \circ f$};
		\node [style=none] (13) at (8.75, 0) {$=$};
		\node [style=none] (14) at (10.25, -2) {};
		\node [style=none] (15) at (10.25, 2) {};
		\node [style=label] (16) at (7, -1.5) {$A$};
		\node [style=label] (17) at (10.25, -2.5) {$A$};
		\node [style=label] (18) at (10.25, 2.25) {$C$};
		\node [style=label] (19) at (7, 1.25) {$C$};
		\node [style=map] (20) at (10.25, -1) {$f$};
		\node [style=map] (21) at (10.25, 1) {$g$};
		\node [style=label] (22) at (10.75, 0) {$B$};
	\end{pgfonlayer}
	\begin{pgfonlayer}{edgelayer}
		\draw (11.center) to (10.center);
		\draw (15.center) to (14.center);
	\end{pgfonlayer}
\end{tikzpicture}
\]
Parallel composition via the tensor $\otimes$ is given by drawing diagrams side-by-side.
\[
\begin{tikzpicture}[tikzfig]
	\begin{pgfonlayer}{nodelayer}
		\node [style=none] (23) at (15.25, -1) {};
		\node [style=none] (24) at (15.25, 1) {};
		\node [style=map] (25) at (15.25, 0) {$f \otimes g$};
		\node [style=label] (26) at (15.25, -1.5) {$A \otimes B$};
		\node [style=label] (27) at (15.25, 1.5) {$C \otimes D$};
		\node [style=none] (28) at (17, 0) {$=$};
		\node [style=none] (29) at (18.5, -1) {};
		\node [style=none] (30) at (18.5, 1) {};
		\node [style=map] (31) at (18.5, 0) {$f$};
		\node [style=label] (32) at (18.5, -1.5) {$A$};
		\node [style=label] (33) at (18.5, 1.25) {$C$};
		\node [style=none] (34) at (20, -1) {};
		\node [style=none] (35) at (20, 1) {};
		\node [style=map] (36) at (20, 0) {$g$};
		\node [style=label] (37) at (20, -1.5) {$B$};
		\node [style=label] (38) at (20, 1.25) {$D$};
	\end{pgfonlayer}
	\begin{pgfonlayer}{edgelayer}
		\draw (24.center) to (23.center);
		\draw (30.center) to (29.center);
		\draw (35.center) to (34.center);
	\end{pgfonlayer}
\end{tikzpicture}
\]
The (identity on the) monoidal unit $I$ is the empty diagram. Morphisms $\omega \colon I \to A$, $e \colon A \to I$ and $r \colon I \to I$ are called \emph{states}, \emph{effects} and \emph{scalars} respectively, depicted with no input, output or neither, respectively. 

Here we consider categories $\catC$ with further structure. First, each object  $A$
will come with a distinguished \emph{discarding} effect denoted $\discard{A}$, which we interpret as ``throwing the system away", with $\discard{I} = \id{I}$ and $\discard{X \otimes Y} = \discard{X} \otimes \discard{Y}$. A morphism $f$ is a \emph{channel} when it preserves discarding, as in left-hand below. A special case is that of a \emph{normalised} state $\omega$, right-hand below.  
\[
\begin{tikzpicture}[tikzfig]
	\begin{pgfonlayer}{nodelayer}
		\node [style=map] (0) at (0.5, 0) {$f$};
		\node [style=upground] (1) at (0.5, 1.5) {};
		\node [style=none] (2) at (0.5, -1) {};
		\node [style=none] (3) at (2, 0) {$=$};
		\node [style=upground] (4) at (3.5, 0.25) {};
		\node [style=none] (5) at (3.5, -1) {};
		\node [style=label] (6) at (-0.25, 0.75) {$B$};
		\node [style=label] (7) at (0.5, -1.5) {$A$};
		\node [style=label] (8) at (3.5, -1.5) {$A$};
	\end{pgfonlayer}
	\begin{pgfonlayer}{edgelayer}
		\draw (1) to (0);
		\draw (2.center) to (0);
		\draw (5.center) to (4);
	\end{pgfonlayer}
\end{tikzpicture}
\qquad 
\qquad 
\qquad 
\begin{tikzpicture}[tikzfig]
	\begin{pgfonlayer}{nodelayer}
		\node [style=map] (0) at (0.5, -0.5) {$\omega$};
		\node [style=upground] (1) at (0.5, 1.25) {};
		\node [style=none] (2) at (2, 0) {$=$};
		\node [style=none] (3) at (3.25, 0) {$1$};
		\node [style=label] (4) at (-0.25, 0.5) {$A$};
	\end{pgfonlayer}
	\begin{pgfonlayer}{edgelayer}
		\draw (1) to (0);
	\end{pgfonlayer}
\end{tikzpicture}
\]
We also assume $\catC$ is enriched in partial orders, so that each homset $\catC(A,B)$ forms a partially ordered set $\leq$, respected by composition. This allows us to generalise inclusions of convex subsets via the following, related to ``comprehensions" \shortcite{introeffectus} and ``compression" maps in quantum reconstructions \shortcite[Chap. 4]{tull2019categorical}. 

\begin{definition} 
A \emph{projection} is a morphism $p \colon A \to A$ with $\discard{} \circ p \leq \discard{}$ and such that for all morphisms $f$ with output (resp. input) $A$ we have: 
\begin{enumerate}
  \item $\discard{} \circ f \leq \discard{} \circ p \implies f = f \circ p$; 
  \item $\discard{} \circ f \leq \discard{} \circ p \circ f \implies f = p \circ f$.
\end{enumerate} 
It follows that $p = p \circ p$. An \emph{embedding} of an object $A$ into $B$ is given by a channel $e \colon A \to B$ and morphism $e^\dagger \colon B \to A$, depicted using triangles below, such that $e^\dagger \circ e = \id{A}$ and $p = e \circ e^\dagger$ is a projection. 
\[
\begin{tikzpicture}
	\begin{pgfonlayer}{nodelayer}
		\node [style=decomp] (0) at (0, -0.75) {};
		\node [style=decompflip] (1) at (0, 0.75) {};
		\node [style=none] (2) at (0, -1.75) {};
		\node [style=none] (3) at (0, 0) {};
		\node [style=label] (4) at (0, -2.25) {$A$};
		\node [style=label] (6) at (0, 2.25) {$A$};
		\node [style=label] (7) at (-1.25, 0) {$B$};
		\node [style=none] (8) at (0, 1.75) {};
		\node [style=none] (9) at (0, 0) {};
		\node [style=none] (10) at (1.75, 0) {$=$};
		\node [style=none] (11) at (3.5, -1.75) {};
		\node [style=none] (12) at (3.5, 1.75) {};
		\node [style=label] (13) at (3.5, -2.25) {$A$};
		\node [style=label] (14) at (3.5, 2.25) {$A$};
	\end{pgfonlayer}
	\begin{pgfonlayer}{edgelayer}
		\draw (3.center) to (2.center);
		\draw (8.center) to (9.center);
		\draw (12.center) to (11.center);
	\end{pgfonlayer}
\end{tikzpicture} \qquad \qquad 
\begin{tikzpicture}[tikzfig]
  \begin{pgfonlayer}{nodelayer}
    \node [style=decompflip] (0) at (0, -0.75) {$e$};
    \node [style=decomp] (1) at (0, 0.75) {$e$};
    \node [style=none] (2) at (0, -1.75) {};
    \node [style=none] (3) at (0, 1.75) {};
    \node [style=label] (4) at (-0.75, 0) {$A$};
    \node [style=label] (5) at (0, 2.25) {$B$};
    \node [style=label] (6) at (0, -2.25) {$B$};
  \end{pgfonlayer}
  \begin{pgfonlayer}{edgelayer}
    \draw (3.center) to (2.center);
  \end{pgfonlayer}
\end{tikzpicture}
\text{       is a projection }
\]
We often call the morphism $e$ the \emph{embedding} and $e^\dagger$ the \emph{projection} of the pair. 
\end{definition}


Any channel which is an isomorphism $A \simeq B$ forms a special case of an embedding, where $e^\dagger = e^{-1}$. Another important special case is an embedding of $I$ into $A$, which we call a \emph{point} of $A$.\footnote{Later we will define instances as special cases of points. Instances and points differ in quantum models, because of entanglement, but coincide classically.} By definition it includes a normalised state $\psi$ with an effect $\psi^\dagger \leq \discard{}$ satisfying:
\begin{equation} \label{eq:point}
\begin{tikzpicture}
	\begin{pgfonlayer}{nodelayer}
		\node [style=decomp] (0) at (0, -1) {$\psi$};
		\node [style=decompflip] (1) at (0, 1) {$\psi$};
		\node [style=none] (2) at (0, -1.25) {};
		\node [style=none] (3) at (0, 0) {};
		\node [style=none] (8) at (0, 1) {};
		\node [style=none] (9) at (0, 0) {};
		\node [style=none] (10) at (2, 0) {$=$};
		\node [style=none] (11) at (3.75, 0) {$1$};
	\end{pgfonlayer}
	\begin{pgfonlayer}{edgelayer}
		\draw (3.center) to (2.center);
		\draw (8.center) to (9.center);
	\end{pgfonlayer}
\end{tikzpicture}
\end{equation}



Embeddings are always closed under composition in the following sense. 

\begin{lemma} \label{lem:embed-subcat}
If $d \colon A \to B$ and $e \colon B \to C$ are embeddings then so is $e \circ d \colon A \to C$, with projection $d^\dagger \circ e^\dagger$.
\end{lemma}

\subsection{Conceptual Models} \label{sec:cat-con-spaces}

Let us now see how each of our earlier features from conceptual space theory can be described in a general category $\catC$ with the structure outlined in Section \ref{sec:cats}. Firstly, monoidal categories immediately allow us to describe the compositions of factors $Z_i$ appearing in a conceptual space, as follows. 

\begin{definition} \label{Def:conceptual-model}
A \emph{conceptual  model}\footnote{Henceforth we use the generic term ``model" rather than ``space" since a conceptual model can be defined in a category without any spatial character.} is given by an object $Z$ along with an indexed collection of objects $Z_1,\dots, Z_n$, called the \emph{factors}, and an embedding of $Z$ into $Z_1 \otimes \dots \otimes Z_n$.
\[
\begin{tikzpicture}
	\begin{pgfonlayer}{nodelayer}
		\node [style=decompwide] (0) at (0, 0) {};
		\node [style=none] (1) at (0, -1.25) {};
		\node [style=none] (2) at (-1, 0.25) {};
		\node [style=none] (3) at (1, 0.25) {};
		\node [style=none] (4) at (-1, 1.25) {};
		\node [style=none] (5) at (1, 1.25) {};
		\node [style=none] (6) at (0, 0.75) {$\dots$};
		\node [style=label] (7) at (-1, 1.75) {$Z_1$};
		\node [style=label] (8) at (1, 1.75) {$Z_n$};
		\node [style=label] (9) at (0, -1.75) {$Z$};
	\end{pgfonlayer}
	\begin{pgfonlayer}{edgelayer}
		\draw (5.center) to (3.center);
		\draw (4.center) to (2.center);
		\draw (1.center) to (0);
	\end{pgfonlayer}
\end{tikzpicture}
\]
\end{definition}

For simplicity we refer to a model as $Z$, with the factors and embedding implicit. Often the embedding is an isomorphism $Z \simeq Z_1 \otimes \dots \otimes Z_n$ exhibiting $Z$ as a product of the factors.

\begin{definition} 
A \emph{concept} of a conceptual model $Z$ is an effect $C$ on $Z$. 
\[
\begin{tikzpicture}
	\begin{pgfonlayer}{nodelayer}
		\node [style=map] (0) at (0, 0.75) {$C$};
		\node [style=none] (1) at (0, -0.25) {};
		\node [style=label] (2) at (0, -0.75) {$Z$};
	\end{pgfonlayer}
	\begin{pgfonlayer}{edgelayer}
		\draw (1.center) to (0);
	\end{pgfonlayer}
\end{tikzpicture}

\]
An \emph{instance} is a point $z$ of $Z$ which forms a product of points $z_i$ of the factors $Z_i$, as below:
\begin{equation} \label{eq:instance-factors}
\begin{tikzpicture}
	\begin{pgfonlayer}{nodelayer}
		\node [style=decompwide] (0) at (1.5, 0) {};
		\node [style=none] (1) at (1.5, -1.25) {};
		\node [style=none] (2) at (0.5, 0.25) {};
		\node [style=none] (3) at (2.5, 0.25) {};
		\node [style=none] (4) at (0.5, 1.25) {};
		\node [style=none] (5) at (2.5, 1.25) {};
		\node [style=none] (6) at (1.5, 0.75) {$\dots$};
		\node [style=label] (7) at (0.5, 1.75) {$Z_1$};
		\node [style=label] (8) at (2.5, 1.75) {$Z_n$};
		\node [style=label] (9) at (4.75, 0) {$=$};
		\node [style=point] (10) at (1.5, -1.5) {$z$};
		\node [style=none] (12) at (7, -0.25) {};
		\node [style=none] (13) at (9, -0.25) {};
		\node [style=none] (14) at (7, 0.75) {};
		\node [style=none] (15) at (9, 0.75) {};
		\node [style=none] (16) at (8, 0.25) {$\dots$};
		\node [style=label] (17) at (7, 1.25) {$Z_1$};
		\node [style=label] (18) at (9, 1.25) {$Z_n$};
		\node [style=point] (19) at (7, -0.5) {$z_1$};
		\node [style=point] (20) at (9, -0.5) {$z_n$};
		\node [style=label] (21) at (0.75, -0.75) {$Z$};
	\end{pgfonlayer}
	\begin{pgfonlayer}{edgelayer}
		\draw (5.center) to (3.center);
		\draw (4.center) to (2.center);
		\draw (1.center) to (0);
		\draw (15.center) to (13.center);
		\draw (14.center) to (12.center);
	\end{pgfonlayer}
\end{tikzpicture}
\end{equation}


\end{definition}

The order structure on morphisms means that the concepts are automatically partially ordered. We interpret $C \leq D$ as stating that  $D$ is a ``more general" concept than $C$. The factorisation property \eqref{eq:instance-factors} generalises the fact that in a conceptual space every instance $z=(z_1,\dots,z_n)$ factors as a product of one instance $z_i$ per factor $Z_i$. Composing a concept $C$ with any input state, in particular any instance $z$, will yield a scalar. For an instance we interpret this as specifying how well the instance fits the concept:
\[
\begin{tikzpicture}
	\begin{pgfonlayer}{nodelayer}
		\node [style=point] (0) at (0, -0.75) {$z$};
		\node [style=map] (1) at (0, 1.25) {$C$};
		\node [style=label] (2) at (-0.5, 0.25) {$Z$};
	\end{pgfonlayer}
	\begin{pgfonlayer}{edgelayer}
		\draw (1) to (0);
	\end{pgfonlayer}
\end{tikzpicture}

\]
We say that an instance $z$ is \emph{prototypical} for a concept $C$ when $C \circ w  \leq C \circ z$ for all instances $w$. It remains for us to identify those concepts which are crisp. 

\begin{definition} 
A concept $C$ on $Z$ is \emph{crisp} when it is of the form
\[
\begin{tikzpicture}
	\begin{pgfonlayer}{nodelayer}
		\node [style=decompflip] (0) at (0, 0) {};
		\node [style=none] (1) at (0, -1) {};
		\node [style=upground] (2) at (0, 1.25) {};
		\node [style=label] (3) at (-0.75, 0.75) {$K$};
		\node [style=label] (4) at (0, -1.5) {$Z$};
		\node [style=map] (5) at (-4, 0.5) {$C$};
		\node [style=none] (6) at (-4, -0.75) {};
		\node [style=none] (7) at (-2, 0) {$=$};
		\node [style=label] (8) at (-4, -1.25) {$Z$};
	\end{pgfonlayer}
	\begin{pgfonlayer}{edgelayer}
		\draw (2) to (1.center);
		\draw (6.center) to (5);
	\end{pgfonlayer}
\end{tikzpicture}
\]
for some projection morphism $Z \to K$ induced by an embedding of $K$ into $Z$. When the projection is given by a point of $Z$ we call $C$ a \emph{pure} concept. 
\end{definition} 

By definition each crisp concept has $C \leq \discard{}$. Intuitively we can identify the crisp concept with object $K$ via its embedding $e$. Indeed by the definition of an embedding, for any instance $z$ of $Z$ we have $C \circ z = 1$ iff $z=e \circ k$ for some point $k$ of $K$. Moreover any concept $D$ with $D \leq C$ restricts to $K$ in that $D =E \circ e^\dagger$ for some effect $E$ on $K$.
A pure concept can be thought of as a ``maximally sharp" concept, being of the form $z=\psi^\dagger$ as in \eqref{eq:point} where $z = \psi$ is in fact a point of $Z$.

\subsection{Classical Conceptual Models} \label{sec:class-models}

Let us now meet our main classical examples of categories and their notions of conceptual model. 

\paragraph{$\Class$: Discrete probability.}

In the category $\Class$ the objects are finite sets and the morphisms $M \colon X \to Y$ are matrices $(M(y \mid x))_{x \in X, y \in Y}$ with values in $\mathbb{R}^+$. Composition is matrix multiplication:
\[
(N \circ M)(x,z) := \sum_{y \in Y} N(z \mid y) M(y \mid x)
\]
Identity morphisms satisfy $\id{X}(y \mid x) = \delta_{x,y}$. $X \otimes Y = X \times Y$, with  $I=\{\star\}$ the singleton set, and $M \otimes N$ the Kronecker product of matrices. We can equate states $\omega$ and effects $e$ of $X$ each with functions $X \to \mathbb{R}^+$. In particular, scalars correspond to positive reals $s \in \mathbb{R}^+$. $\discard{}$ is the function $x \mapsto 1$ for all $x \in X$. 

A state $\omega$ of $X$ is normalised iff it describes a probability distribution, with $\sum_{x \in X} \omega(x) = 1$. More generally, a morphism $M \colon X \to Y$ is a channel iff it is a finite probability channel (\emph{Stochastic} matrix) with $\sum_{y \in Y} M(y \mid x) = 1$ for each $x \in X$. $\leq$ is the element-wise ordering from $\mathbb{R}^+$. The points of $X$ are precisely the point distributions $\delta_x$ for $x \in X$. An embedding $X \hookrightarrow Y$ is given by an inclusion of a subset $X \subseteq Y$ via $x \mapsto \delta_x$, and its projection $Y \to X$ is given by $y \mapsto \delta_y$ when $y \in X$ and $y \mapsto 0$ otherwise.

A conceptual model in $\Class$ is thus a finite set $Z$ given as a subset $Z \subseteq Z_1 \times \dots \times Z_n$. A concept is an arbitrary function $C \colon Z \to \mathbb{R}^+$, ordered point-wise. An instance is any element $z=(z_1,\dots,z_n) \in Z$, with \eqref{eq:instance-factors} holding automatically. Applying a concept $C$ to an instance $z$ gives $C(z) \in \mathbb{R}^+$. Crisp concepts are the indicator functions $1_K$ of arbitrary subsets $K \subseteq Z$, while pure concepts are those of instances $z \in Z$. 

\paragraph{$\Prob$: Measure-theoretic probability.}

In $\Prob$ the objects are measurable spaces $(X, \Sigma_X)$. A morphism $f \colon X \to Y$ is a \emph{Markov (sub)kernel}, a function sending each $x \in X$ to a sub-probability measure $f(x)$ over $Y$, in a ``measurable" way \shortcite{panangaden1998probabilistic,cho2019disintegration}. Composition of $f \colon X \to Y$ and  $g \colon Y \kto Z$ is via integration:
 \[
(g \circ f)(x,A) := \int_{y \in Y} g(y,A) df(x)(y)
\] 
for each $x \in X, A \in \Sigma_Z$. The identity sends each $x$ to the point measure $\delta_x$. We set $X \otimes Y = X \times Y$, with $I$ being the singleton set, and define $f \otimes g$ to send each pair $(x,y)$ to the \emph{product measure} of the measures $f(x)$ and $g(y)$. States of $X$ may be identified with sub-probability measures $\omega$ over $X$, and are normalised iff they form a probability measure, with $\omega(X)=1$. Effects correspond to measurable functions $e \colon X \to [0,1]$. $\discard{}$ is the constant function at $1$. Scalars are probabilities $p \in [0,1]$. Composing a state with an effect yields the expectation value $e \circ \omega = \int_{x \in X} e(x) d\omega(x) \in \mathbb{R}^+$.

A morphism  $f \colon X \to Y$ is a channel iff it sends each $x \in X$ to a probability measure. Then $f \leq g$ whenever $f(x,A) \leq g(x,A)$ for all $x \in X, A \in \Sigma_Y$. An embedding $X \hookrightarrow Y$ is an inclusion of a subset $X \subseteq Y$ via $x \mapsto \delta_x$ for $x \in X$, with the projection $Y \to X$ given by $y \mapsto \delta_y$ when $y \in X$ and $y \mapsto 0$ otherwise.

A conceptual model in $\Prob$ is thus a measurable space given as a measurable subset $Z \subseteq Z_1 \times \dots \times Z_n$ of spaces $Z_i$. Concepts are measurable functions $C \colon Z \to [0,1]$, instances and pure concepts correspond to points $z \in Z$, crisp concepts $1_K$ correspond to arbitrary measurable subsets $K \subseteq Z$.

\paragraph{$\ConSp$: Conceptual spaces.}

The category $\ConSp$ \shortcite{tull2021categorical} is defined just like $\Prob$ except that the objects are now convex spaces and morphisms are (sub)kernels $f$ which are \emph{log-concave}, meaning that 
\begin{equation} \label{eq:log-conc-han}
f(p x + (1-p)y, p A + (1-p) B) \geq f(x,A)^p f(y,B)^{1-p}
\end{equation}
for all $p \in [0,1], x, y \in X$ and $A, B \in \Sigma_Y$. Here $X \otimes Y = X \times Y$ is the product of convex spaces, with element-wise convex operations. 

A conceptual model in $\ConSp$ is precisely a conceptual space, i.e. a convex space viewed as a convex subset $Z \subseteq Z_1 \times \dots \times Z_n$ of convex spaces $Z_i$. Instances are points $z \in Z$. Crisp concepts are precisely those of Definition \ref{def:crisp-concept}, namely the indicator functions $1_K$ of convex measurable subsets $K \subseteq Z$, with pure concepts being the indicator functions $1_z$ of points $z \in Z$. Concepts are fuzzy concepts $C \colon Z \to [0,1]$ in the sense of Definition \ref{def:fuzzy-concept}. 


\subsection{Quantum Conceptual Models} \label{sec:quantum-models2}

We can now define our quantum model of concepts inspired by the conceptual space framework. To do so we will simply unpack our definitions from Section \ref{sec:cat-con-spaces} in the following category of quantum processes. 

\paragraph{$\Quant{}$: Quantum processes.}
In the category $\Quant{}$ the objects are finite dimensional Hilbert spaces, and morphisms $f \colon \hilbH \to \hilbK$ are \emph{completely positive} (CP) maps $f \colon L(\hilbH) \to L(\hilbK)$, where $L(\hilbH)$ is the space of linear operators on $\hilbH$.  Such a map $f$ is linear, and such that for any $\hilbH'$ the map $g = f \otimes \id{\hilbH'}$ is \emph{positive} in that whenever $a$ is a positive operator then $g(a)$ is also. We set $f \leq g$ whenever $g -f$ is CP. 

Here $\otimes$ is the usual tensor of Hilbert spaces and linear maps, with $I = \mathbb{C}$. In particular, states $\omega$ and effects $e$ on $\hilbH$ may both be identified with positive operators $a \in L(\hilbH)$ via $a=\omega(1)$ and $e(b) = \Tr(ab)$, respectively, where $\Tr$ denotes the trace.  Scalars are again positive reals $r \in \mathbb{R}^+$.  Discarding is the functional $\discard{}(a) = \Tr(a)$, corresponding to the identity operator $\id{\hilbH}$. 

A morphism $f$ is a channel iff it is a completely positive trace-preserving (CPTP) map, with $\Tr(f(a)) = \Tr(a)$ for all $a \in L(\hilbH)$. A state $\rho$ is normalised precisely when it is a density matrix, with $\Tr(\rho) = 1$.

A special class of morphisms are the \emph{pure} CP maps $\hat f \colon L(\hilbH) \to L(\hilbK)$, given by $\hat f(a) = f \circ a \circ  f^\dagger$ for some linear map $f \colon \hilbH \to \hilbK$. All other morphisms are called \emph{mixed}. Embedding morphisms are the pure maps induced by inclusions $i \colon \hilbK \hookrightarrow \hilbH$ of subspaces into $\hilbH$. The corresponding projection is the pure map induced by the linear projection $i^\dagger$ onto $\hilbK$. A point of $\hilbH$ may be identified with a pure quantum state $\ket{\psi} \bra{\psi}$ for some unit vector $\psi \in \hilbH$.\footnote{Here we use the standard ``bra-ket" notation whereby vectors and linear functionals on $\hilbH$ are written in the form $\ket{\psi}$, $\bra{\phi}$ respectively. Then for a unit vector $\psi \in \hilbH$, $\ket{\psi}\bra{\psi}$ is the density operator of the corresponding pure state on $\hilbH$.} 

We now arrive at our quantum adaptation of the conceptual space framework.

\begin{definition} 
A \emph{quantum conceptual model} is a conceptual model in $\Quant{}$:
\[
\begin{tikzpicture}
	\begin{pgfonlayer}{nodelayer}
		\node [style=decompwide] (0) at (0, 0) {};
		\node [style=none] (1) at (0, -1.25) {};
		\node [style=none] (2) at (-1, 0.25) {};
		\node [style=none] (3) at (1, 0.25) {};
		\node [style=none] (4) at (-1, 1.25) {};
		\node [style=none] (5) at (1, 1.25) {};
		\node [style=none] (6) at (0, 0.75) {$\dots$};
		\node [style=label] (7) at (-1, 1.75) {$\hilbH_1$};
		\node [style=label] (8) at (1, 1.75) {$\hilbH_n$};
		\node [style=label] (9) at (0, -1.75) {$\hilbH$};
	\end{pgfonlayer}
	\begin{pgfonlayer}{edgelayer}
		\draw (5.center) to (3.center);
		\draw (4.center) to (2.center);
		\draw (1.center) to (0);
	\end{pgfonlayer}
\end{tikzpicture}
\]
\end{definition} 

Thus a quantum conceptual model is a Hilbert space $\hilbH$ given as a subspace of a tensor product of Hilbert spaces $\hilbH \subseteq \hilbH_1 \otimes \dots \otimes \hilbH_n$. A quantum concept is then precisely a quantum effect, i.e. a positive operator $C \in L(\hilbH)$, ordered via $C \leq D$ whenever $D -C$ is positive. An instance is a pure state $\ket{\psi}\bra{\psi}$ given by a unit vector $\psi \in \hilbH$, which furthermore factorises as 
\begin{equation} \label{eq:pureproduct}
\psi = \psi_1 \otimes \dots \otimes \psi_n
\end{equation}
for unit vectors $\psi_i \in \hilbH_i$, giving it a well-defined pure state value on each factor $\hilbH_i$. All instances are pure, with mixed states $\rho$ interpreted as states of uncertainty (i.e. probabilistic mixtures) over pure states such as instances. In contrast concepts may be mixed or pure. The application of a quantum concept $C$ to an instance $\psi$ is given by 
\[
\begin{tikzpicture}
	\begin{pgfonlayer}{nodelayer}
		\node [style=map] (0) at (0, 1) {$C$};
		\node [style=point] (1) at (0, -1) {$\psi$};
		\node [style=label] (2) at (-0.75, 0) {$\hilbH$};
	\end{pgfonlayer}
	\begin{pgfonlayer}{edgelayer}
		\draw (0) to (1);
	\end{pgfonlayer}
\end{tikzpicture}
 \ \  = \ \  \bra{\psi} C \ket{\psi} \in \mathbb{R}^+
\]
More generally applying $C$ to a mixed state $\rho$ yields $\Tr(C \rho) \in \mathbb{R}^+$.

Crisp concepts correspond to subspaces $\hilbK \subseteq \hilbH$. More precisely, any such subspace defines a crisp concept via the projection operator $P$ onto $\hilbK$ with $P(\psi) = \psi$ for $\psi$ in $\hilbK$ and $P(\psi) = 0$ for $\psi$ in $\hilbK^\bot$.

Pure quantum concepts are precisely those crisp quantum concepts which are themselves pure as effects. For these, $\hilbK$ is given by a one-dimensional subspace $\langle \psi \rangle$ spanned by some unit vector $\psi \in \hilbH$. Thus pure quantum concepts are precisely effects of the form $\ket{\psi}\bra{\psi}$ where $\psi$ is any unit vector (not necessarily an instance). Such a concept sends each instance $\phi$ to $|\braket{\psi | \phi}|^2 \in [0,1]$.

\begin{example} \label{ex:HSL}
A quantum conceptual model $\hilbH = \hilbH_{\mathrm{Hue}} \otimes \hilbH_\mathrm{Sat} \otimes \hilbH_{Light}$ for \textsc{colour} with factors \textsc{hue}, \textsc{saturation} and \textsc{lightness} is given in \shortciteA{yan2021qhsl}, where \textsc{hue} is encoded on a single qubit, represented on the Bloch sphere. In particular each instance (colour) is taken to be a tensor of pure states over each of the factors. 
\end{example}

We will meet further examples of quantum conceptual models in Section \ref{sec:quantum-models}. 


\subsection{Entangled Concepts} \label{sec:entangled-concepts}

It is natural to wonder what advantages, if any, quantum concepts might possess over classical ones. One feature distinguishing quantum models from classical ones is the presence of pure  \emph{entangled} concepts. For the following, we restrict to categories with scalars given by $\mathbb{R}^+$, as in all of our examples here. 

\begin{definition} \label{def:prod-sep}
A concept $C$ is a \emph{product} concept when there are effects $C_1, \dots, C_n$ such that 
\begin{equation} \label{eq:prod-concept}
\begin{tikzpicture}
	\begin{pgfonlayer}{nodelayer}
		\node [style=map] (0) at (-2.5, 0.75) {$C$};
		\node [style=none] (1) at (-2.5, 0.5) {};
		\node [style=none] (3) at (-2.5, -0.75) {};
		\node [style=label] (5) at (-2.5, -1.25) {$Z$};
		\node [style=none] (7) at (-0.5, 0) {$=$};
		\node [style=none] (9) at (1.75, 1.5) {};
		\node [style=none] (10) at (3.75, 1.5) {};
		\node [style=none] (11) at (1.75, 0.25) {};
		\node [style=none] (12) at (3.75, 0.25) {};
		\node [style=none] (13) at (2.75, 1) {$\dots$};
		\node [style=decompwide] (15) at (2.75, 0) {};
		\node [style=map] (16) at (1.75, 2) {$C_1$};
		\node [style=map] (17) at (3.75, 2) {$C_n$};
		\node [style=none] (18) at (2.75, -0.5) {};
		\node [style=none] (19) at (2.75, -1.5) {};
		\node [style=label] (20) at (2.75, -2) {$Z$};
		\node [style=label] (21) at (1, 0.75) {$Z_1$};
		\node [style=label] (22) at (4.5, 0.75) {$Z_n$};
	\end{pgfonlayer}
	\begin{pgfonlayer}{edgelayer}
		\draw (3.center) to (1.center);
		\draw (12.center) to (10.center);
		\draw (11.center) to (9.center);
		\draw (18.center) to (19.center);
	\end{pgfonlayer}
\end{tikzpicture}
\end{equation}
A concept $C$ is \emph{separable} when its value on instances is equal to a convex mixture of product concepts. That is, there are product concepts $C^{(1)},\dots,C^{(k)}$ such that $C \circ z = \sum^k_{j=1} C^{(j)} \circ z$ for all instances $z$, where the sum is taken in $\mathbb{R}^+$. If a concept $C$ is not separable we say that it is \emph{entangled}. 
\end{definition}

A product concept treats the factors independently, applying a fixed concept to each. Entangled concepts capture correlations between factors which cannot be reduced to any mixture over such product concepts. $\Class, \Prob$ and $\ConSp$ contain product concepts as well as separable (but non-product) concepts. Nonetheless in $\Class$ every concept is separable. However, these categories do not contain any pure entangled concepts, since every point of a model $Z \subseteq Z_1 \times \dots \times Z_n$ forms an instance $z=(z_1,\dots, z_n)$ and hence every pure concept is a product of pure effects $z_i^\dagger$ on each factor.

In contrast, quantum models $\hilbH$ contain both entangled and pure entangled concepts.  For any unit vector $\psi \in \hilbH$ which is entangled in the usual sense, i.e. not of the form \eqref{eq:pureproduct}, the point $\ket{\psi}\bra{\psi}$ is not an instance, and its corresponding pure concept on $\hilbH$ is entangled.

\begin{example} \label{Ex:Bell-effect}
Consider a Hilbert space $\hilbH$ with orthonormal basis $\{\ket{i}\}^{n-1}_{i=0}$. An  entangled pure concept on $\hilbH \otimes \hilbH$ is given by the \emph{Bell effect}, induced by the (unnormalised) vector $\sum^{n-1}_{i=0} \ket{i 
 \ i}$ (where sum denotes superposition), with operator 
$\sum^{n-1}_{i,j=0}\ket{i \ i}\bra{j \  j}$. 
\end{example}

\begin{remark} \label{rem:cts-entanglement}
The finite sum in Def. \ref{def:prod-sep} should ultimately be replaced with an integral, so that each concept in $\Prob$ is separable. It would be interesting to explore whether entanglement exists in $\ConSp$. 
\end{remark}

\subsection{Quantum and Classical Concept Combinations} \label{sec:combinations}

To compare classical and quantum concepts, and to demonstrate the role of entangled concepts in quantum models, let us now consider the ways in which we may ``combine" (crisp) concepts in each of our example categories. Given a collection of crisp concepts $(C_i)^n_{i=1}$, by a \emph{combination} we mean a new (crisp) concept $C$ such that every prototypical instance of one of the $C_i$ is a prototypical instance of $C$.\footnote{In this article ``combination" of concepts is always meant in this sense. However there are many distinct meaningful operations on concepts which could also be called their combination, such as the more conjunction-like notion of combining ``pet" and ``fish" into ``pet fish" \shortcite{aerts_gabora2005}. }

We will focus in particular on the natural scenario in which we are given a model $Z$ and wish to combine (the pure concepts induced by) a collection of instances $z_1,\dots,z_n$. The result is a concept $C$ with the $z_1, \dots, z_n$ as prototypical instances, which we think of as learned from these exemplars. 

A starting point is to observe that crisp concepts in each category are closed under intersections $\bigcap_{i \in I} C_i$ (of arbitrary, measurable, convex, linear subsets respectively). They hence form a complete lattice with top element $ \ \discard{}$ (and so may be viewed as a \emph{Formal Concept Lattice} in the sense of \shortciteA{ganter1999formal}). This means that one way to combine crisp concepts is via their \emph{disjunction} or least upper bound   
$C = \bigvee_{i \in I} C_i$. 

\paragraph{Classical combinations} 
  In $\Class$ and $\Prob$, the disjunction is given by the union of subsets $C_i$. In fact this is seemingly the only natural way to combine concepts. Indeed here any crisp concept may be identified with its set of prototypical instances, so that any combination $C$ satisfies $\bigvee^n_{i=1}C_i \leq C$. In particular the classical combination $z_1^\dagger \vee \dots \vee z_n^\dagger$ of instances $z_1,\dots,z_n$ is the subset $\{z_1,\dots,z_n\}$. 

\paragraph{Spatial combinations} 
In $\ConSp$ the disjunction is given by the \emph{convex closure} $C=\mathrm{Conv}(\bigcup_{i \in I} C_i)$  of the convex subsets $C_i$,  the smallest convex subset containing all of them. Again any combination $C$ has $\bigvee^n_{i=1}C_i \leq C$. The spatial combination of $z_1,\dots,z_n$ now includes any convex combination of them.

\begin{example} \label{ex:banana-running}
Consider a model with factors $C=\textsc{colour}$ and $T=\textsc{taste}$ and a concept $B$ for \emph{banana} which combines two instances: a yellow $(Y)$ sweet $(S)$ banana, and a green $(G)$ bitter $(B)$ banana.
\[
\begin{tikzpicture}[tikzfig]
	\begin{pgfonlayer}{nodelayer}
		\node [style=point] (1) at (1, -0.75) {$Y$};
		\node [style=point] (2) at (2.5, -0.75) {$S$};
		\node [style=none] (3) at (1, 0.75) {};
		\node [style=none] (4) at (2.5, 0.75) {};
		\node [style=point] (6) at (8.75, -0.75) {$G$};
		\node [style=point] (7) at (10.25, -0.75) {$B$};
		\node [style=none] (8) at (8.75, 0.75) {};
		\node [style=none] (9) at (10.25, 0.75) {};
		\node [style=medium map] (10) at (1.75, 1.25) {$B$};
		\node [style=medium map] (11) at (9.5, 1.25) {$B$};
		\node [style=none] (12) at (4.5, 0) {$=$};
		\node [style=none] (13) at (7, 0) {$=$};
		\node [style=none] (14) at (5.75, 0) {$1$};
		\node [style=label] (15) at (0.5, 0) {$C$};
		\node [style=label] (16) at (3, 0) {$T$};
		\node [style=label] (17) at (10.75, 0) {$T$};
		\node [style=label] (18) at (8.25, 0) {$C$};
	\end{pgfonlayer}
	\begin{pgfonlayer}{edgelayer}
		\draw (3.center) to (1);
		\draw (4.center) to (2);
		\draw (8.center) to (6);
		\draw (9.center) to (7);
	\end{pgfonlayer}
\end{tikzpicture}
\]
For simplicity, suppose yellow and green are ``orthogonal" in that $Y^\dagger \circ G = 0$. The classical combination yields the crisp concept whose only points are the two instances $(Y,S), (G,B)$ themselves, which by orthogonality can be equivalently written as a sum using element-wise addition of matrices in $\Class$:
\begin{equation} \label{eq:classical-comb}
\begin{tikzpicture}[tikzfig]
	\begin{pgfonlayer}{nodelayer}
		\node [style=medium map] (0) at (0.75, 0.75) {$D$};
		\node [style=none] (1) at (0, 0.25) {};
		\node [style=none] (2) at (1.5, 0.25) {};
		\node [style=none] (3) at (0, -0.5) {};
		\node [style=none] (4) at (1.5, -0.5) {};
		\node [style=label] (5) at (0, -1) {$C$};
		\node [style=label] (6) at (1.5, -1) {$T$};
		\node [style=none] (7) at (2.75, 0) {$=$};
		\node [style=copoint] (8) at (4.25, 0.5) {$Y$};
		\node [style=copoint] (9) at (5.5, 0.5) {$S$};
		\node [style=copoint] (10) at (8, 0.5) {$G$};
		\node [style=copoint] (11) at (9.25, 0.5) {$B$};
		\node [style=none] (12) at (4.25, -0.5) {};
		\node [style=none] (13) at (5.5, -0.5) {};
		\node [style=none] (14) at (8, -0.5) {};
		\node [style=none] (15) at (9.25, -0.5) {};
		\node [style=label] (16) at (4.25, -1) {$C$};
		\node [style=label] (17) at (5.5, -1) {$T$};
		\node [style=label] (18) at (8, -1) {$C$};
		\node [style=label] (19) at (9.25, -1) {$T$};
		\node [style=none] (20) at (6.75, 0) {$+$};
	\end{pgfonlayer}
	\begin{pgfonlayer}{edgelayer}
		\draw (4.center) to (2.center);
		\draw (3.center) to (1.center);
		\draw (15.center) to (11);
		\draw (14.center) to (10);
		\draw (13.center) to (9);
		\draw (12.center) to (8);
	\end{pgfonlayer}
\end{tikzpicture} 
\end{equation}
The classical combination is depicted left-hand below. The spatial combination instead corresponds to the line connecting the two points (right-hand below).
\begin{equation} \label{eq:class-and-spatial}
\begin{tikzpicture}
	\begin{pgfonlayer}{nodelayer}
		\node [style=none] (0) at (-4, -2) {};
		\node [style=none] (1) at (-3.75, 2) {};
		\node [style=none] (2) at (1, -2) {};
		\node [style=none] (3) at (-3.75, -2.5) {};
		\node [style=none] (4) at (-3, -1.25) {\ding{53}};
		\node [style=none] (5) at (0, 1.25) {\ding{53}};
		\node [style=label] (6) at (0, 0.5) {y, s};
		\node [style=label] (7) at (-3, -0.5) {g, b};
		\node [style=label] (8) at (-3.75, 2.5) {\text{Colour}};
		\node [style=label] (9) at (0, -2.5) {\text{Taste}};
		\node [style=none] (10) at (-3, 1.25) {};
		\node [style=none] (11) at (0, -1.25) {};
	\end{pgfonlayer}
	\begin{pgfonlayer}{edgelayer}
		\draw (0.center) to (2.center);
		\draw (3.center) to (1.center);
	\end{pgfonlayer}
\end{tikzpicture}
\qquad 
\qquad 
\qquad 
\begin{tikzpicture}
	\begin{pgfonlayer}{nodelayer}
		\node [style=none] (0) at (-4, -2) {};
		\node [style=none] (1) at (-3.75, 2) {};
		\node [style=none] (2) at (1, -2) {};
		\node [style=none] (3) at (-3.75, -2.5) {};
		\node [style=none] (4) at (-3, -1.25) {\ding{53}};
		\node [style=none] (5) at (0, 1.25) {\ding{53}};
		\node [style=label] (6) at (0, 0.5) {y, s};
		\node [style=label] (7) at (-3, -0.5) {g, b};
		\node [style=label] (8) at (-3.75, 2.5) {\text{Colour}};
		\node [style=label] (9) at (0, -2.5) {\text{Taste}};
		\node [style=none] (10) at (-3, 1.25) {};
		\node [style=none] (11) at (0, -1.25) {};
	\end{pgfonlayer}
	\begin{pgfonlayer}{edgelayer}
		\draw (0.center) to (2.center);
		\draw (3.center) to (1.center);
		\draw (4.center) to (5.center);
	\end{pgfonlayer}
\end{tikzpicture}
\end{equation}
\end{example} 

\paragraph{Quantum combinations}

In $\Quant{}$, the disjunction is given by the linear closure $C=\mathrm{Lin}(\bigcup_{i \in I} C_i)$ of all the subspaces $C_i$, the smallest subspace containing all of them. This yields a mixed quantum concept which we may interpret as their ``coarse-graining", and again refer to as their \emph{classical combination}. Crucially, however, in a quantum conceptual model  there are in fact \emph{many} possible ways to combine crisp concepts, aside from the disjunction, even into a pure concept. That is, there are combinations $C$ of the crisp concepts $C_i$ which do not satisfy $\bigvee_i C_i \leq C$. 

\begin{definition} 
In a quantum conceptual model, by a \emph{quantum combination} of instances $\psi_1,\dots,\psi_n$ we mean a pure concept $C=\phi^\dagger$ with these instances as prototypical. 
\end{definition} 

The presence of quantum combinations is closely related to entanglement, coming from the fact that instances are only a subset of the points in a quantum model, since they are non-entangled. Indeed any quantum combination of two or more instances will be entangled.

\begin{example} 
The Bell effect in Example \ref{Ex:Bell-effect} is a pure concept with prototypical instances being precisely those of the form $\ket{\psi^*} \otimes \ket{\psi}$ for unit vectors $\ket{\psi}$, where $\ket{\psi^*}$ denotes the conjugate vector with respect to the given basis. Thus it forms a pure quantum combination of any such instances.
\end{example}

\begin{example}
Consider again the setting of the \emph{banana} concept combination from Example \ref{ex:banana-running}. In $\Quant$ we can form the classical combination of instances which is again of the form \eqref{eq:classical-comb}, where $+$ is now the sum of CP maps. Alternatively, we may form a  a quantum combination $\ket{\psi}\bra{\psi}$ where: 
\begin{equation} \label{eq:psi-concept}
\psi = \ket{Y,S} + \ket{G,B} \in C \otimes T
\end{equation}
More generally any linear map $f \colon C \to T$ such that $f(\ket{Y}) = \ket{S}$, $f(\ket{G}) = \ket{B}$ defines a suitable entangled concept $E = \tinycap \circ (f \otimes \id{})$, where $\tinycap$ denotes the Bell Effect from Example \ref{Ex:Bell-effect}. 
Consider the case where $C = T = \mathbb{C}^2$, $\ket{Y} = \ket{S} = \ket{0}$ and $\ket{G} = \ket{B} = \ket{1}$.  A quantum combination $E$ is now given by the Bell effect. The classical and quantum combinations $D, E$ act on instances as follows:
\begin{align*} 
\begin{tikzpicture}[tikzfig]
	\begin{pgfonlayer}{nodelayer}
		\node [style=medium map] (0) at (1, 0.75) {$D$};
		\node [style=none] (1) at (0.25, 0.25) {};
		\node [style=none] (2) at (1.75, 0.25) {};
		\node [style=none] (3) at (0.25, -0.5) {};
		\node [style=none] (4) at (1.75, -0.5) {};
		\node [style=point] (15) at (0.25, -0.75) {$\psi$};
		\node [style=point] (16) at (1.75, -0.75) {$\phi$};
	\end{pgfonlayer}
	\begin{pgfonlayer}{edgelayer}
		\draw (4.center) to (2.center);
		\draw (3.center) to (1.center);
	\end{pgfonlayer}
\end{tikzpicture} 
 \ \ 
 & =  \ \  \sum^{1}_{i=0} |\braket{i|\psi}|^2 |\braket{i|\phi}|^2
\\
\begin{tikzpicture}[tikzfig]
  \begin{pgfonlayer}{nodelayer}
    \node [style=medium map] (4) at (1.25, 0.75) {$E$};
    \node [style=none] (5) at (0.5, 0.25) {};
    \node [style=none] (6) at (2, 0.25) {};
    \node [style=none] (7) at (0.5, -0.75) {};
    \node [style=none] (8) at (2, -0.75) {};
    \node [style=point] (9) at (0.5, -0.75) {$\psi$};
    \node [style=point] (10) at (2, -0.75) {$\phi$};
  \end{pgfonlayer}
  \begin{pgfonlayer}{edgelayer}
    \draw (8.center) to (6.center);
    \draw (5.center) to (7.center);
  \end{pgfonlayer}
\end{tikzpicture}
  \ \ &= \ \  |\braket{\psi^* \mid \phi}|^2 
\end{align*}

The classical combination $D$ simply compares any input to the two instances, with no further prototypical instances besides those given. As a result the structure of each space ``between" $\ket{0}$ and $\ket{1}$ is lost, with the orthogonal states  $\ket{\pm} = \frac{1}{\sqrt{2}}(\ket{0} \pm \ket{1})$ treated identically and ${D(\ket{+} \otimes \ket{-})} = \frac{1}{2}$. In contrast the quantum combination $E$ can be seen to encode a structural relationship between the factors, generalising from $\ket{00}, \ket{11}$. Any instance $\ket{\phi^*} \otimes \ket{\phi}$ is prototypical, and conversely, tensors of (conjugate) orthogonal points will not fit the concept, e.g. $E(\ket{+} \otimes \ket{-}) = 0$. 
\end{example}

In the above example we see that entangled quantum concept combinations can encode relationships between factors, rather than simply (weighted) collections of exemplars. Indeed any pure entangled concept on $C \otimes T$ corresponds to a pure linear map $f \colon C \to T$. We can understand this as a generalisation from the instances into a structural relationship between the factors, akin to a concept of the form $\{(x,f(x)) \mid x \in C \}$ where $f$ is now affine (convexity-preserving). 

As such, quantum combinations share the benefits of spatial combinations on a conceptual space, in that one may form structured concepts by generalising from a small set of instances, as on the right of \eqref{eq:class-and-spatial}. However, in the quantum case this can be encoded even within a \emph{single} pure concept.  Our conclusion is that entanglement provides an effective way for concepts to encode relationships between factors in quantum conceptual models.


\subsection{Is a Quantum Model a Conceptual Space?} \label{sec:Is-Quantum-Conceptual}

In comparing conceptual spaces with quantum models, it is natural to ask whether we may view the latter as an instance of the former, while our  discussion of entangled concepts in the previous section suggested they should be considered distinct. We now discuss this question in detail. We begin with the case of a model with only a single factor, described by a Hilbert space $\hilbH$. 

\paragraph{Hilbert space as a convex space.}
Naively we can first observe that, as a complex vector space, $\hilbH$ does indeed count as a convex space according to Definition  \ref{def:convex-space}. However, arbitrary vectors in $\hilbH$ do not have a direct physical interpretation as states, but only the unit vectors $\psi$ (after identification up to global phase) which form the pure states. These pure states do not straightforwardly form a convex space in the sense of Def. \ref{def:convex-space}, since convex combinations of unit vectors are not unit vectors in general. 

\paragraph{Pure states as a betweenness space.}

\begin{figure} 
\centering 
\includegraphics[scale=0.14]{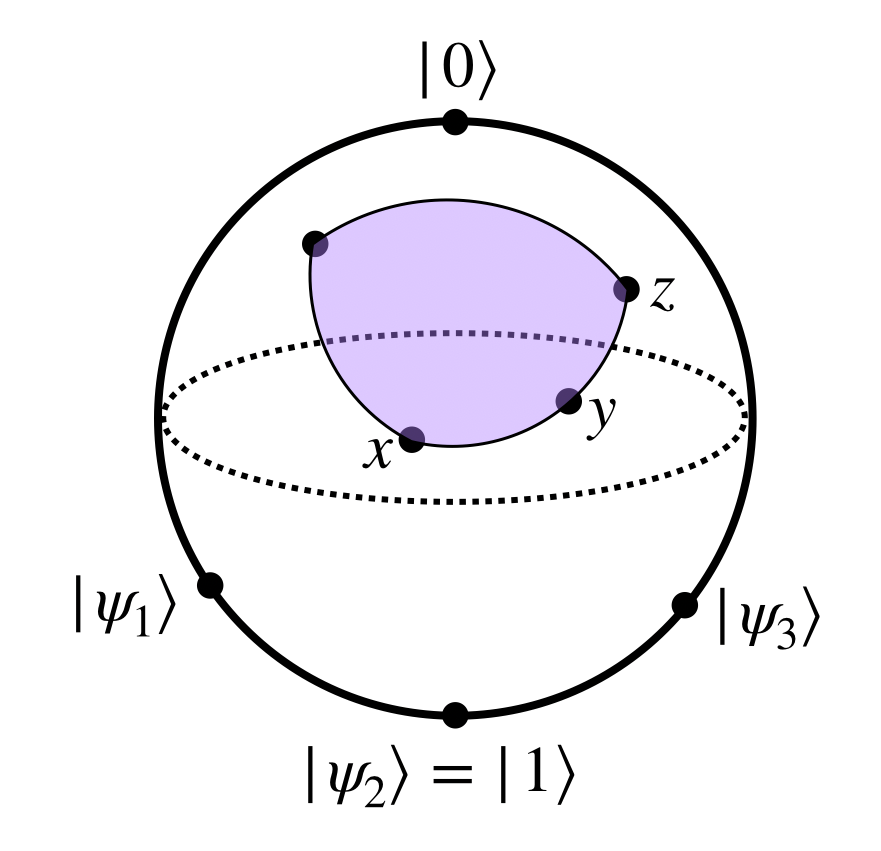}
\caption{The Bloch sphere as a Betweenness space, with marked examples of betweenness $B(x,y,z)$, and a convex region shown in purple. The states $\ket{\psi_i}\bra{\psi_i}$ are used to show that $\ket{0}\bra{0}$ is not quasi-concave. }
\label{fig:bloch-convex}
\end{figure}

We can nonetheless view the pure states as a geometric space, in a different way. This is most evident for a qubit $\hilbH=\mathbb{C}^2$, whose pure states are visualised via the surface of the \emph{Bloch sphere}. Though the surface of the sphere does not come with convex mixing in the sense of Definition \ref{def:convex-space}, it forms an instance of a broader notion of convex space which may be used to formalise conceptual spaces, known as a \emph{Betweenness space} \shortcite{gardenfors2004conceptual,gardenfors2014,aisbett2001general}. This is a set $Z$ along with a ternary operation $B(x,y,z)$ which says that the point $y$ is ``in-between" $x$ and $z$. A subset $S$ is then \emph{convex} if whenever $x, z \in S$ and $B(x,y,z)$, then $y \in S$ also. The Bloch sphere forms a Betweenness space when defining $B(x,y,z)$ whenever a geodesic from $x$ to $z$ passes through $y$; see Figure \ref{fig:bloch-convex}.

We now ask: is the quantum model of concepts on $\mathbb{C}^2$ the same as that given by the Bloch sphere as a Betweenness space? In fact the sets of concepts in each model are distinct.  Firstly, crisp concepts in the quantum model correspond to subspaces, which on the Bloch sphere are either single points (dimension 1) or the entire sphere (dimension 2). So most convex regions on the sphere, the crisp concepts in the Betweenness space $Z$, are not valid quantum concepts. Conversely, most quantum concepts are not valid fuzzy concepts in the Betweenness space $Z$. As argued in \shortciteA{tull2021categorical} and mentioned before in Def.~\ref{def:fuzzy-concept}, a fuzzy concept $C \colon Z \to [0,1]$ should at least satisfy the notion of \emph{quasi-concavity}, which states that if $C(x),C(z) \geq t$ then the same holds for any $y$ with $B(x,y,z)$. Example \ref{ex:quasi-conc-fails} below demonstrates that quantum concepts can fail to satisfy this condition. 

\begin{example} \label{ex:quasi-conc-fails}
Consider the pure concept $C = \ket{0}\bra{0}$. Let $\ket{\psi_i} = \cos(\frac{\theta_i}{2}) \ket{0} + \sin(\frac{\theta_i}{2}) \ket{1}$ for $i=1,2,3$, as in Figure \ref{fig:bloch-convex}. Setting $\theta_1 = \frac{2 \pi}{3}$, $\theta_2 = \pi$, $\theta_3 = \frac{4 \pi}{3}$ then $\ket{\psi_2}\bra{\psi_2} = \ket{1}\bra{1}$ is between $\ket{\psi_1} \bra{\psi_1}$ and $\ket{\psi_3} \bra{\psi_3}$, making $C$ not quasi-concave, since 
$
C(\ket{\psi_1}\bra{\psi_1}) = 
C(\ket{\psi_3}\bra{\psi_3}) =
\frac{1}{4} > 0 =
C(\ket{\psi_2}\bra{\psi_2})
$.
\end{example}

\paragraph{Spaces of mixed states.} One may be tempted to instead view a quantum conceptual model as a different convex space, namely the space $Z=\St(
\hilbH)$ of (pure and mixed) density matrices on $\hilbH$, so that these form the instances $z \in Z$. Indeed it follows from linearity that quantum concepts $C$ do satisfy quasi-concavity on this space. However, since density matrices are interpreted as states of uncertainty over pure quantum states, it is more natural to view them as the analogues of \emph{distributions over} a conceptual space, rather than instances. Finally, even if one attempts to view a quantum model as a convex space $\St(\hilbH)$, the manner in which we compose such models via the tensor is fundamentally different, making both classes of models distinct:  
\[
\St(\hilbH \otimes \hilbK) = \St(\hilbH) \otimes \St(\hilbK) \neq \St(\hilbH) \times \St(\hilbK) 
\]   

In summary, due to their treatment of entangled concepts and the arguments above, it is most natural to view quantum models as distinct from conceptual spaces. Nonetheless they possess the same benefits for learnability, replacing convex by linear subspaces, and thanks to entanglement may be even more natural for describing correlated concepts.

\section{Classical Implementation: The Conceptual VAE}
\label{sec:classical_impl}

Our first implementation comes from instantiating the framework using the $\ConSp$ category from Section~\ref{sec:class-models}, and implementing fuzzy concepts as Gaussians, as described in Example~\ref{ex:gaussian-fuzzy-concept}. There is already an existing literature on learning Gaussian representations of concepts, using a tool from machine learning called the \emph{Variational Autoencoder (VAE)} \shortcite{beta-vae}. Here we show how to extend that work by defining a new VAE model which provides explicit representations of concepts which fit our framework.

\subsection{VAEs for Concept Modelling}
\label{sec:conceptual_vae}

The Variational Autoencoder (VAE) \shortcite{kingma14,rezende14} provides a framework for the generative modeling of data, where the data potentially lives in some high-dimensional space. It uses the power of neural networks to act as arbitrary function approximators to capture complex dependencies in the data (e.g. between the pixels in an image). The VAE uses a latent space $\mathbf{Z}$ which acts as a bottleneck, compressing the high-dimensional data into a lower dimensional space.\footnote{In this section we use bold font for variables, e.g. the conceptual space $\mathbf{Z}$, to be consistent with the machine learning literature.} The question we  investigate is whether the VAE model can be adapted so that $\mathbf{Z}$ has desirable properties from a conceptual space perspective, such as interpretable dimensions which contain neatly separated, labelled concepts from individual domains. First we describe the standard VAE model before describing how to adapt it in order to incorporate labelled concepts. 

\subsubsection{The Vanilla VAE}

\begin{figure}
    \centering
        \[
        \begin{tikzpicture}[scale=0.8]
	\begin{pgfonlayer}{nodelayer}
		\node [style=seen] (0) at (2.5, -3) {$\mathbf{x}$};
		\node [style=unseen] (1) at (2.5, 0) {$\mathbf{z}$};
		\node [style=none] (2) at (1.25, -4.5) {};
		\node [style=none] (3) at (1.25, 1.25) {};
		\node [style=none] (4) at (3.75, 1.25) {};
		\node [style=none] (5) at (3.75, -4.5) {};
		\node [style=none] (6) at (3.75, 0.75) {};
		\node [style=none] (7) at (4.5, -3) {$\theta$};
		\node [style=none] (8) at (4.25, -3) {};
		\node [style=seen] (9) at (18.75, -3) {$\mathbf{x}$};
		\node [style=unseen] (10) at (18.75, 0) {$\mathbf{z}$};
		\node [style=seen] (11) at (18.75, 3) {$\mathbf{c}$};
		\node [style=seen] (12) at (11, 0) {$\mathbf{c}$};
		\node [style=none] (13) at (0.75, 0) {};
		\node [style=none] (14) at (0.5, 0) {$\phi$};
		\node [style=none] (15) at (3.25, -4) {$N$};
		\node [style=none] (16) at (20.5, 0) {};
		\node [style=none] (17) at (20.75, 0) {$\psi$};
		\node [style=none] (18) at (17, 0) {};
		\node [style=none] (19) at (16.75, 0) {$\phi$};
		\node [style=none] (20) at (20.75, -3) {$\theta$};
		\node [style=none] (21) at (20.5, -3) {};
		\node [style=seen] (22) at (9.5, -3) {$\mathbf{x}$};
		\node [style=unseen] (23) at (9.5, 0) {$\mathbf{z}$};
		\node [style=none] (24) at (8.25, -4.5) {};
		\node [style=none] (25) at (8.25, 1.25) {};
		\node [style=none] (26) at (12, 1.25) {};
		\node [style=none] (27) at (12, -4.5) {};
		\node [style=none] (28) at (12, 0.75) {};
		\node [style=none] (29) at (12.75, -3) {$\theta$};
		\node [style=none] (30) at (12.5, -3) {};
		\node [style=none] (31) at (7.75, 0) {};
		\node [style=none] (32) at (7.5, 0) {$\phi$};
		\node [style=none] (33) at (11.5, -4) {$N$};
		\node [style=none] (34) at (17.5, -4.5) {};
		\node [style=none] (35) at (17.5, 3.75) {};
		\node [style=none] (36) at (20, 3.75) {};
		\node [style=none] (37) at (20, -4.5) {};
		\node [style=none] (38) at (20, 3.25) {};
		\node [style=none] (39) at (19.5, -4) {$N$};
	\end{pgfonlayer}
	\begin{pgfonlayer}{edgelayer}
		\draw [style=full] (1) to (0);
		\draw (5.center) to (2.center);
		\draw (2.center) to (3.center);
		\draw (6.center) to (5.center);
		\draw (6.center) to (4.center);
		\draw (4.center) to (3.center);
		\draw [style=full] (8.center) to (0);
		\draw [style=full] (10) to (9);
		\draw [style=full] (11) to (10);
		\draw [style=full] (16.center) to (10);
		\draw [style=full] (21.center) to (9);
		\draw [style=full] (23) to (22);
		\draw (27.center) to (24.center);
		\draw (24.center) to (25.center);
		\draw (28.center) to (27.center);
		\draw (28.center) to (26.center);
		\draw (26.center) to (25.center);
		\draw [style=full] (30.center) to (22);
		\draw [style=densely dashed, ->] (31.center) to (23);
		\draw [style=full] (12) to (22);
		\draw (37.center) to (34.center);
		\draw (34.center) to (35.center);
		\draw (38.center) to (37.center);
		\draw (38.center) to (36.center);
		\draw (36.center) to (35.center);
		\draw [style=densely dashed, ->] (13.center) to (1);
		\draw [style=densely dashed, ->, bend left] (0) to (1);
		\draw [style=densely dashed, ->, bend left] (9) to (10);
		\draw [style=densely dashed, ->, bend left] (22) to (23);
		\draw [style=densely dashed, ->] (18.center) to (10);
	\end{pgfonlayer}
\end{tikzpicture}
        \] 
    \caption{Graphical models for the VAE (left), conditional VAE (centre) and the Conceptual VAE (right). Grey nodes represent observed variables and white nodes hidden variables.}
    \label{fig:vae}
\end{figure}
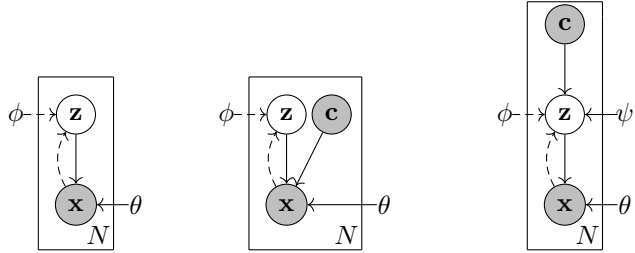

Fig.~\ref{fig:vae} (left) shows the graphical model for the VAE. In terms of the generative story, which is represented by the solid arrows in the plate diagram, first a point $\z$ in the latent space $\Z$ is sampled according to the prior $p(\z)$, and then a data point $\x$ is generated according to the likelihood $p_\theta(\x|\z)$. The dashed arrows denote the variational approximation $q_\phi(\mathbf{z}|\mathbf{x})$ to the intractable posterior $p_\theta(\mathbf{z}|\mathbf{x})$.
The prior is assumed to be a centered isotropic multivariate Gaussian $p(\mathbf{z}) = \mathcal{N}(\mathbf{z};\mathbf{0},\mathbf{1})$ \shortcite{kingma14}. The approximate posterior $q_\phi(\mathbf{z}|\mathbf{x})$ is also assumed to be a multivariate Gaussian with a diagonal covariance matrix, but with means and variances  predicted by a neural network with learnable parameters $\phi$. In our case, since $\mathbf{X}$ is a dataset of images, $q_\phi$ will be instantiated by a convolutional neural network (CNN), which is referred to as the \emph{encoder}. Similarly, $p_\theta$ will be instantiated by a de-convolutional neural network (de-CNN), and referred to as the \emph{decoder}.

The function that is optimised during training is the RHS of the following equation \shortcite{doersch}:
 \begin{equation}
 \begin{aligned}
\log p(\x) - \mathcal{D}(q(\z|\x),p(\z|\x)) =\\
    \mathbb{E}_{\z\sim q(\z|\x)}[\log p(\x|\z)] - \mathcal{D}(q(\z|\x),p(\z))
    \label{eqn:elbo_zx}
    \end{aligned}
\end{equation}
\noindent
where $\mathcal{D}$ is the KL divergence. Note that, since the KL on the LHS is positive, the equation provides a lower bound on the likelihood, known as the \emph{evidence lower bound} (ELBO).
The advantage of this formulation is that the RHS can be maximised using gradient-based optimisation techniques. Since the KL on the RHS is between two multivariate Gaussians, there is an analytical expression for calculating this quantity, and estimate of the expectation can be obtained using numerical methods, in particular Monte Carlo sampling (together with the \emph{reparametrisation trick} \shortcite{kingma14}).


Are the latent representations induced by a VAE in any way \emph{conceptual}? First, note that there is no pressure within the model to induce the sorts of factored representations in which the dimensions of $\Z$ correspond to conceptual domains. \shortciteA{beta-vae} attempt to address this problem by introducing a weighting factor on the KL loss. Second, there is currently no mechanism in the model which allows concepts to be referred to using their names (e.g. \emph{blue square}).

\subsubsection{The Conceptual VAE}


One feature that we would like in the model is an explicit representation of the words or symbols that are used to refer to a concept (which we'll call the concept \emph{label}). 
The obvious way to include the concept label in the model is as an explicit random variable $\con$. We could use a conditional VAE \shortcite{doersch}, with the label acting as an additional input into the decoder, so that when the decoder generates a  data instance $\x$, it does so conditioned on $\con$ as well as a point from the latent space $\z$ (Figure~\ref{fig:vae}; centre).
However, with this model there is no explicit representation of a concept (beyond its symbolic label). The key to the conceptual VAE is to introduce a new random variable for a concept label, $\con$, but introduce it at the very top of the graphical model (Figure~\ref{fig:vae}; right). The difference with the conditional VAE is that each concept $\con$ now has an explicit set of parameters associated with it, which acts as $\con$'s representation.


 In terms of the generative story, first a concept label $\con$ is generated, and then a point $\z$ in the latent conceptual space is generated, \emph{conditioned on} $\con$; after that the generative story is the same as for the vanilla VAE: an instance $\x$ is generated conditioned on $\z$. In this work we assume a uniform prior over the concept labels (more specifically a uniform prior over the atomic labels corresponding to each conceptual domain $\con_i$), and $\con$ can effectively be thought of as a fixed input to the model, as provided by the data.   
 
How do we model $p(\z|\con)$? As before we use multivariate Gaussians with diagonal covariance matrices, but now the means and variances are \emph{learnable parameters} $\psi$. We will sometimes refer to $p_\psi(\z|\con)$ for a given concept $\con$ as a \emph{conceptual ``prior"} (since these Gaussians replace the unit normal prior in the vanilla VAE), as well as $\con$'s learned representation. Since $\con$ is factored, each $\con_i$ has its own (univariate) Gaussian distribution; e.g., \emph{red} will have its own mean and variance which define a Gaussian on the dimension corresponding to the \textsc{colour} domain. It is this Gaussian which provides the anwser to the question ``what is the conceptual representation for \emph{red}?". 

The ELBO equation now takes the following form:
\begin{equation}
\begin{aligned}
\log p(\x|\con) - \mathcal{D}(q(\z|\x),p(\z|\x,\con)) =\\
    \mathbb{E}_{\z\sim q(\z|\x)}[\log p(\x|\z)] - \mathcal{D}(q(\z|\x),p(\z|\con))
    \label{eqn:elbo_con}
    \end{aligned}
\end{equation}
How is this model trained, and what are the pressures that lead to conceptual representations being learned? For a training instance $\x$ labelled with a concept $\con$, the training proceeds as before for the vanilla VAE: the encoder predicts a Gaussian $q(\z|\x)$; this is sampled from (using the reparametrisation trick) to give a sample $\z_s$; and $-\log p(\x|\z_s)$ is calculated to give the reconstruction loss. The key difference is in the calculation of the KL loss. Suppose that $\con$ = (\emph{green, medium, triangle, bottom}). The KL is calculated for each dimension, relative to the Gaussian for the particular atomic label for that dimension. For example, for the \textsc{colour} domain (dimension 0), the KL would be between $q_\phi(\z_0|\x)$ and $p_\psi(\z_0|\mbox{\emph{green}})$. So note that the supervision regarding the domains is provided here in the calculation of the KL.\footnote{The question of whether, and how, the level of supervision could be reduced and the domains  learned automatically is an ongoing debate \shortcite{beta-vae,Locatello2019}.}
Unlike the vanilla VAE, the conceptual ``priors" depend on the learned parameters $\psi$, which are the means and variances of the individual (univariate) Gaussians. We expect these learned means and variances to result in a neat separation along a dimension, since this will make it easier for the model to fit $q$ to the conceptual representations, leading to a lower KL.

\paragraph{Conceptual space description}
Explicitly, in terms of our framework from Section \ref{sec:cat-con-spaces}, 
our conceptual model is given in the category $\ConSp$, i.e. by a conceptual space. The model is $\Z = \mathbb{R}^n$, viewed as a product of $n$ one-dimensional domains $\Z = \prod^n_{i=1} \Z_i$ with $\Z_i = \mathbb{R}$. An instance is a vector $\z=(z_1,\dots,z_n) \in \Z$. In particular for each image $\x \in X$ we obtain an instance via the (deterministic) encoder $q_\phi(\x)$. Each concept label $\con=(\con_1,\dots,\con_n)$ defines a Gaussian fuzzy concept $\con(\z) = p (\z \mid \con)$ with diagonal covariance matrix, as in Example \ref{ex:gaussian-fuzzy-concept}. It forms a product concept over the domains as in \eqref{eq:prod-concept}, via:
\[
\con(\z) = \prod^n_{i=1} \con_i(\z_i) 
\]
where each $\con_i(\z_i) = \con_i(\z_i; \mu_i, \sigma_i^2)$ is a one-dimensional Gaussian concept for concept label $\con_i$ on $\Z_i$, with mean $\mu_i$ and variance $\sigma_i^2$ as trainable parameters.

\subsubsection{A Concept Classifier}

\begin{figure*}
\centering
    \includegraphics[]{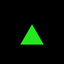}
    \includegraphics[]{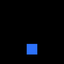}
    \includegraphics[]{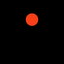}
    \includegraphics[]{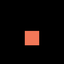}
    \includegraphics[]{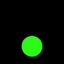}
    \caption{Example shapes: \small{(\emph{green, large, triangle, centre})}; (\emph{blue, small, square, bottom}); 
    (\emph{red, medium, circle, top});
    (\emph{red, medium, square, centre});
    (\emph{green, large, circle, bottom}).}
    \label{fig:shapes}
\end{figure*}

Here we show how the model can be adapted to act as a concept classifier. Note that, from a computer vision perspective, the classification task is trivial, and one that we would expect a well-trained CNN to solve. The classification task is being used here as a test of whether the induced conceptual representations can be employed in a useful way.

From a probabilistic perspective, the goal is to find the most probable concept $\con'$ given an input image $\x$:
\begin{eqnarray}
    \con' & = &\arg\max_\con p(\con|\x) \label{bayes} \\
    \label{bayes2}
          & = & \arg\max_\con p(\x|\con)\\ \label{bayes3}
          & \approx & \arg\max_\con\; - \mathcal{D}(q(\z|\x),p(\z|\con)) + \mbox{recon\_loss} \\
          \label{bayes4}
          & = & \arg\max_\con\;  - \mathcal{D}(q(\z|\x),p(\z|\con))
\end{eqnarray}
Line (\ref{bayes2}) follows from (\ref{bayes}) because of the assumed uniform prior over concepts, and we use the ELBO from (\ref{eqn:elbo_con}) as an approximation to the likelihood in going from (\ref{bayes2}) to (\ref{bayes3}) (where recon\_loss is the remaining part of the loss after the KL). The reconstruction loss is independent of $\con$ and so we end up with the satisfying form of the classifier in (\ref{bayes4}), in which the most likely concept for an input $\x$ is the one with the smallest KL relative to the encoding of $\x$, as provided by $q$.

\subsection{Experiments}
\label{sec:expts-vae}

We use the Spriteworld software \shortcite{spriteworld19} to generate simple images. These consist of coloured shapes of particular sizes in particular positions in a 2D box, against a black background.  For the main dataset, 
there are three shapes: \{\emph{square, triangle, circle}\}; three colours: \{\emph{red, green, blue}\}; three sizes: \{\emph{small, medium, large}\}; and three (vertical) positions: \{\emph{bottom, centre, top}\} (see Fig.~\ref{fig:shapes}). We ran the sampler to generate a training set of 3,000 instances, and development and test sets with 300 instances each. 
Appendix~\ref{sec:app_shapes} contains the parameters used in the Spriteworld software to generate the main dataset. 



The encoder, which takes an image $\x$ as input, is instantiated as a CNN, with 4 convolutional layers followed by a fully-connected layer. A final layer predicts the means and variances of the multivariate Gaussian $q_\phi(\z|\x)$. The ReLU activation function is used throughout.
The decoder, which takes a latent point $\z$ as input, is instantiated as a de-CNN, with essentially the mirrored architecture of the encoder. The reconstruction loss we use on the decoder for predicting the pixel values in an image $\x$ is the MSE loss.

The implementation was in Tensorflow. The full set of parameters to be learned is $\theta \cup \phi \cup \psi$, where $\theta$ is the set of parameters in the encoder, $\phi$ the parameters in the decoder, and $\psi$ the means and variances for the conceptual representations (12 each for the main dataset). The training was run for 200 epochs (unless stated otherwise), with a batch size of 32, and the Adam optimizer was used. 
Finally, we added 2 ``slack" dimensions to the latent space $\Z$, in addition to the 4 dimensions for the conceptual domains. These slack dimensions are intended to capture any remaining variability in the images, beyond that contained in the concepts themselves. Appendix~\ref{sec:app_neural_nets} contains more details of the neural architectures used in our experiments, including the various hyper-parameter choices.

\subsubsection{Clustering Effects and Classification Accuracy}

\begin{figure}
    \includegraphics[scale=0.24]{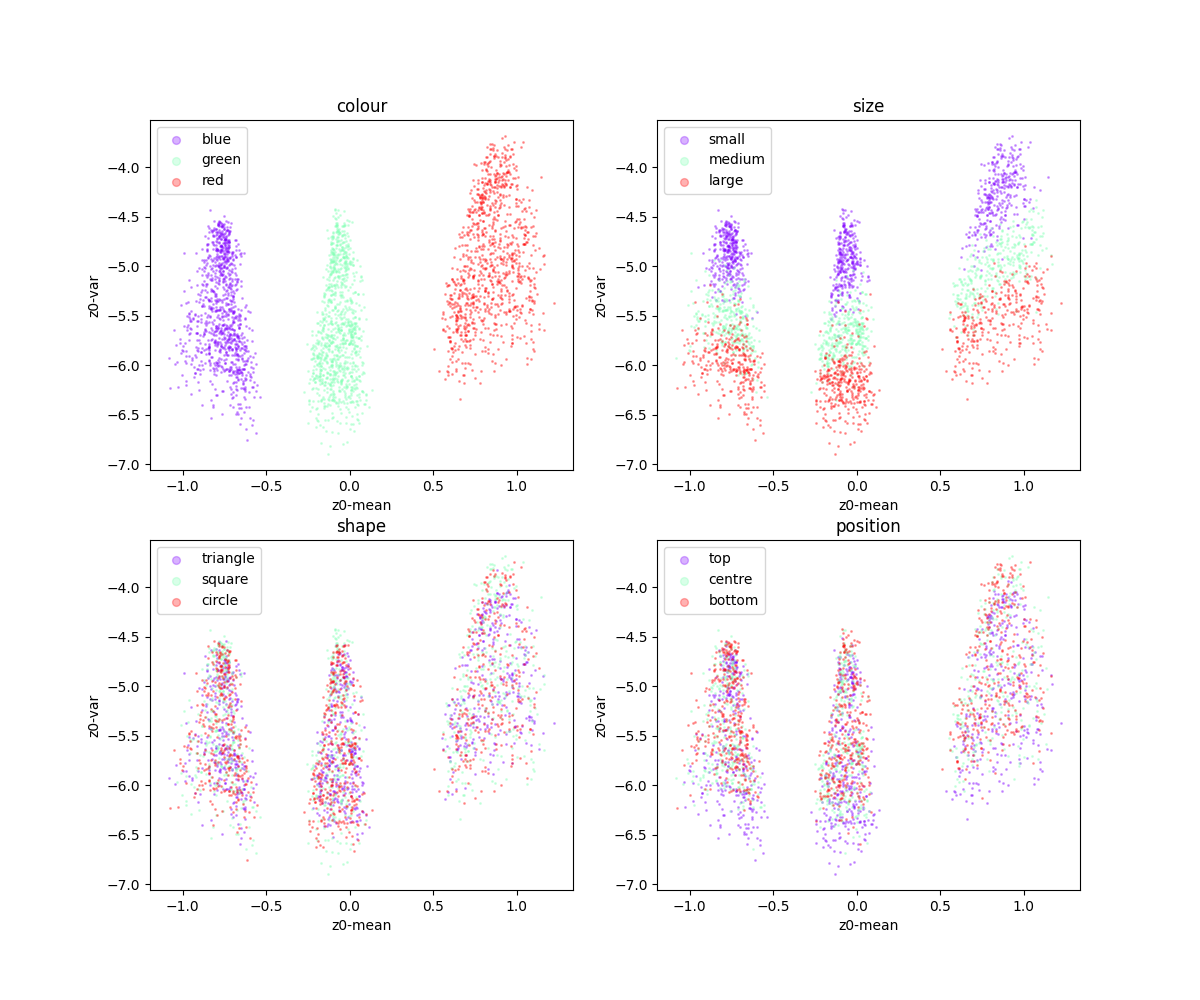}
    \vspace*{-0.2cm}
    \includegraphics[scale=0.24]{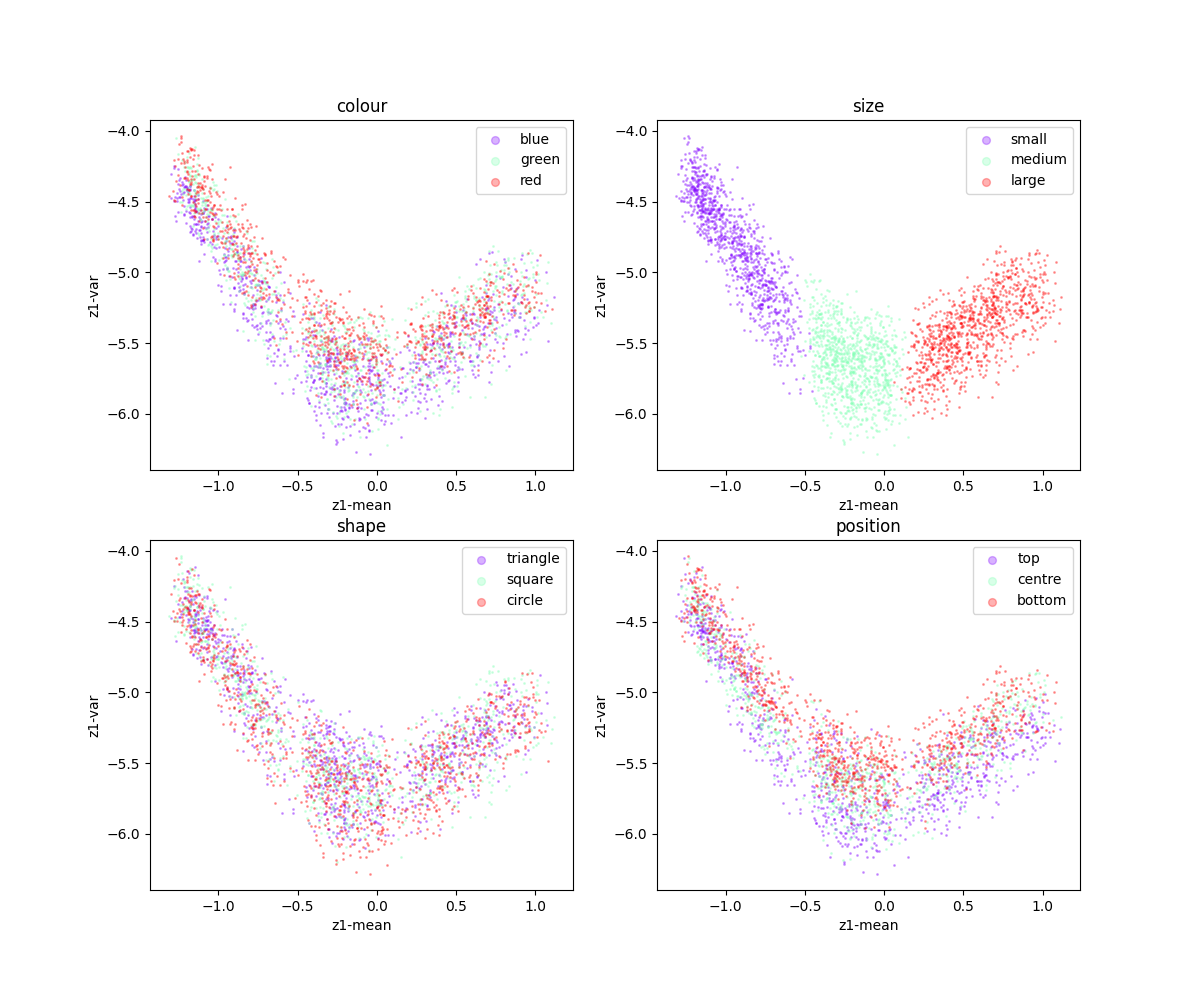}
    \vspace*{-0.2cm}
    \includegraphics[scale=0.24]{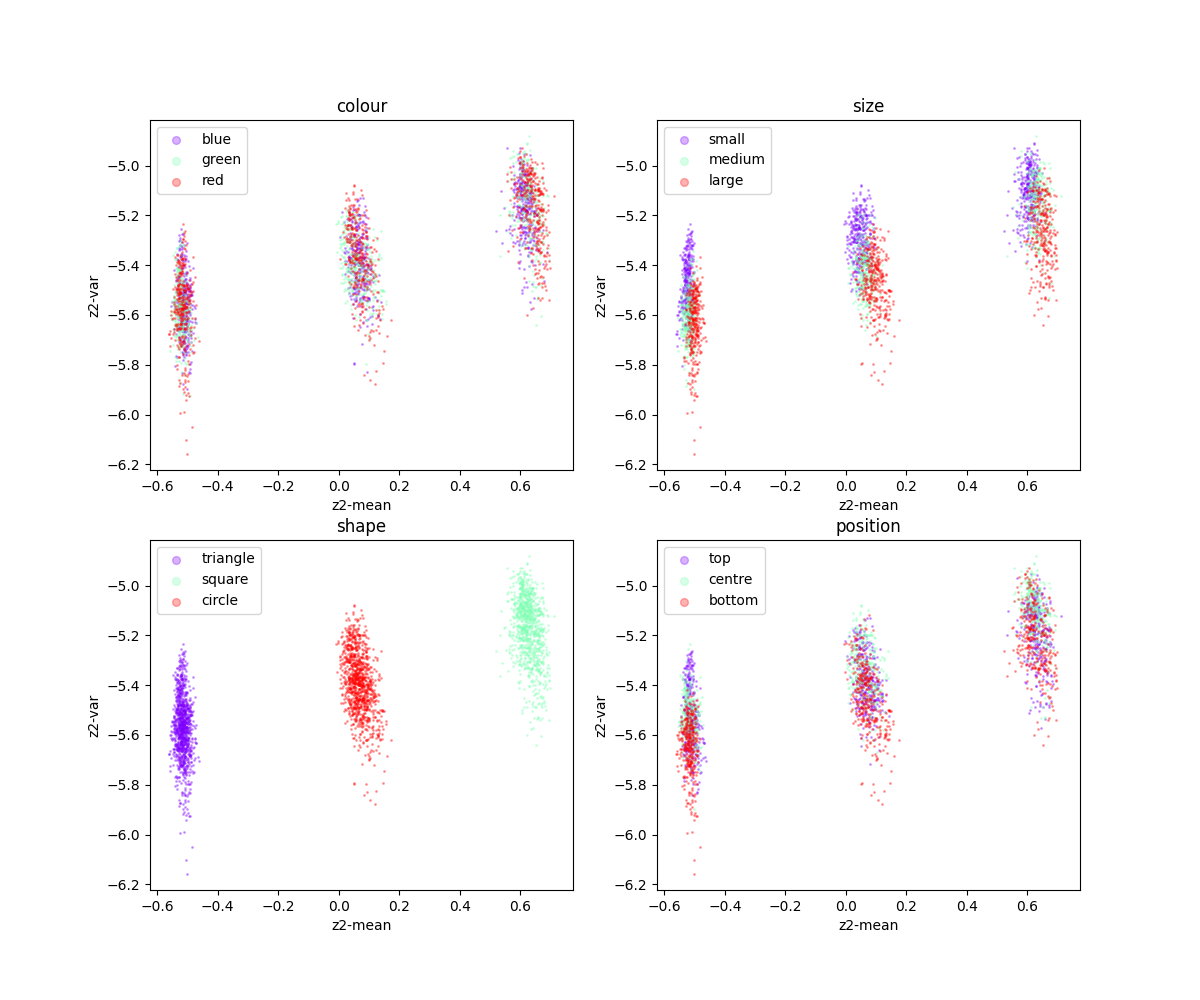}
    \vspace*{-0.2cm}
    \includegraphics[scale=0.24]{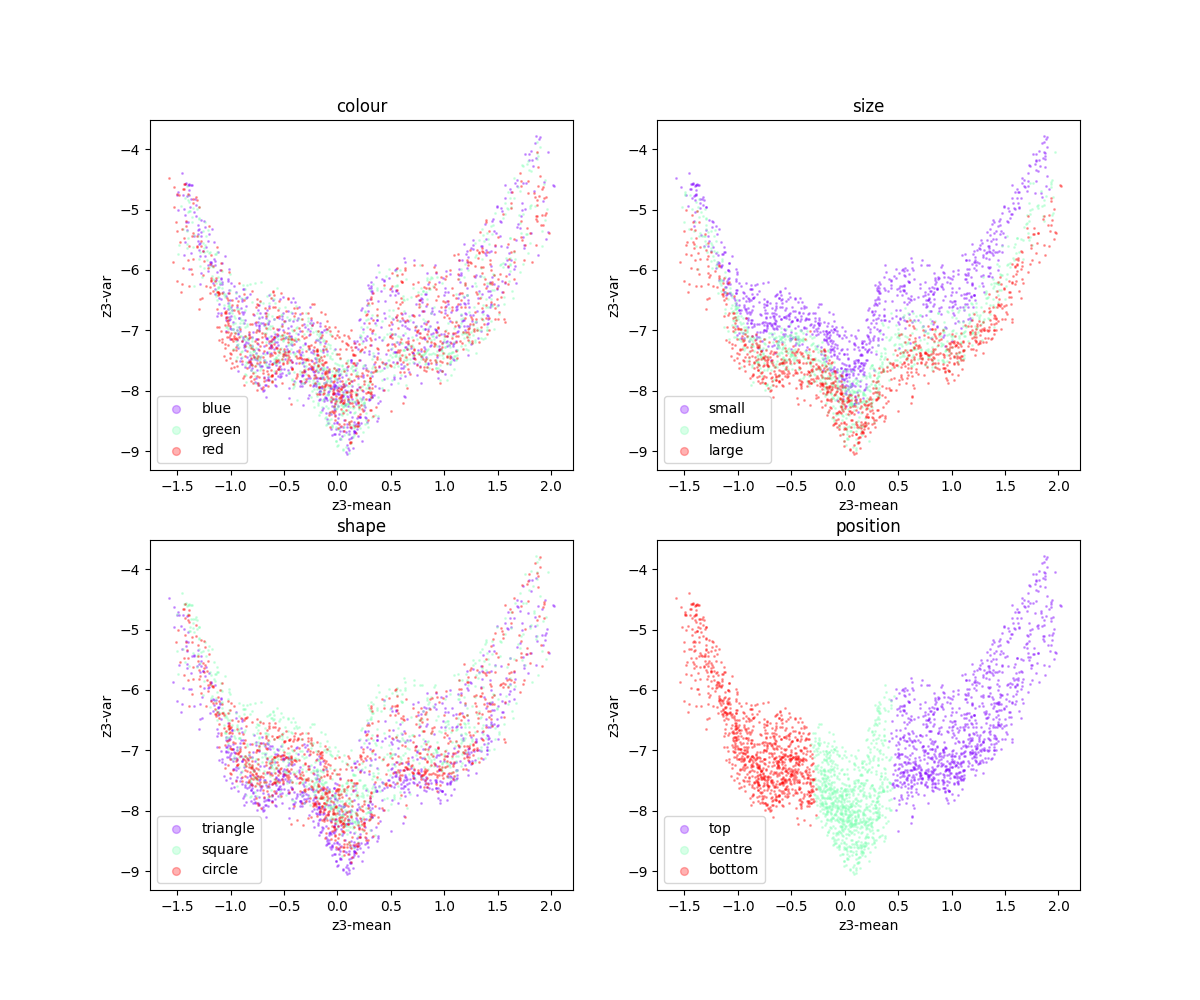}
    \hspace*{-0.2cm}
    \includegraphics[scale=0.24]{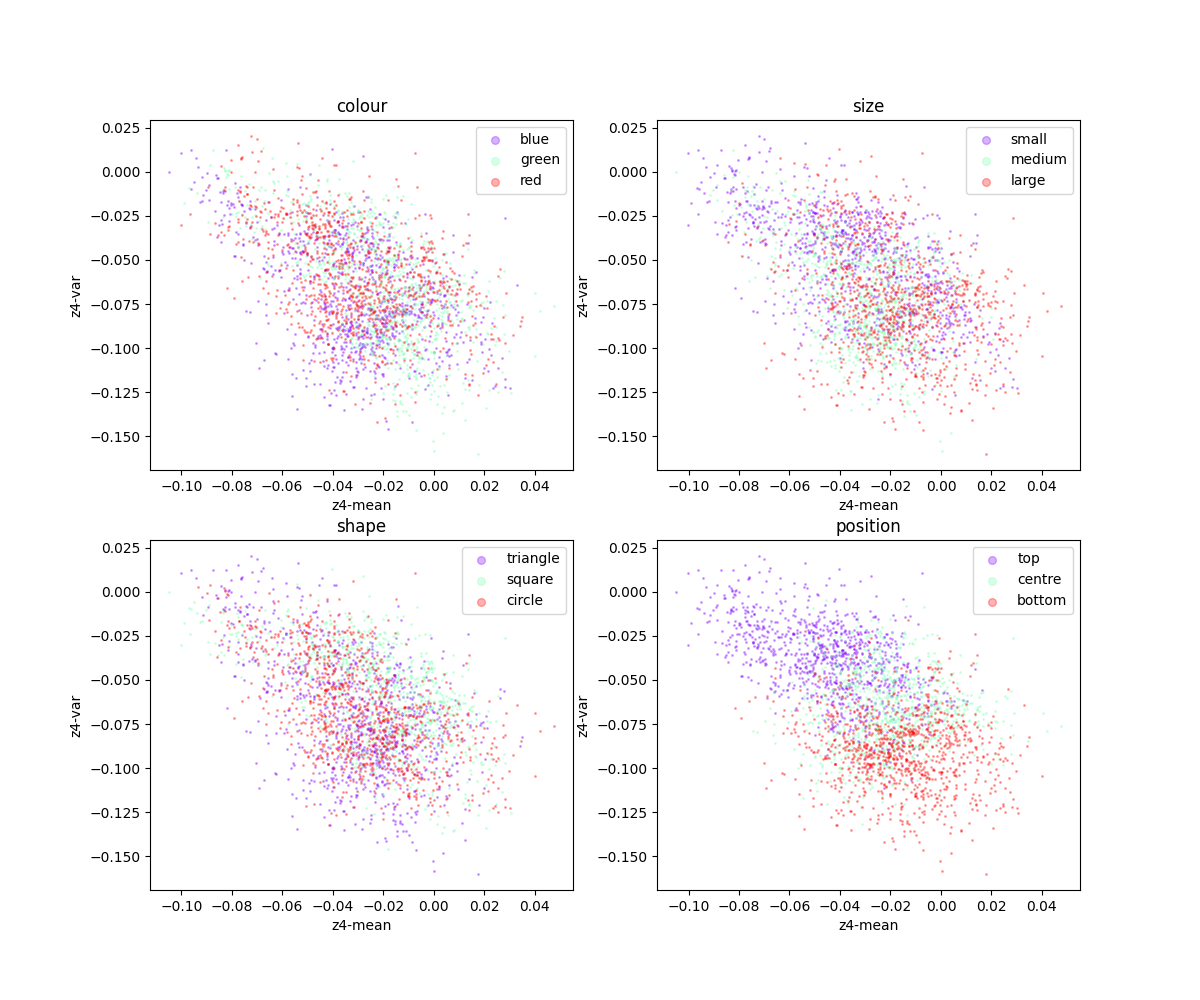}
    \includegraphics[scale=0.24]{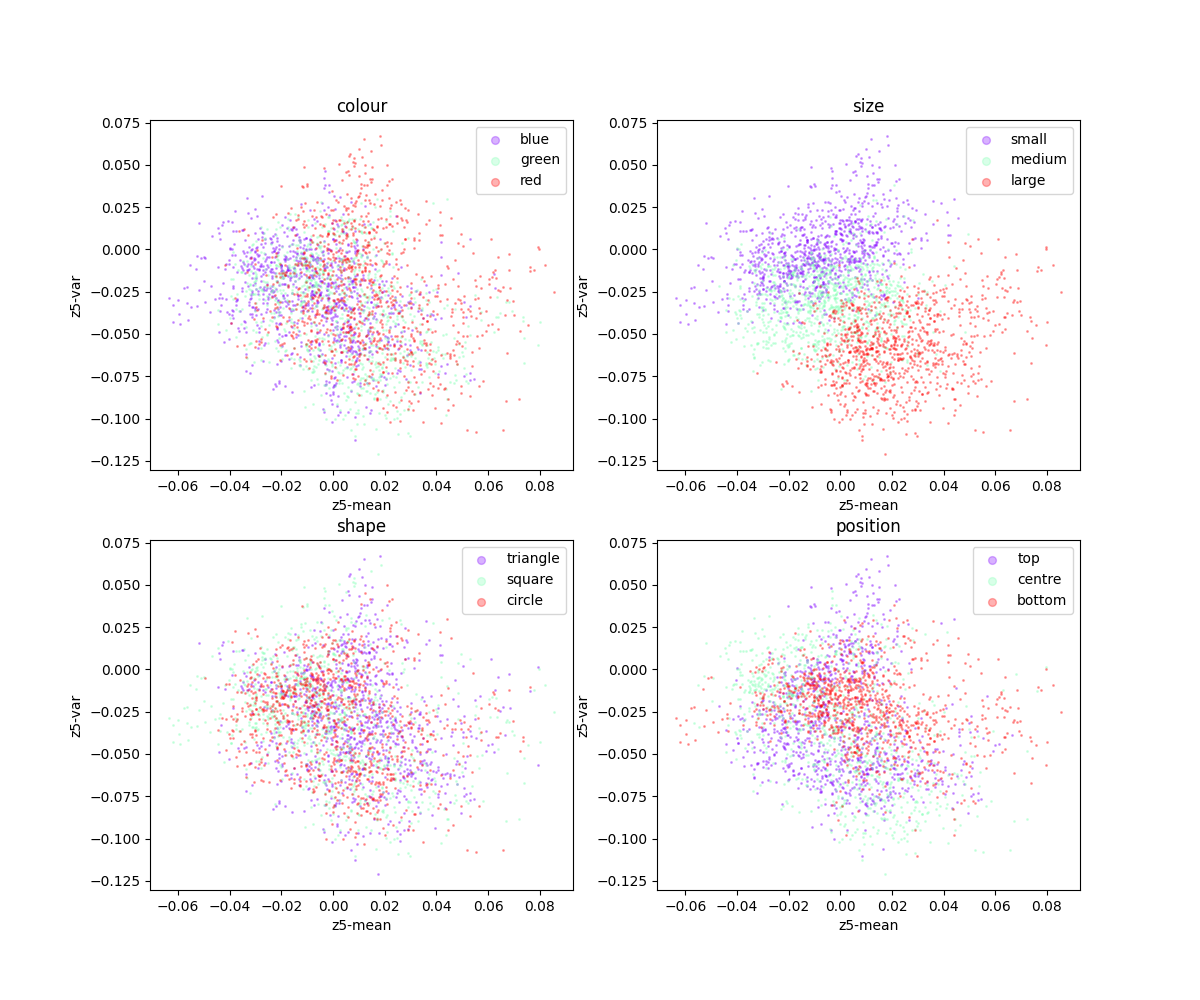}
    \caption{The means and log-variances for each dimension predicted by the encoder, for a set of instances; means on the x-axis, log-variances on the y-axis. Colour-coding, from top-left clockwise:
\textsc{colour}, \textsc{size}, \textsc{position}, \textsc{slack-dim-2}, \textsc{slack-dim-1}, \textsc{shape}.}
    \label{fig:clustering}
\end{figure}

Figure~\ref{fig:clustering} shows the means and log-variances predicted by the encoder for each dimension, for a set of instances, with the colour-coding indicating the atomic concept labels from the different domains. For example, in the set of 4 plots at the top left, the means and log-variances for dimension 0 are plotted; and in the top-left of those 4 plots, each point is colour-coded with the colour of the corresponding instance. What this plot shows is the neat separation for the means along the \textsc{colour} dimension, for each of the 3 colours. The other 3 plots contain the same set of points, but colour-coded with atomic labels from the remaining domains of \textsc{size}, \textsc{shape} and \textsc{position}. With the 3 remaining plots we expect to see no discerning pattern, since we would like the first dimension to encode \textsc{colour} only (although note that, in this particular training run, dimension 1---corresponding to \textsc{size}---does appear to be encoding some information about the colour).

The plots were created using the model evaluated on classification accuracy below, which performed well on the development data. The instances were taken from the training data.\footnote{The same patterns were observed on the development data. We used the training data since this gives denser plots.} The plots in the top-right are for dimension 1 (corresponding to \textsc{size}), and again we obtain a neat separation for the means, when colour-coded with the size of the instance, with instances labelled \emph{medium} sitting in the middle. The second-row (right) plots are for dimension 3 (\textsc{position}), and again we see a neat separation of the means with instances labelled \emph{centre} sitting between those labelled \emph{top} and \emph{bottom}. The second-row (left) plots are for dimension 2 (\textsc{shape}). Here we see a clear separation with the predicted means occupying a short range, which reflects the discrete nature of these concepts. The plots for the slack dimensions are in the bottom row, with no discernible pattern (except perhaps in the \textsc{size} dimension bottom-right).




We evaluated the same model as a classifier, using the formulation in (\ref{bayes})-(\ref{bayes4}) above. The accuracy on the development data for the \textsc{colour} and \textsc{shape} domains was 100\%, with accuracies above 98\% for the other two domains. These high accuracies transferred over to the test data.



\subsubsection{Continuity within Domains}

\begin{figure*}
    \centering
    \includegraphics[scale=2.0]{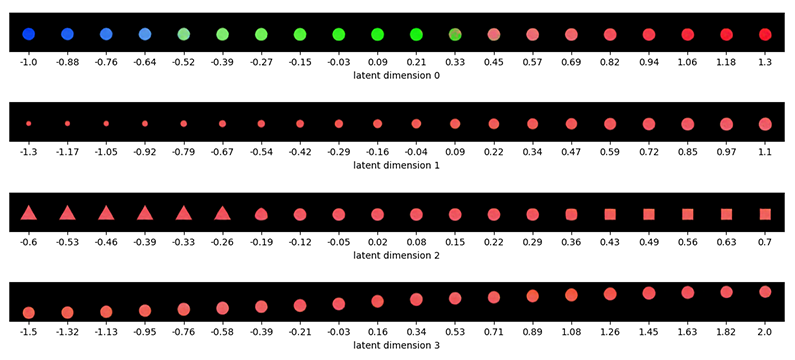}
    \caption{Traversals along each latent dimension for a large red circle in the centre.}
    \label{fig:traversal_example_1}
\end{figure*}

\begin{figure*}
    \centering
    \includegraphics[scale=2.0]{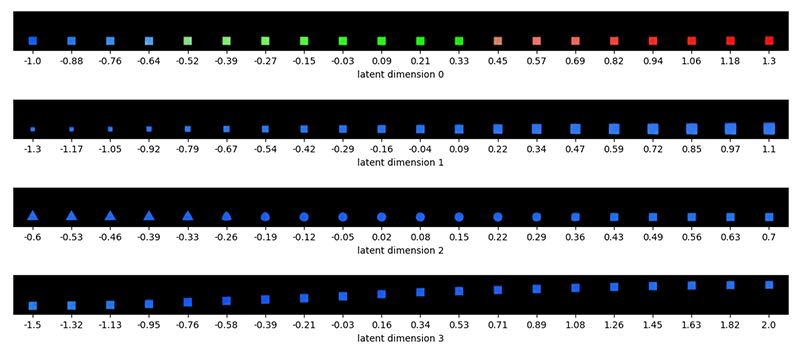}
\caption{Traversals along each latent dimension for a medium-sized blue square at the bottom.}
    \label{fig:traversal_example_2}
\end{figure*}

Figures~\ref{fig:traversal_example_1} and \ref{fig:traversal_example_2} provide  further qualitative demonstration of how the conceptual domains are neatly represented on each dimension. An instance of a large red circle in the centre and a medium-sized blue square at the bottom are passed through the encoder, giving a mean for each of the 4 dimensions. Then, the mean value is systematically varied for one of the dimensions only (through regular increases and decreases), keeping the other 3 fixed. All resulting combinations of the 4 mean values are then input to the decoder, giving the images in the figure.\footnote{The idea of plotting transitions along a dimension is taken from \shortciteA{beta-vae}.} 
What the transitions clearly demonstrate is not only how one latent dimension encodes just one domain, but also how the concepts smoothly vary along one dimension. Note how in both examples dimension 2 encodes a shape somewhere between a \emph{triangle} and a \emph{circle}, and  also a  shape somewhere between a \emph{circle} and a \emph{square}. Dimension 1 shows a smooth transition from \emph{small} to \emph{medium} to \emph{large}, and dimension 3 from \emph{bottom} to \emph{center} to \emph{top}.

\begin{figure}
    \begin{center}
    \hspace*{-0.0cm}\includegraphics[scale=0.3]{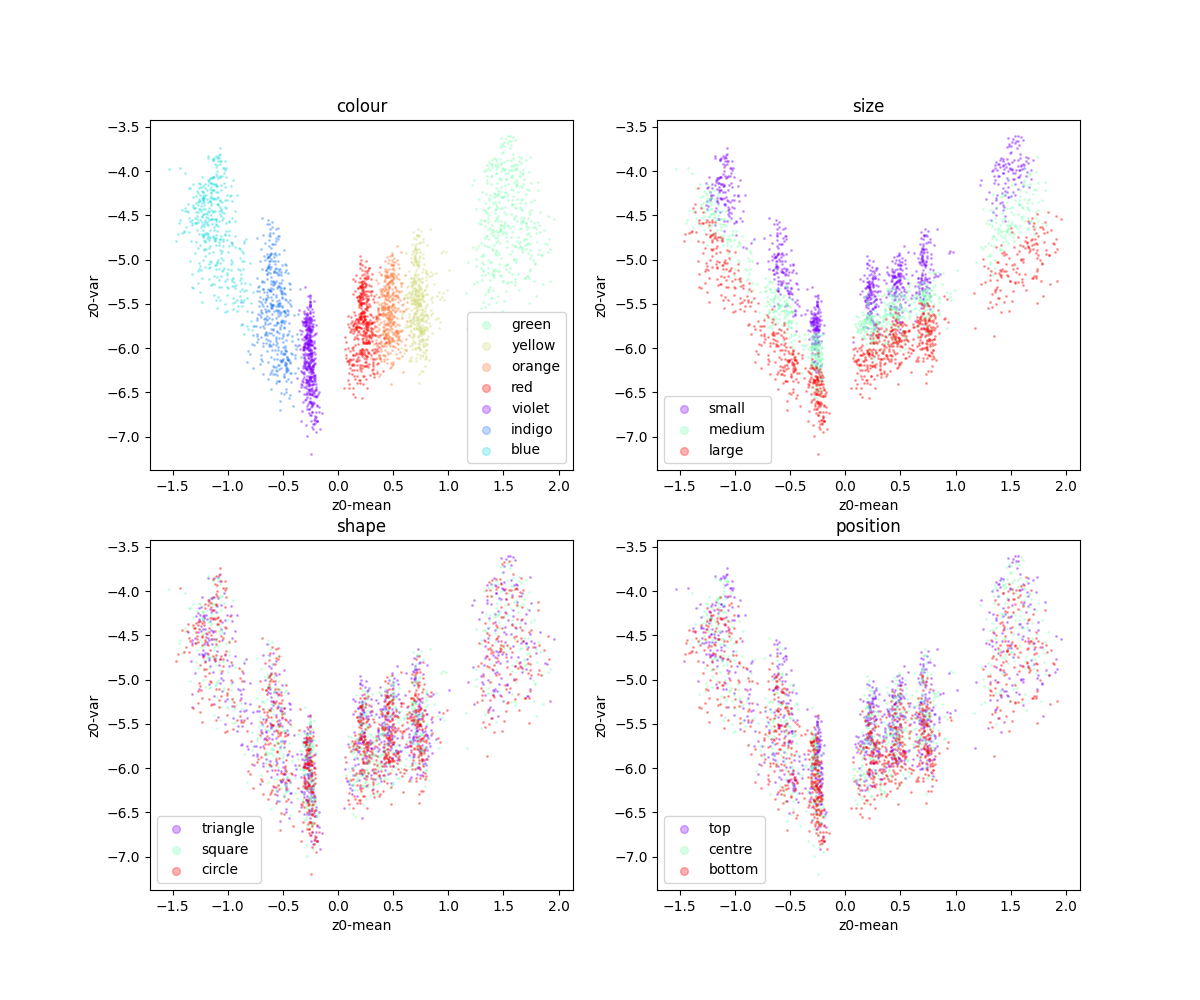}
    \caption{The means and log-variances for \textsc{colour} predicted by the encoder, for the rainbow colour set.}
    \label{fig:clustering_rainbow}
\end{center}
\end{figure}

\begin{figure*}
    \centering
    \includegraphics[scale=0.45]{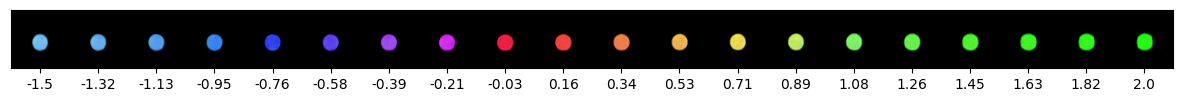}
    \includegraphics[scale=0.45]{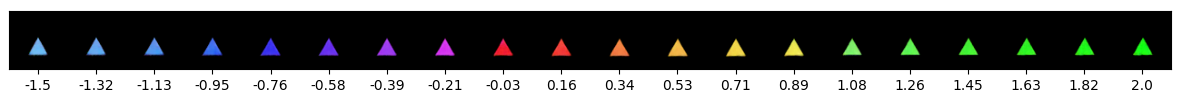} 
    \caption{Traversals along the \textsc{colour} dimension for two examples from the colour-extended dataset.}
    \label{fig:traversal_example_rainbow}
\end{figure*}


In order to investigate these ordering effects further, we created a new dataset 
which contains all the colours of the rainbow, with the same shapes, sizes and positions. The continuous colour ranges now cover a much larger proportion of the range of possible values (see Appendix~\ref{sec:ext_shapes}), with the occasional gap (e.g. between \emph{green} and \emph{blue}). The training data again consisted of 3,000 randomly generated instances, with a development set of 300 instances.

Again we chose a trained model which performed well on the classification task on the development data (with accuracies well into the 90s for all domains), and plotted the colour-coded means and variances as predicted by the encoder. Figure~\ref{fig:clustering_rainbow} again shows a neat separation for the 
\textsc{colour} domain, with very similar patterns for the other domains (not shown), and to those exhibited in Figure~\ref{fig:clustering}.
Looking carefully at the plot in the top-left, we see that the colours are not only neatly separated along the \textsc{colour} dimension, but also that the ordering of the rainbow is faithfully represented: \emph{blue, indigo, violet, red, orange, yellow, green}. 
Fig.~\ref{fig:traversal_example_rainbow} shows an example traversal along the \textsc{colour} dimension only, for the colour-extended dataset, again demonstrating an ordering consistent with a rainbow.

\section{Quantum Implementation: A Hybrid Network with PQCs} \label{sec:quantum-models}

In this section we also set up a probabilistic learning objective in order to induce conceptual representations, but using a discriminative classifier rather than a generative model. In addition, the classifier is implemented as a hybrid network consisting of a classical convolutional neural network (CNN) \shortcite[Ch.9]{deep_learning} followed by a parameterised quantum circuit (PQC) \shortcite{Benedetti2019}. We use the network to classify the same set of images and labels from the classical experiments in Section~\ref{sec:classical_impl}.

\subsection{The Hybrid Network}
An input image is first processed by a CNN which outputs classical parameters which are fed into a PQC. This PQC we call the \emph{encoder PQC}; it implements a quantum state $z$ which is the representation of the image in our model. Given a concept $C$, a separate \emph{concept PQC} implements a quantum effect corresponding to $C$ which can be applied to the instance $z$, as described in Sections~\ref{sec:cat-con-spaces} and \ref{sec:quantum-models2}. 
We assume that the factorisation of the model into the domains $\hilbH_1,\dots,\hilbH_n$ is known by the model; in our experiments these will be the four domains \textsc{shape}, \textsc{colour}, \textsc{size}, \textsc{position}. The overall setup is shown below, with thin wires denoting classical data and each thick wire denoting a Hilbert space given by some number of qubits. 


\begin{center}
\begin{tikzpicture}[tikzfig]
	\begin{pgfonlayer}{nodelayer}
		\node [style=map] (0) at (4, -0.75) {CNN};
		\node [style=map] (1) at (4, 1.75) {Encoder PQC};
		\node [style=large map] (2) at (5.5, 4.75) { \ \ \ \ Concept PQC \ \  \ \ };
		\node [style=none] (3) at (4, -2.5) {};
		\node [style=none] (4) at (4, -1) {};
		\node [style=label] (5) at (4, -3) {Image};
		\node [style=none] (6) at (4, -0.25) {};
		\node [style=none] (7) at (4, 1.5) {};
		\node [style=label] (8) at (3.25, 0.5) {$\theta$};
		\node [style=none] (9) at (2.5, 2.25) {};
		\node [style=none] (10) at (2.5, 4.25) {};
		\node [style=none] (11) at (8, 4.25) {};
		\node [style=none] (12) at (8, 2.5) {};
		\node [style=label] (13) at (8, 2) {Concept };
		\node [style=label] (14) at (8, 1.25) {params};
		\node [style=label] (15) at (3.75, 3.25) {$\dots$};
		\node [style=none] (16) at (4, 7.75) {};
		\node [style=none] (17) at (5.5, 2.25) {};
		\node [style=none] (18) at (5.5, 4.25) {};
		\node [style=none] (19) at (4, 8.5) {};
		\node [style=none] (20) at (4, 7.75) {};
		\node [style=none] (21) at (4, 8.5) {};
		\node [style=label] (22) at (4, 9) {Classification};
		\node [style=label] (23) at (2, 3.25) {$\hilbH_1$};
		\node [style=label] (24) at (5, 3.25) {$\hilbH_n$};
		\node [style=none] (25) at (2.5, 5.25) {};
		\node [style=none] (26) at (2.5, 7) {};
		\node [style=label] (27) at (4, 6.25) {$\dots$};
		\node [style=none] (28) at (5.5, 5.25) {};
		\node [style=none] (29) at (5.5, 7) {};
		\node [style=map] (32) at (4, 7.25) {Measurement};
	\end{pgfonlayer}
	\begin{pgfonlayer}{edgelayer}
		\draw (4.center) to (3.center);
		\draw (7.center) to (6.center);
		\draw [style=thick] (10.center) to (9.center);
		\draw [in=-90, out=90] (12.center) to (11.center);
		\draw [style=thick] (18.center) to (17.center);
		\draw (21.center) to (20.center);
		\draw [style=thick] (26.center) to (25.center);
		\draw [style=thick] (29.center) to (28.center);
	\end{pgfonlayer}
\end{tikzpicture}
\label{fig:PQCs}
\end{center}

Given an input image and the parameters encoding a concept, a single run of the circuit produces a ``yes" or ``no" to determine whether the concept has been deemed to fit the image. The probability of each outcome is obtained either by sampling the circuit many times (on a physical device) or direct calculation (in simulation). With the probabilities for each concept, one can then classify which concept best fits the input image.  

In more detail, each instance $z$ is a pure quantum state given by passing an image $X$ into the CNN and then using the resulting parameters in the encoder PQC network:
\[
\begin{tikzpicture}[tikzfig]
	\begin{pgfonlayer}{nodelayer}
		\node [style=map] (0) at (3.75, -0.75) {CNN};
		\node [style=map] (1) at (3.75, 0.75) {Encoder PQC};
		\node [style=none] (3) at (3.75, -2) {};
		\node [style=none] (4) at (3.75, -1) {};
		\node [style=map] (5) at (3.75, -2.25) {$X$};
		\node [style=none] (6) at (3.75, -0.25) {};
		\node [style=none] (7) at (3.75, 0.5) {};
		\node [style=none] (9) at (2.25, 1.25) {};
		\node [style=none] (10) at (2.25, 2.25) {};
		\node [style=label] (15) at (3.75, 2) {$\dots$};
		\node [style=none] (17) at (5.25, 1.25) {};
		\node [style=none] (18) at (5.25, 2.25) {};
		\node [style=label] (23) at (2.25, 2.75) {$\hilbH_1$};
		\node [style=label] (24) at (5.25, 2.75) {$\hilbH_n$};
		\node [style=none] (25) at (-2.75, -0.5) {};
		\node [style=wide point] (26) at (-2.75, -0.75) {$z$};
		\node [style=none] (27) at (-3.5, -0.5) {};
		\node [style=none] (28) at (-3.5, 1) {};
		\node [style=label] (29) at (-2.75, 0.25) {$\dots$};
		\node [style=none] (30) at (-2, -0.5) {};
		\node [style=none] (31) at (-2, 1) {};
		\node [style=label] (32) at (-3.5, 1.5) {$\hilbH_1$};
		\node [style=label] (33) at (-2, 1.5) {$\hilbH_n$};
		\node [style=none] (34) at (-0.25, 0) {$=$};
	\end{pgfonlayer}
	\begin{pgfonlayer}{edgelayer}
		\draw (4.center) to (3.center);
		\draw (7.center) to (6.center);
		\draw [style=thick] (10.center) to (9.center);
		\draw [style=thick] (18.center) to (17.center);
		\draw [style=thick] (28.center) to (27.center);
		\draw [style=thick] (31.center) to (30.center);
	\end{pgfonlayer}
\end{tikzpicture}
\]

Each specific concept $C$ can be understood as a measurement with two outcomes ``yes" and ``no", such that outcome ``yes" means the concept has been deemed to fit the instance. The measurement is given by a Pauli-Z measurement on each qubit, with the overall outcome ``yes" identified with obtaining outcome $0$ on every qubit individually, and all other outcomes labelled as ``no". Diagrammatically this is expressed as follows:
\begin{equation} \label{eq:concept-impl}
\begin{tikzpicture}
	\begin{pgfonlayer}{nodelayer}
		\node [style=map] (0) at (9.5, 0) {\ \ \ \ \ Concept PQC \ \ \ \ \ };
		\node [style=none] (1) at (7, -1.75) {};
		\node [style=none] (2) at (7, -0.5) {};
		\node [style=none] (3) at (12.25, -0.5) {};
		\node [style=none] (4) at (12.25, -1.75) {};
		\node [style=label] (5) at (14.25, -1.25) {};
		\node [style=none] (6) at (8.75, -1.25) {$\dots$};
		\node [style=none] (7) at (7, 0.5) {};
		\node [style=none] (8) at (10.5, -1.75) {};
		\node [style=none] (9) at (10.5, -0.5) {};
		\node [style=none] (10) at (7, 1.5) {};
		\node [style=none] (11) at (7, 0.5) {};
		\node [style=none] (12) at (7, 1.5) {};
		\node [style=label] (14) at (7, -2.25) {$\hilbH_1$};
		\node [style=label] (15) at (10.5, -2.25) {$\hilbH_n$};
		\node [style=large map] (16) at (1.75, 0) {$C$};
		\node [style=none] (17) at (0.5, -1.75) {};
		\node [style=none] (18) at (0.5, -0.5) {};
		\node [style=none] (19) at (1.75, -1.25) {$\dots$};
		\node [style=none] (20) at (3, -1.75) {};
		\node [style=none] (21) at (3, -0.5) {};
		\node [style=label] (22) at (0.5, -2.25) {$\hilbH_1$};
		\node [style=label] (23) at (3, -2.25) {$\hilbH_n$};
		\node [style=none] (24) at (4.75, 0) {$:=$};
		\node [style=copoint] (25) at (7, 1.75) {$0$};
		\node [style=point] (26) at (12.25, -2) {$\thetac$};
		\node [style=none] (27) at (10.5, 0.75) {};
		\node [style=none] (28) at (10.5, 1.5) {};
		\node [style=none] (29) at (10.5, 0.5) {};
		\node [style=none] (30) at (10.5, 1.5) {};
		\node [style=copoint] (31) at (10.5, 1.75) {$0$};
	\end{pgfonlayer}
	\begin{pgfonlayer}{edgelayer}
		\draw [style=thick] (2.center) to (1.center);
		\draw [in=-90, out=90] (4.center) to (3.center);
		\draw [style=thick] (9.center) to (8.center);
		\draw [style=thick] (12.center) to (11.center);
		\draw [style=thick] (18.center) to (17.center);
		\draw [style=thick] (21.center) to (20.center);
		\draw [style=thick] (30.center) to (29.center);
	\end{pgfonlayer}
\end{tikzpicture}
\end{equation}
where $\thetac$ are the parameters encoding the concept $C$. Each concept $C$ can be either pure or mixed, depending on whether a pure or mixed circuit is chosen for the concept PQC, which we discuss in Section~\ref{cnn_pqcs}.


\subsubsection{The CNN and PQCs}
\label{cnn_pqcs}

We use the same CNN from the classical experiments in Section~\ref{sec:expts-vae} for the image processing. For the classical experiments the CNN predicted the means and variances of a multivariate Gaussian, whereas here the CNN predicts the parameters of the encoder PQC.
The PQCs make use of the parameterised circuit ansatz shown below, defined over any finite collection of qubits. The ansatz $U(\theta)$ is given by performing parameterised $X, Y, Z$ rotations on each qubit, followed by entangling pairs of adjacent qubits using controlled $Z$ gates (with an additional gate operating on the two outermost qubits to complete the chain). Multiple layers of this ansatz can be composed to give a more complex circuit.
We define another ansatz $V(\theta)$ in the same way but with initial rotations in the reverse order $Z, Y, X$. An important special case is that, when given on a single qubit, $U(\theta)$ is simply equal to sequential parameterised $X, Y$ and $Z$ rotations. Similarly $V(\theta)$ on a single qubit amounts to rotating in the order $Z, Y, X$. 

\begin{equation}
\begin{tikzpicture}[scale=0.75]
	\begin{pgfonlayer}{nodelayer}
		\node [style=none] (6) at (5.75, -6.25) {};
		\node [style=none] (7) at (5.75, -0.25) {};
		\node [style=map] (10) at (5.75, -5) {$\RX{1}$};
		\node [style=map] (11) at (5.75, -3.25) {$\RY{1}$};
		\node [style=map] (12) at (5.75, -1.5) {$\RZ{1}$};
		\node [style=none] (22) at (-3, -2.25) {};
		\node [style=none] (23) at (-3, 0.75) {};
		\node [style=large map] (24) at (-1, -0.75) {$ \ \ \  \ \ \ \ U(\theta) \ \ \  \ \ \ \ $};
		\node [style=none] (25) at (1, -2.25) {};
		\node [style=none] (26) at (1, 0.75) {};
		\node [style=none] (27) at (3, -0.75) {$:=$};
		\node [style=none] (28) at (5.75, -0.25) {};
		\node [style=none] (29) at (5.75, 5.25) {};
		\node [style=none] (30) at (17.25, -0.25) {};
		\node [style=none] (31) at (17.25, 5.25) {};
		\node [style=none] (32) at (8.5, -0.25) {};
		\node [style=none] (33) at (8.5, 5.25) {};
		\node [style=blackdot] (34) at (8.5, 0.75) {};
		\node [style=blackdot] (35) at (5.75, 0.75) {};
		\node [style=none] (36) at (-1, 0.25) {$\dots$};
		\node [style=none] (37) at (-1, -1.75) {$\dots$};
		\node [style=blackdot] (38) at (11.25, 1.5) {};
		\node [style=blackdot] (39) at (8.5, 1.5) {};
		\node [style=blackdot] (40) at (17.25, 3.25) {};
		\node [style=blackdot] (41) at (14.5, 3.25) {};
		\node [style=none] (42) at (13, 2.25) {$\dots$};
		\node [style=none] (43) at (11.25, 5.25) {};
		\node [style=none] (44) at (11.25, -0.25) {};
		\node [style=none] (45) at (14.5, 5.25) {};
		\node [style=none] (46) at (14.5, -0.25) {};
		\node [style=blackdot] (47) at (17.25, 4.25) {};
		\node [style=blackdot] (48) at (5.75, 4.25) {};
		\node [style=none] (49) at (17.25, -6.25) {};
		\node [style=none] (50) at (17.25, -0.25) {};
		\node [style=none] (54) at (14.5, -6.25) {};
		\node [style=none] (55) at (14.5, -0.25) {};
		\node [style=none] (59) at (8.5, -6.25) {};
		\node [style=none] (60) at (8.5, -0.25) {};
		\node [style=none] (64) at (11.25, -6.25) {};
		\node [style=none] (65) at (11.25, -0.25) {};
		\node [style=map] (69) at (8.5, -5) {$\RX{2}$};
		\node [style=map] (70) at (8.5, -3.25) {$\RY{2}$};
		\node [style=map] (71) at (8.5, -1.5) {$\RZ{2}$};
		\node [style=map] (72) at (11.25, -5) {$\RX{3}$};
		\node [style=map] (73) at (11.25, -3.25) {$\RY{3}$};
		\node [style=map] (74) at (11.25, -1.5) {$\RZ{3}$};
		\node [style=map] (75) at (14.5, -5) {$\RX{n-1}$};
		\node [style=map] (76) at (14.5, -3.25) {$\RY{n-1}$};
		\node [style=map] (77) at (14.5, -1.5) {$\RZ{n-1}$};
		\node [style=map] (78) at (17.25, -5) {$\RX{n}$};
		\node [style=map] (79) at (17.25, -3.25) {$\RY{n}$};
		\node [style=map] (80) at (17.25, -1.5) {$\RZ{n}$};
		\node [style=none] (81) at (13, -3.25) {$\dots$};
	\end{pgfonlayer}
	\begin{pgfonlayer}{edgelayer}
		\draw [style=thick] (7.center) to (6.center);
		\draw [style=thick] (23.center) to (22.center);
		\draw [style=thick] (26.center) to (25.center);
		\draw [style=thick] (29.center) to (28.center);
		\draw [style=thick] (31.center) to (30.center);
		\draw [style=thick] (33.center) to (32.center);
		\draw (35) to (34);
		\draw (39) to (38);
		\draw (41) to (40);
		\draw [style=thick] (43.center) to (44.center);
		\draw [style=thick] (46.center) to (45.center);
		\draw (48) to (47);
		\draw [style=thick] (50.center) to (49.center);
		\draw [style=thick] (55.center) to (54.center);
		\draw [style=thick] (60.center) to (59.center);
		\draw [style=thick] (65.center) to (64.center);
	\end{pgfonlayer}
\end{tikzpicture}
\label{fig:ansatz}
\end{equation}
In the above each $\theta_j = (\theta_{j,X},\theta_{j,Y},\theta_{j,Z})$ consisting of three single parameters passed respectively to the $X, Y, Z$ rotations on qubit $j = 1,\dots,n$. All are in turn contained in the parameters vector $\theta$. In fact this ansatz is universal in that with sufficient layers of the form $U(\theta)$ one may implement any unitary circuit.\footnote{The entangling layer is self-inverse, so that two layers allow us to implement a rotation on any qubit. A swap operation on any pair of qubits can be implemented using three layers, and from this any CX gate. Hence we may implement the universal gate set given by single-qubit phase and Clifford gates; see, for example, \shortciteA{van2021constructing}.}

Now let us describe the encoder and concept PQCs in more detail. Both consist of some number of qubits per domain $\hilbH_i$. The form of the encoder PQC is as follows: 
\[
\begin{tikzpicture}
	\begin{pgfonlayer}{nodelayer}
		\node [style=none] (0) at (-1, 0) {};
		\node [style=none] (1) at (-1, 1.5) {};
		\node [style=large map] (2) at (0.5, 0) {Encoder PQC};
		\node [style=label] (3) at (-1, 2) {$\hilbH_1$};
		\node [style=none] (5) at (2, 0) {};
		\node [style=none] (6) at (2, 1.5) {};
		\node [style=label] (7) at (2, 2) {$\hilbH_n$};
		\node [style=none] (10) at (3.75, 0) {$=$};
		\node [style=none] (11) at (6.5, -1.25) {};
		\node [style=none] (12) at (6.5, 1.5) {};
		\node [style=map] (13) at (6.5, 0) {$U(\theta_1)$};
		\node [style=label] (14) at (6.5, 2) {$\hilbH_1$};
		\node [style=none] (16) at (11.5, -1.25) {};
		\node [style=none] (17) at (11.5, 1.5) {};
		\node [style=label] (18) at (11.5, 2) {$\hilbH_n$};
		\node [style=map] (20) at (11.5, 0) {$U(\theta_n)$};
		\node [style=none] (21) at (9, 0) {$\dots$};
		\node [style=none] (22) at (0.5, 1.25) {$\dots$};
		\node [style=none] (23) at (0.5, -0.5) {};
		\node [style=none] (24) at (0.5, -1) {};
		\node [style=point] (25) at (0.5, -1.25) {$\theta$};
		\node [style=point] (26) at (6.5, -1.5) {$0$};
		\node [style=point] (27) at (11.5, -1.5) {$0$};
	\end{pgfonlayer}
	\begin{pgfonlayer}{edgelayer}
		\draw [style=thick] (1.center) to (0.center);
		\draw [style=thick] (6.center) to (5.center);
		\draw [style=thick] (12.center) to (11.center);
		\draw [style=thick] (17.center) to (16.center);
		\draw (24.center) to (23.center);
	\end{pgfonlayer}
\end{tikzpicture} 
\]
More generally we can compose multiple layers of such $U$ circuits on each domain. Here the $\ket{0}$ states denote product states $\ket{0 \dots 0}$ on each $\hilbH_i$. Thus by construction the encoder never involves entanglement across domains, and can be viewed as a single encoder per domain. Since the ansatz $U$ is universal, the encoder is able to prepare an arbitrary quantum instance.

In the initial basic setup used, beginning in Section~\ref{sec:states_effects}, we only have one qubit per domain $\hilbH_i$, and only use one layer in the encoder. In this case the encoder simply carries an $X, Y$ and $Z$ rotation per qubit, involving no entanglement. In this  basic setup, the concept PQC also involves no entanglement, taking the following form. 
\begin{equation} \label{eq:concept-initial}
\begin{tikzpicture}[tikzfig]
	\begin{pgfonlayer}{nodelayer}
		\node [style=none] (0) at (0.5, -1.5) {};
		\node [style=none] (1) at (0.5, 1.5) {};
		\node [style=large map] (2) at (2.25, 0) {\ \ Concept PQC \ \  };
		\node [style=label] (3) at (0.5, 2) {$\hilbH_1$};
		\node [style=none] (4) at (3.5, -1.5) {};
		\node [style=none] (5) at (3.5, 1.5) {};
		\node [style=label] (6) at (3.5, 2) {$\hilbH_n$};
		\node [style=none] (7) at (6.5, 0) {$:=$};
		\node [style=none] (8) at (9, -1.5) {};
		\node [style=none] (9) at (9, 1.5) {};
		\node [style=map] (10) at (9, 0) {$V(\thetac_1)$};
		\node [style=label] (11) at (9, 2) {$\hilbH_1$};
		\node [style=none] (12) at (13.5, -1.5) {};
		\node [style=none] (13) at (13.5, 1.5) {};
		\node [style=label] (14) at (13.5, 2) {$\hilbH_n$};
		\node [style=map] (15) at (13.5, 0) {$V(\thetac_n)$};
		\node [style=none] (16) at (11.25, 0) {$\dots$};
		\node [style=none] (17) at (2, 1) {$\dots$};
		\node [style=none] (18) at (2, -1) {$\dots$};
		\node [style=label] (19) at (0.5, -2) {$\hilbH_1$};
		\node [style=label] (20) at (3.5, -2) {$\hilbH_n$};
		\node [style=label] (21) at (9, -2) {$\hilbH_1$};
		\node [style=label] (22) at (13.5, -2) {$\hilbH_n$};
		\node [style=none] (23) at (4.75, 0) {};
		\node [style=none] (24) at (4.75, -1.25) {};
		\node [style=point] (25) at (4.75, -1.5) {$\thetac$};
	\end{pgfonlayer}
	\begin{pgfonlayer}{edgelayer}
		\draw [style=thick] (1.center) to (0.center);
		\draw [style=thick] (5.center) to (4.center);
		\draw [style=thick] (9.center) to (8.center);
		\draw [style=thick] (13.center) to (12.center);
		\draw [in=-90, out=90, looseness=1.25] (24.center) to (23.center);
	\end{pgfonlayer}
\end{tikzpicture}
\end{equation}

Concretely, with four domains and one qubit per domain, in this setup the application of a concept $C$ to an instance $z$ amounts to the (probability of the) circuit shown below with post-selection, where $\theta$ is the encoding of the image from the CNN, $\thetac$ are the learned concept parameters and each wire is a single qubit. 

\begin{equation}
\begin{tikzpicture}
	\begin{pgfonlayer}{nodelayer}
		\node [style=none] (0) at (6.5, 6) {};
		\node [style=none] (1) at (6.5, 0) {};
		\node [style=map] (2) at (6.5, 4.75) {$\RXC{1}$};
		\node [style=map] (3) at (6.5, 3) {$\RYC{1}$};
		\node [style=map] (4) at (6.5, 1.25) {$\RZC{1}$};
		\node [style=none] (5) at (6.5, 0) {};
		\node [style=none] (6) at (9.25, 0) {};
		\node [style=none] (7) at (12, 0) {};
		\node [style=none] (8) at (14.75, 0) {};
		\node [style=none] (9) at (14.75, 6) {};
		\node [style=none] (10) at (14.75, 0) {};
		\node [style=none] (11) at (9.25, 6) {};
		\node [style=none] (12) at (9.25, 0) {};
		\node [style=none] (13) at (12, 6) {};
		\node [style=none] (14) at (12, 0) {};
		\node [style=map] (15) at (9.25, 4.75) {$\RXC{2}$};
		\node [style=map] (16) at (9.25, 3) {$\RYC{2}$};
		\node [style=map] (17) at (9.25, 1.25) {$\RZC{2}$};
		\node [style=map] (18) at (12, 4.75) {$\RXC{3}$};
		\node [style=map] (19) at (12, 3) {$\RYC{3}$};
		\node [style=map] (20) at (12, 1.25) {$\RZC{3}$};
		\node [style=map] (21) at (14.75, 4.75) {$\RXC{4}$};
		\node [style=map] (22) at (14.75, 3) {$\RYC{4}$};
		\node [style=map] (23) at (14.75, 1.25) {$\RZC{4}$};
		\node [style=copoint] (24) at (6.5, 6.25) {$0$};
		\node [style=copoint] (25) at (9.25, 6.25) {$0$};
		\node [style=copoint] (26) at (12, 6.25) {$0$};
		\node [style=copoint] (27) at (14.75, 6.25) {$0$};
		\node [style=none] (28) at (4, 0) {$=$};
		\node [style=wide copoint] (29) at (1.25, 1.75) {$C$};
		\node [style=none] (30) at (1, 1.5) {};
		\node [style=none] (31) at (2, 1.5) {};
		\node [style=none] (32) at (1.5, 1.5) {};
		\node [style=none] (33) at (0.5, 1.5) {};
		\node [style=none] (34) at (1, 0) {};
		\node [style=none] (35) at (2, 0) {};
		\node [style=none] (36) at (1.5, 0) {};
		\node [style=none] (37) at (0.5, 0) {};
		\node [style=none] (38) at (6.5, -6) {};
		\node [style=none] (39) at (6.5, 0) {};
		\node [style=map] (40) at (6.5, -4.75) {$\RX{1}$};
		\node [style=map] (41) at (6.5, -3) {$\RY{1}$};
		\node [style=map] (42) at (6.5, -1.25) {$\RZ{1}$};
		\node [style=none] (43) at (6.5, 0) {};
		\node [style=none] (44) at (9.25, 0) {};
		\node [style=none] (45) at (12, 0) {};
		\node [style=none] (46) at (14.75, 0) {};
		\node [style=none] (47) at (14.75, -6) {};
		\node [style=none] (48) at (14.75, 0) {};
		\node [style=none] (49) at (9.25, -6) {};
		\node [style=none] (50) at (9.25, 0) {};
		\node [style=none] (51) at (12, -6) {};
		\node [style=none] (52) at (12, 0) {};
		\node [style=map] (53) at (9.25, -4.75) {$\RX{2}$};
		\node [style=map] (54) at (9.25, -3) {$\RY{2}$};
		\node [style=map] (55) at (9.25, -1.25) {$\RZ{2}$};
		\node [style=map] (56) at (12, -4.75) {$\RX{3}$};
		\node [style=map] (57) at (12, -3) {$\RY{3}$};
		\node [style=map] (58) at (12, -1.25) {$\RZ{3}$};
		\node [style=map] (59) at (14.75, -4.75) {$\RX{4}$};
		\node [style=map] (60) at (14.75, -3) {$\RY{4}$};
		\node [style=map] (61) at (14.75, -1.25) {$\RZ{4}$};
		\node [style=point] (62) at (6.5, -6.25) {$0$};
		\node [style=point] (63) at (9.25, -6.25) {$0$};
		\node [style=point] (64) at (12, -6.25) {$0$};
		\node [style=point] (65) at (14.75, -6.25) {$0$};
		\node [style=wide point] (67) at (1.25, -1.75) {$z$};
		\node [style=none] (68) at (1, -1.5) {};
		\node [style=none] (69) at (2, -1.5) {};
		\node [style=none] (70) at (1.5, -1.5) {};
		\node [style=none] (71) at (0.5, -1.5) {};
		\node [style=none] (72) at (1, 0) {};
		\node [style=none] (73) at (2, 0) {};
		\node [style=none] (74) at (1.5, 0) {};
		\node [style=none] (75) at (0.5, 0) {};
		\node [style=none] (76) at (-0.5, 0) {};
		\node [style=none] (77) at (3, 0) {};
		\node [style=none] (78) at (5.5, 0) {};
		\node [style=none] (79) at (16, 0) {};
	\end{pgfonlayer}
	\begin{pgfonlayer}{edgelayer}
		\draw [style=thick] (1.center) to (0.center);
		\draw [style=thick] (10.center) to (9.center);
		\draw [style=thick] (12.center) to (11.center);
		\draw [style=thick] (14.center) to (13.center);
		\draw [style=thick] (33.center) to (37.center);
		\draw [style=thick] (30.center) to (34.center);
		\draw [style=thick] (32.center) to (36.center);
		\draw [style=thick] (31.center) to (35.center);
		\draw [style=thick] (39.center) to (38.center);
		\draw [style=thick] (48.center) to (47.center);
		\draw [style=thick] (50.center) to (49.center);
		\draw [style=thick] (52.center) to (51.center);
		\draw [style=thick] (71.center) to (75.center);
		\draw [style=thick] (68.center) to (72.center);
		\draw [style=thick] (70.center) to (74.center);
		\draw [style=thick] (69.center) to (73.center);
		\draw [dash pattern=on 3pt off 3pt] (76.center) to (77.center);
		\draw [dash pattern=on 3pt off 3pt] (79.center) to (78.center);
	\end{pgfonlayer}
\end{tikzpicture}
\label{fig:basic-setup-concrete}
\end{equation}

In order to capture mixed and entangled concepts, in  Section \ref{sec:mixedconcepts} we use a richer form for the concept PQC. Entanglement is provided by using the full ansatz $V(\theta)$ over all domains. To introduce mixing, we use an ancilliary copy of each domain $\hilbH_1,\dots\hilbH_n$, prepared in initial state $\ket{0}$, and then discard the original domains as in the following circuit:
\begin{equation} \label{eq:concepts-pqc-general}
\begin{tikzpicture}[tikzfig]
	\begin{pgfonlayer}{nodelayer}
		\node [style=none] (0) at (0.5, -1.5) {};
		\node [style=none] (1) at (0.5, 1.5) {};
		\node [style=large map] (2) at (2.25, 0) {\ \  Concept PQC \ \ };
		\node [style=label] (3) at (0.5, 2) {$\hilbH_1$};
		\node [style=none] (4) at (3.5, -1.5) {};
		\node [style=none] (5) at (3.5, 1.5) {};
		\node [style=label] (6) at (3.5, 2) {$\hilbH_n$};
		\node [style=none] (7) at (6.25, 0) {$:=$};
		\node [style=none] (8) at (8.5, -1.5) {};
		\node [style=upground] (9) at (8.5, 1.75) {};
		\node [style=label] (10) at (8.5, 2.25) {$\hilbH_1$};
		\node [style=none] (11) at (11, -1.5) {};
		\node [style=upground] (12) at (11, 1.75) {};
		\node [style=label] (13) at (11, 2.25) {$\hilbH_n$};
		\node [style=none] (14) at (9.75, -1) {$\dots$};
		\node [style=none] (15) at (2, 1) {$\dots$};
		\node [style=none] (16) at (2, -1) {$\dots$};
		\node [style=label] (17) at (0.5, -2) {$\hilbH_1$};
		\node [style=label] (18) at (3.5, -2) {$\hilbH_n$};
		\node [style=label] (19) at (8.5, -2) {$\hilbH_1$};
		\node [style=label] (20) at (11, -2) {$\hilbH_n$};
		\node [style=none] (21) at (4.5, 0) {};
		\node [style=none] (22) at (4.5, -1.25) {};
		\node [style=point] (23) at (4.5, -1.5) {$\thetac$};
		\node [style=medium map] (24) at (11.5, 0) {$ \  \ \ \ \ \ \ \ \ \ \ V(\thetac) \ \ \ \ \ \ \ \ \ \  \ $};
		\node [style=point] (25) at (12, -2.75) {$0$};
		\node [style=none] (26) at (12, 1.75) {};
		\node [style=label] (27) at (12, 2.25) {$\hilbH_1$};
		\node [style=label] (28) at (12.75, -2) {$\hilbH_1$};
		\node [style=none] (29) at (14.25, -2.75) {};
		\node [style=none] (30) at (14.25, 1.75) {};
		\node [style=label] (31) at (14.25, 2.25) {$\hilbH_n$};
		\node [style=label] (32) at (14.75, -2) {$\hilbH_n$};
		\node [style=none] (33) at (9.75, 1) {$\dots$};
		\node [style=none] (34) at (13.25, 1) {$\dots$};
		\node [style=none] (35) at (13.25, -1) {$\dots$};
		\node [style=point] (36) at (14.25, -2.75) {$0$};
	\end{pgfonlayer}
	\begin{pgfonlayer}{edgelayer}
		\draw [style=thick] (1.center) to (0.center);
		\draw [style=thick] (5.center) to (4.center);
		\draw [style=thick] (9) to (8.center);
		\draw [style=thick] (12) to (11.center);
		\draw [in=-90, out=90, looseness=1.25] (22.center) to (21.center);
		\draw [style=thick] (26.center) to (25);
		\draw [style=thick] (30.center) to (29.center);
	\end{pgfonlayer}
\end{tikzpicture}
\end{equation}
More generally one can include multiple $V$ layers prior to discarding. Note that since this ansatz is universal we can implement any unitary with sufficient layers of the form $V$, and thus any (sub-normal) quantum concept. 


\subsubsection{Discriminative Training}

The classical concepts model from earlier is a generative model consisting of an encoder and a decoder. Here we choose to train the quantum model to perform binary classification; hence the basic model is a discriminative model with an encoder only.\footnote{In Section~\ref{sec:decoder} below we investigate how the addition of a decoder can affect the instance and concept representations.}
The loss function is the standard binary cross entropy (BCE) loss for binary classification. The full set of parameters to be learned is $\psi \cup \phi$, where $\psi$ is the set of parameters in the classical encoder CNN and $\phi$ is the set of PQC parameters associated with the set of 12 basic concepts. 

The training data contains the 3,000 positive examples from Section~\ref{sec:expts-vae} and an additional 3,000 negative examples.
Each negative example is created from a positive one by randomly sampling an incorrect concept for each domain; for example, if the positive example is (\emph{green, large, triangle, centre}) then a negative example could be (\emph{blue, medium, square, bottom}). Since we are effectively learning each domain independently in the basic model, a negative example disagrees on every domain. Later models  will use variations on this data. 


The implementation is in Tensorflow Quantum, and the whole hybrid network---both the quantum and the classical parts---are trained end-to-end in simulation on a GPU. The training was run for 100 epochs (unless stated otherwise), with a batch size of 64 (32 images, each with a postive and negative label), and the Adam optimizer was used.

\subsection{Instance States and Concept Effects}
\label{sec:states_effects}

We trained a quantum model, using the circuit shown in (\ref{fig:basic-setup-concrete}) above, and tested it on the 300 examples in the development set. The model was trained to perform binary classification, but at test time we choose the concept for each domain which has the highest probability of applying to the input image.
The classification model performed with almost perfect accuracy, obtaining 100\% on the \textsc{colour} and \textsc{shape} domains, and 99\% and 97\% on the \textsc{position} and \textsc{size} domains, respectively.
This high accuracy carried over to the 300 examples in the test set, obtaining 100\% on the \textsc{colour} and \textsc{shape} domains, and 96\% and 97\% on the \textsc{position} and \textsc{size} domains, respectively.

\begin{figure*}[t!]
\begin{center}
    \includegraphics[scale=0.32]{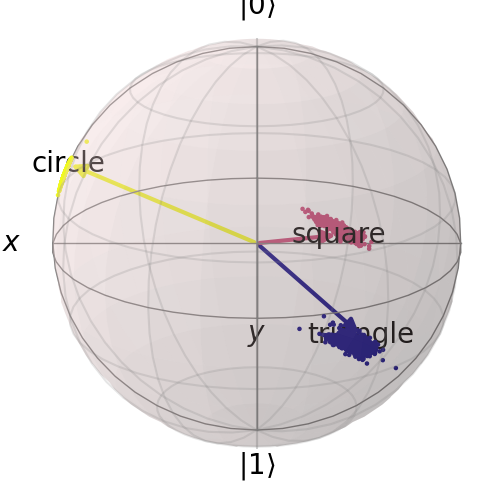}
    \includegraphics[scale=0.32]{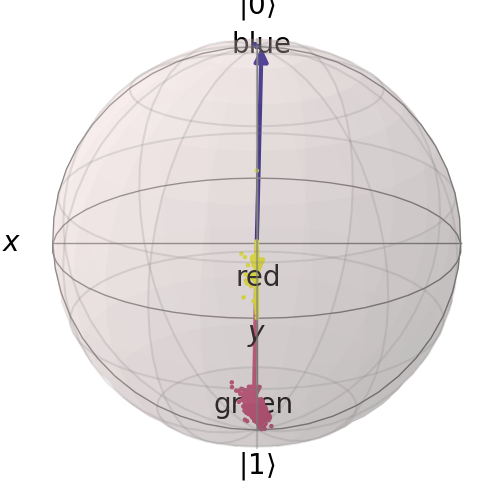}
    \includegraphics[scale=0.32]{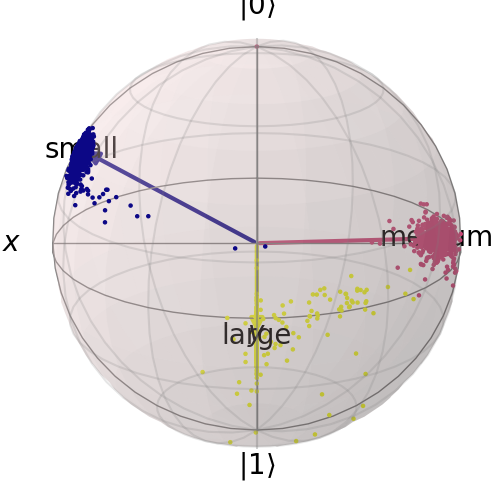}
    \includegraphics[scale=0.32]{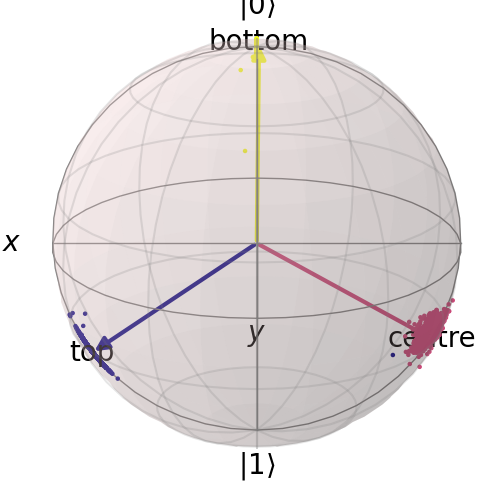}
    \end{center}
    \caption{Visualisation of the pure concept effects and instance states on the Bloch sphere, for \textsc{shape}, \textsc{colour}, \textsc{size} and \textsc{position}.}
    \label{fig:basic_bloch}
\end{figure*}

\begin{figure*}[t!]
\begin{center}
    \hspace*{-0.0cm}
    \includegraphics[scale=0.22]{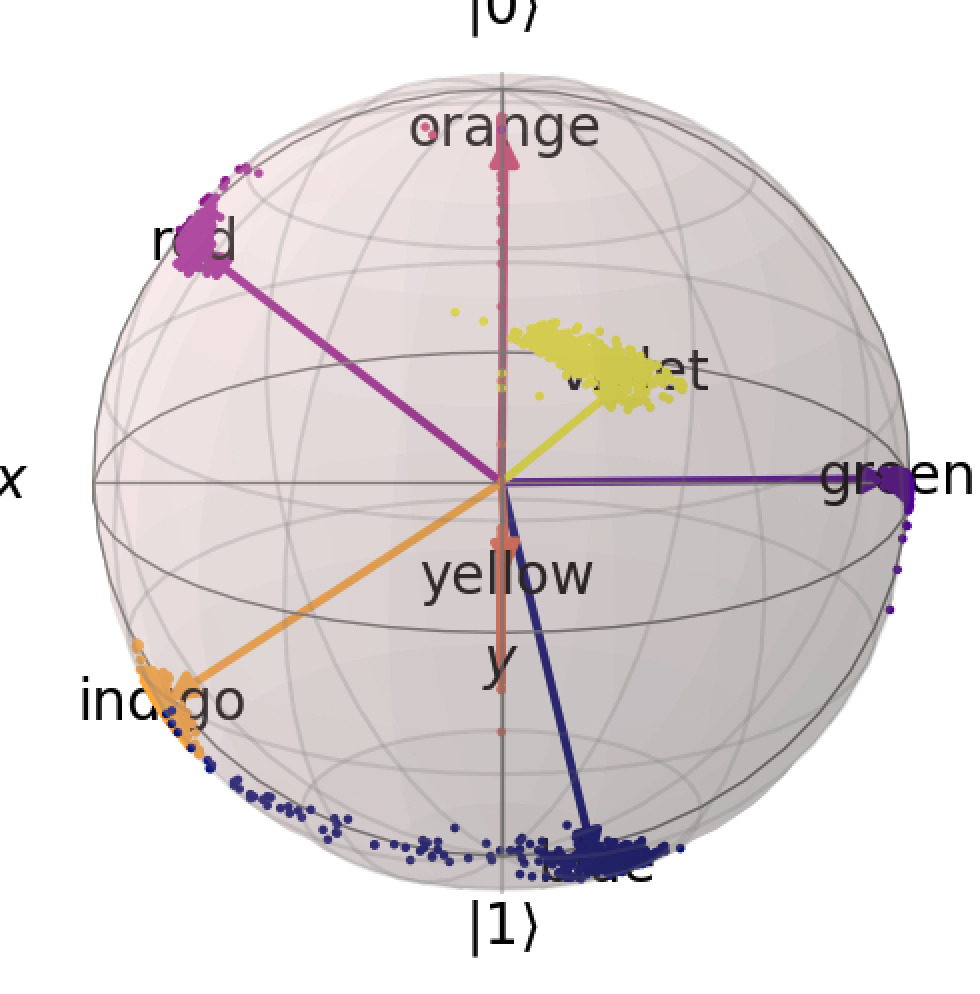}
    \includegraphics[scale=0.22]{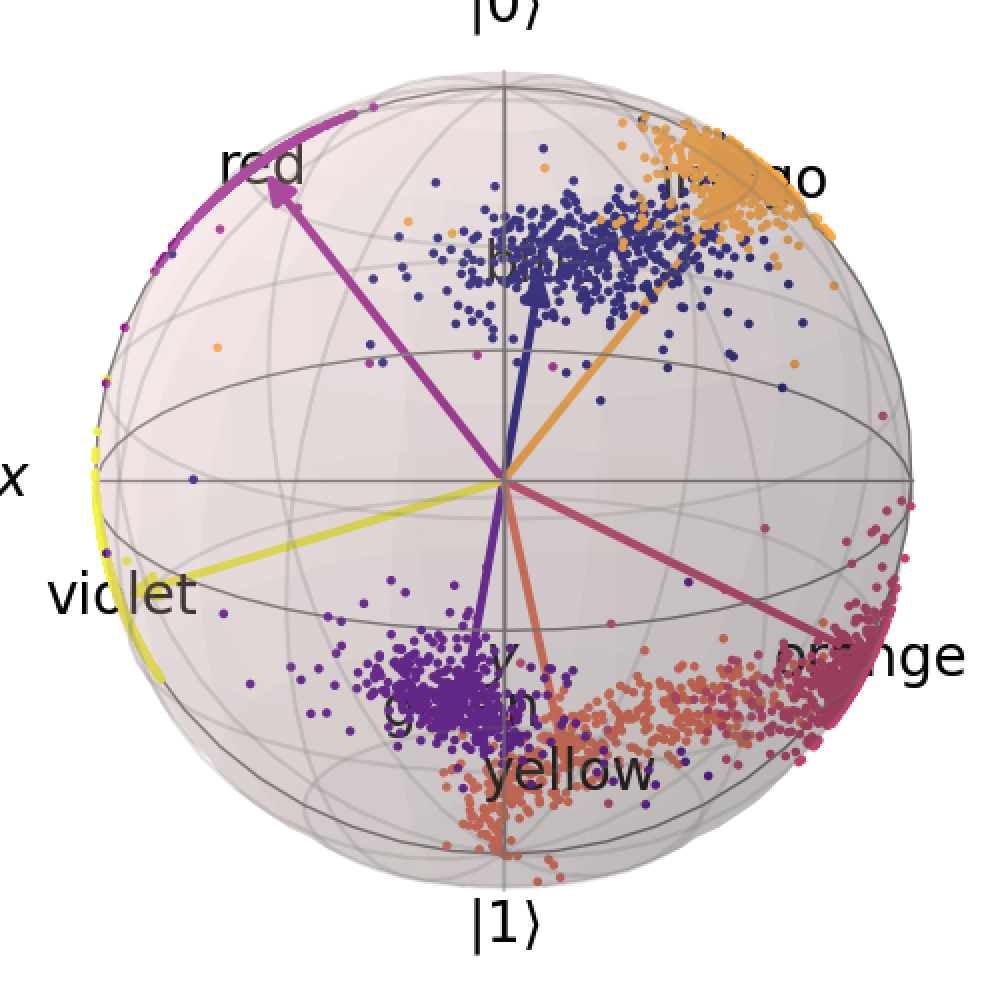}
    \includegraphics[scale=0.26]{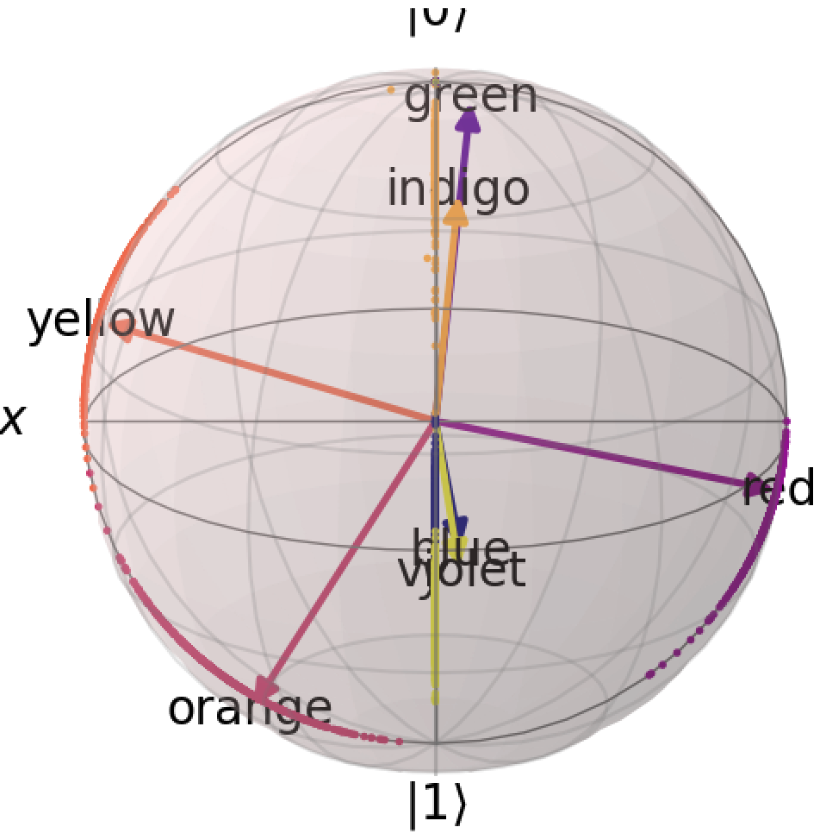}
    \caption{Visualisation of the concept effects and instance states on the Bloch sphere, for 3 trained models, for the \textsc{colour} domain on the rainbow dataset.}
    \label{fig:rainbow_bloch}
    \end{center}
\end{figure*}

Figure~\ref{fig:basic_bloch} visualises the pure effects for each of the 3 concepts on the 4 domains, by plotting the corresponding pure states on a Bloch sphere. We are able to perform the visualisation for this basic model since only one qubit is being used per domain, with no entanglement. The clusters of dots around each concept are the corresponding instances (pure states) in the training data. This visualisation is for the model which performs as described above on the classification task; a model trained from a different random initialisation would have the concepts and instances distributed differently around the sphere, but this visualisation is representative in terms of how the concepts are typically separated and the instances clustered. Note how the 3 concepts on each domain are being pushed apart (strikingly so in the case of the \textsc{position} domain) and how the concepts sit neatly in the centre of each cluster of instances. 


\subsubsection{The Rainbow Dataset}
\label{sec:rainbow}

In order to test our model further, we used the rainbow dataset from Section~\ref{sec:expts-vae}, and in order to train the discriminative model, we added a further 3,000 negative examples (for each epoch) to the 3,000 positive ones, randomly generated as before.
Perhaps unsurprisingly, it was more difficult with this data to obtain a clean separation of the colours on a single qubit.\footnote{Of course there is nothing to prevent us from using more than one qubit per domain, in order to provide a larger Hilbert space in which to represent the additional colours, but the visualisation is harder with more qubits.} However, with a weighting of 0.5 applied to the negative examples in the binary cross-entropy loss, and running the training for 200 epochs, we were able to obtain the distribution of colours around the Bloch sphere shown in Figure~\ref{fig:rainbow_bloch} (with instances again taken from the training data). The three visualisations are for three separately trained models (i.e. with three different random initialisations of the model parameters).

In terms of accuracy on the development data, the classification model for the Bloch sphere on the far left achieved similar scores on the non-colour domains as before, and an overall accuracy of 95\% on \textsc{colour}, with F1-scores ranging from 91\% to 100\% for the individual colours. The Bloch sphere in the middle is for a model with similar performance, and is shown to demonstrate the variation in models. The example on the far right is cherry-picked as an example of how the training is able to neatly represent the various colours on the Bloch sphere: note how the \emph{yellow}, \emph{orange} and \emph{red} instances are beautifully placed on the circumference of a circle, with the \emph{red} instances leading into \emph{orange} and then \emph{yellow}.

\subsubsection{Adding a Decoder Loss}
\label{sec:decoder}

One notable feature of the visualisations in Figure~\ref{fig:basic_bloch} is how ``tight" the instance clusters are, despite the variation in the images for a single concept (for example the variation in red shapes in Figure~\ref{fig:shapes}). There may be use-cases where we would like the representation of instances to better reflect the variation in the underlying images, for example in order to better capture correlations across domains (see Section~\ref{sec:correlations} below). 


\begin{figure*}[t!]
\begin{center}
    \includegraphics[scale=0.25]{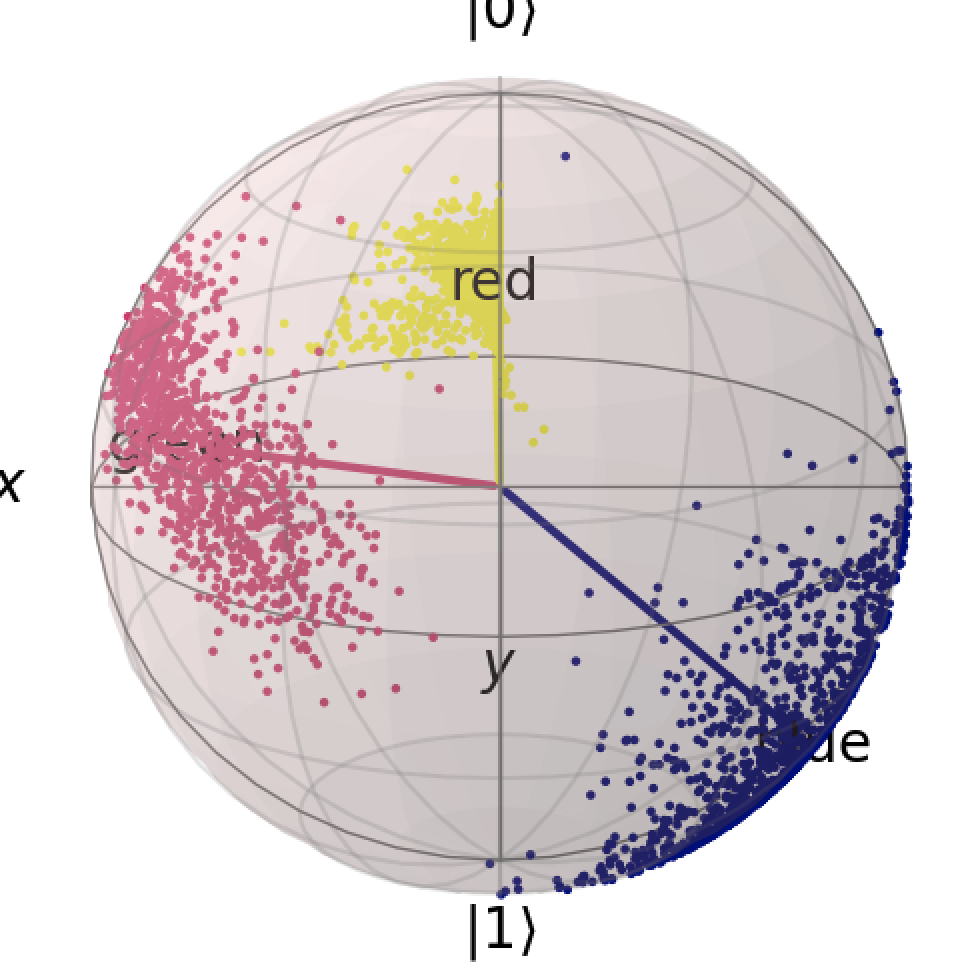}
    \includegraphics[scale=0.25]{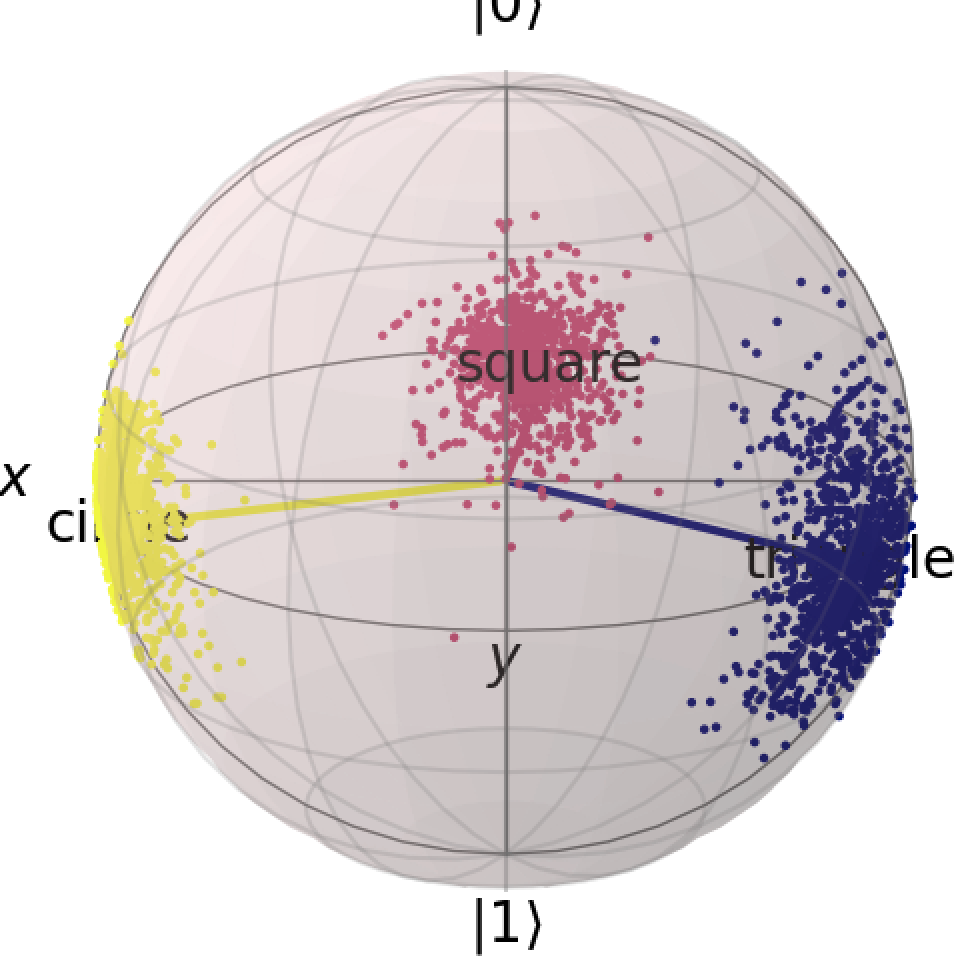}
    \includegraphics[scale=0.34]{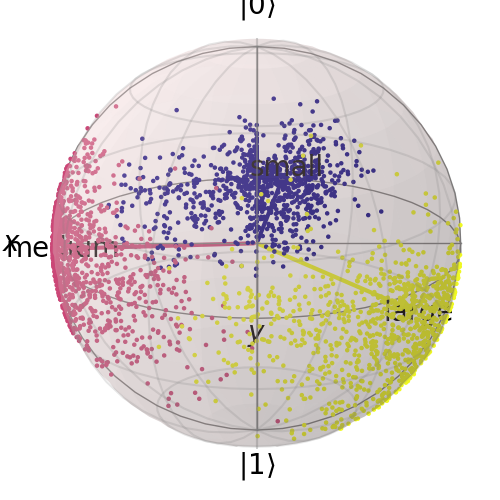}
    \includegraphics[scale=0.34]{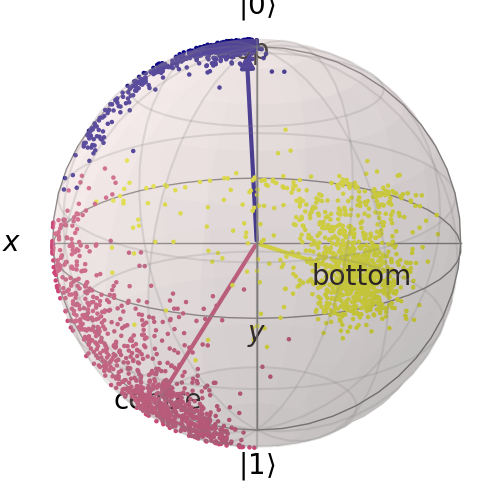}
    \end{center}
    \caption{Visualisation of the concept effects and instance states for all 4 domains, for the basic dataset with an additional decoder loss.}
    \label{fig:decoder_bloch}
\end{figure*}

In order to provide more of a ``spread" of the instances, we experimented with an additional decoder loss in the loss function below, where BCE is the binary cross-entropy loss, $D$ is the data with $N$ instances $\{X_i\}_i$, and $\psi$ and $\phi$ are the parameters of the encoder network:
\begin{multline}
    \text{Loss}(D, \psi, \phi, \chi) = \text{BCE}(D, \psi, \phi) + \frac{\lambda}{N} \sum_i \text{SE}(\text{DeCNN}(\chi,\text{CNN}(\psi,X_i)),X_i)
    \label{eqn:loss_dec}
\end{multline}
The decoder is a deconvolutional neural network (DeCNN), with parameters $\chi$, which essentially is the CNN ``in reverse": it takes as input the angles output by the CNN, given an image $X_i$, and outputs RGB values for each pixel in the image. SE is the sum of squared errors across all RGB values in the image, and $\lambda$ is a weighting term in the overall loss. The intuition is that, in order to obtain a low SE loss, the encoder CNN has to output angles which are sufficiently informative in order for the DeCNN to accurately reconstruct the original image. 
Now the model is similar to the Conceptual VAE (albeit without the generative model interpretation), in that it has both ``encoder" and ``decoder" parts to the loss.\footnote{One possibility for future work is to develop and implement a ``quantum VAE" \shortcite{khoshaman2018quantum} for concept modelling, and have a generative model in which all parts of the model are quantum.}

Figure~\ref{fig:decoder_bloch} shows how the instances can be distributed more broadly around the Bloch sphere, using the additional decoder loss (with $\lambda = 0.1$). This model still performs well as a classification model on the development data, achieving 
98\% accuracy on \textsc{size}, 99\% on \textsc{colour}, 100\% on \textsc{shape}, and 98\% on \textsc{position}. As a qualitative demonstration of this approach, note how the instances for \emph{centre} and \emph{top} start to merge into each other (blue and red instance dots bottom right), and also for \emph{medium} and \emph{small} (blue and red instance dots bottom left), which is what we would expect for a less discrete representation. 

\subsection{Capturing Correlations}
\label{sec:correlations}

\begin{figure}[t!]
\begin{center}
    \hspace*{-0.0cm}
    \includegraphics[scale=0.35]{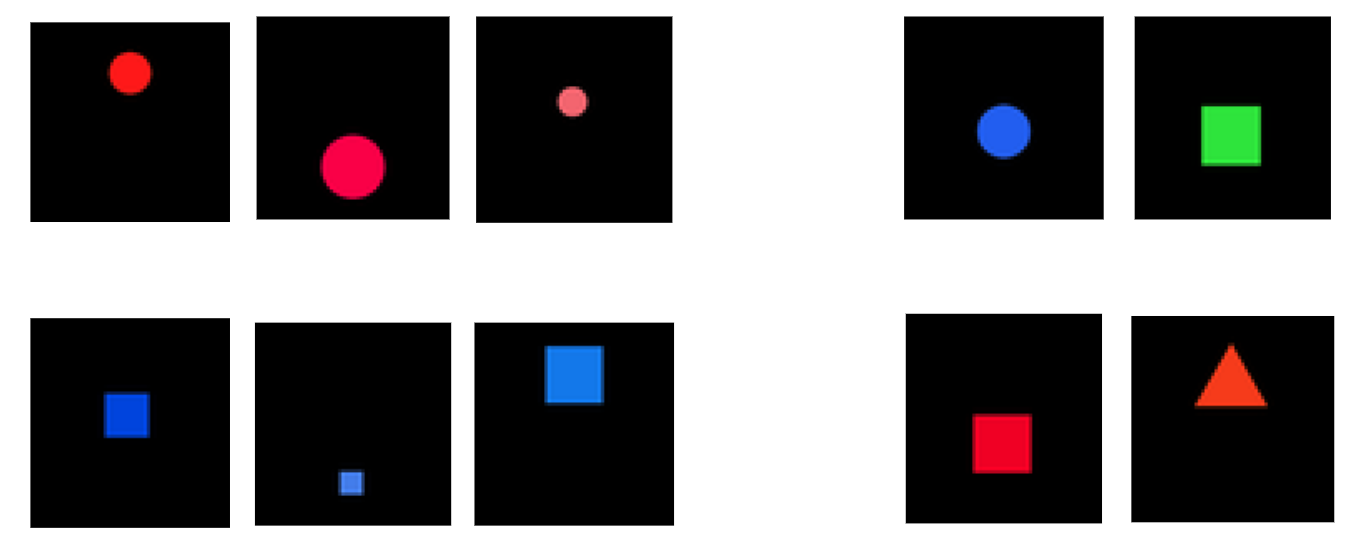}
    \caption{Examples of twikes (on the left) and non-twikes (on the right).}
    \label{fig:twike}
    \end{center}
\end{figure}

Here we show how one of the characteristic features of quantum theory, namely entanglement, can be used to capture correlations across domains. In order to test whether our model can handle concepts which contain correlations, we define a new concept which we call \emph{twike}, which is defined as \emph{(red and circle) or (blue and square)} (i.e. it applies to images containing red circles or blue squares). Figure~\ref{fig:twike} shows some examples of twikes and non-twikes.

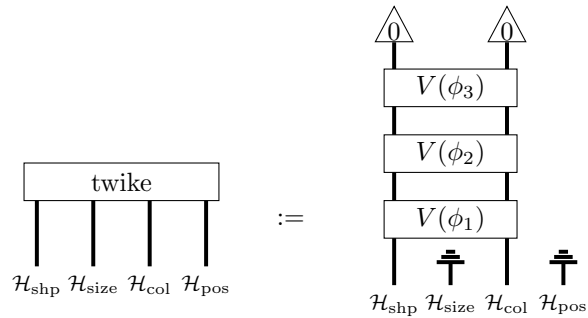
\begin{figure}
\begin{center}
\begin{tikzpicture}
	\begin{pgfonlayer}{nodelayer}
		\node [style=none] (0) at (-2.25, -2.25) {};
		\node [style=none] (1) at (-2.25, -0.25) {};
		\node [style=large map] (2) at (0, 0) {$ \ \ \ \ \ \ $ twike $ \ \ \ \ \ \ $};
		\node [style=none] (3) at (2.25, -2.25) {};
		\node [style=none] (4) at (2.25, -0.25) {};
		\node [style=none] (5) at (4.5, -1) {$:=$};
		\node [style=none] (6) at (7.25, -2.75) {};
		\node [style=copoint] (7) at (7.25, 4) {$0$};
		\node [style=none] (10) at (8.75, -2.75) {};
		\node [style=upground] (11) at (8.75, -2) {};
		\node [style=copoint] (21) at (10.25, 4) {$0$};
		\node [style=none] (22) at (10.25, -2.75) {};
		\node [style=medium map] (27) at (8.75, -1) {$ \ \ \ V(\phi_1) \ \ \ $};
		\node [style=none] (35) at (11.75, -2.75) {};
		\node [style=upground] (36) at (11.75, -2) {};
		\node [style=none] (37) at (-0.75, -2.25) {};
		\node [style=none] (38) at (-0.75, -0.25) {};
		\node [style=none] (39) at (0.75, -2.25) {};
		\node [style=none] (40) at (0.75, -0.25) {};
		\node [style=medium map] (41) at (8.75, 0.75) {$ \ \ \ V(\phi_2) \ \ \ $};
		\node [style=medium map] (42) at (8.75, 2.5) {$ \ \ \ V(\phi_3) \ \ \ $};
		\node [style=label] (43) at (-2.25, -2.75) {$\Hdom{shp}$};
		\node [style=label] (44) at (-0.75, -2.75) {$\Hdom{size}$};
		\node [style=label] (45) at (0.75, -2.75) {$\Hdom{col}$};
		\node [style=label] (46) at (2.25, -2.75) {$\Hdom{pos}$};
		\node [style=label] (47) at (7.25, -3.25) {$\Hdom{shp}$};
		\node [style=label] (48) at (8.75, -3.25) {$\Hdom{size}$};
		\node [style=label] (49) at (10.25, -3.25) {$\Hdom{col}$};
		\node [style=label] (50) at (11.75, -3.25) {$\Hdom{pos}$};
	\end{pgfonlayer}
	\begin{pgfonlayer}{edgelayer}
		\draw [style=thick] (1.center) to (0.center);
		\draw [style=thick] (4.center) to (3.center);
		\draw [style=thick] (7) to (6.center);
		\draw [style=thick] (11) to (10.center);
		\draw [style=thick] (21) to (22.center);
		\draw [style=thick] (36) to (35.center);
		\draw [style=thick] (38.center) to (37.center);
		\draw [style=thick] (40.center) to (39.center);
	\end{pgfonlayer}
\end{tikzpicture}
\end{center}
\caption{Encoder PQC for learning \emph{twike}, here shown with 3 layers of the rotation and entangling $V$ ansatz.}
\label{fig:twike-pqc}
\end{figure}

The concept PQCs we have considered so far, of the form in (\ref{eq:concept-initial}), are unable to learn the concept \emph{twike}, since the domains have been treated independently, with each of the 4 domains effectively containing its own independent concept. 
In order to create connections between the domains in the concept PQC, we can apply our full ansatz $V$ from Section~\ref{cnn_pqcs}, involving controlled-Z gates between wires, across multiple domains. In this first experiment we assume knowledge of the fact that, for the twike concept, the correlations are across the \textsc{shape} and \textsc{colour} domains, with entangling gates only between the qubits for \textsc{shape} and \textsc{colour}. (This assumption will be relaxed for some of the experiments below.) We also assume that the remaining domains are not relevant and so are not measured, thus effectively being discarded in the concept. We apply potentially multiple layers of ansatz $V$ to the relevant domains, and so the resulting form of the twike concept over the four domains is as shown in Figure~\ref{fig:twike-pqc}, where $\phi$ are the learned parameters for the twike concept.

The training of this model only updates the rotation parameters of the concept PQC; the parameters of the encoder (i.e. the CNN) are kept fixed from the earlier training of the basic model. The loss function is binary cross entropy, as before, with the 3,000 examples from Section~\ref{sec:expts-vae} used as training data. Roughly 20\% of these instances are positive examples of \emph{twike}, with the remaining being negative examples. We trained this model for 50 epochs, using 2 layers of the rotation and entangling $V$ ansatz for the concept PQC, and obtained 100\% accuracy on the unseen test examples. It was only through the introduction of the entangling gates that we were able to learn the \emph{twike} concept at all.

In terms of the discussion of entanglement and classical correlation in Section \ref{sec:entangled-concepts}, we can say that the twike concept can be naturally described without entanglement, as a classical combination of the pure concepts \emph{red circle} and \emph{blue square} (at least in the case where these pure effects are orthogonal). However, such correlations are not always immediately implementable in many conventional classical models. 
In terms of the Conceptual VAE, it would be possible to capture correlations using the covariance matrix of the multivariate Gaussian. However, a standard assumption in VAEs is to assume a multivariate Gaussian with a diagonal covariance matrix (and so no correlations across domains). Whether a concept like \emph{twike} could be easily modelled using the Conceptual VAE, especially as the number of domains is increased, is left as a question for future work.

\subsection{Learning General Mixed and Entangled Concepts} \label{sec:mixedconcepts}

One assumption made above in the \emph{twike} experiments was that the relevant domains---in this case \textsc{shape} and \textsc{colour}---are known in advance, so that the concept PQC can effectively ignore the wires corresponding to the other domains. One interesting question is whether the concept PQC could also learn which domains are relevant, as well as which of those domains should be correlated, if provided with all 4 wires as input.
To allow for such correlations between arbitrary domains, the concept PQC should allow for entanglement between any of its domains. Furthermore, to enable discarding of domains, we require mixed quantum effects.
Both of these features can be included by using our most general form of the concept PQC \eqref{eq:concepts-pqc-general}.


In order to test the learning of these general concepts, we set up a similar experiment to \emph{twike}, but this time with just \emph{red} as the concept to be learned. Of course the encoder had already learned \emph{red} when trained to perform classification in the basic setup, but in this experiment we remove knowledge of which wire the \textsc{colour} domain is on, and see whether a new concept PQC can learn \emph{red}, given red and non-red instances as input.

Again the training of this model only updates the rotation parameters of the concept PQC; the parameters of the CNN are kept fixed. The loss function is again binary cross entropy, with the usual 3,000 examples as training data. Roughly 33\% of these instances are positive examples of \emph{red}, with the remaining being negative examples. We trained this model for 50 epochs, using 2 layers of rotation and entangling gates for the concept PQC, and obtained 100\% accuracy on the unseen test examples. It was only through the introduction of the discarding (plus entangling gates) that we were able to obtain these high accuracies.


\subsection{Concepts containing Logical Operators}
\label{sec:logical_operators}

For one final set of experiments, we investigated whether the entangling and discarding PQC \eqref{eq:concepts-pqc-general} could learn concepts built from logical operators, with concepts such as \emph{red or blue}. 


\subsubsection{Conjunction across Domains}

The first concept with a logical operator that we consider is \emph{red and circle}, firstly with the knowledge of which domains are relevant for the concept (in this case \textsc{colour} and \textsc{shape}). The encoder PQC is the simple one from (\ref{fig:basic-setup-concrete}), but with only the \textsc{colour} and \textsc{shape} wires (so the other two are effectively discarded). We used the same 3,000 training examples, of which roughly 17\% are positive examples and 83\% negative examples. In this case the learning is particularly easy, and the model obtains 100\% accuracy with only a single layer of rotations for the PQC, without any entangling gates or discarding of any ancilliary qubits. The reason is that the factorisation of the domains through the tensor product has effectively provided all the structure required to use conjunction.

When the knowledge of which domains are relevant is removed, and the more general encoder PQC in (\ref{eq:concepts-pqc-general}) is used, learning becomes harder but an encoder PQC with 4 layers of rotation and entangling gates is able to learn the concept with 100\% accuracy.

\subsubsection{Disjunction within Domains}

Next we consider disjunction, but \emph{within} rather than \emph{across} domains, with the concept to be learned being \emph{red or blue}. Of the 3,000 training examples, 61\% are positive examples and 39\% negative. Again, when knowledge of which domains are relevant is provided to the concept PQC, the learning is easy, with 100\% accuracy obtained with a single layer of rotations. 

If each point on the Bloch sphere were to correspond to an instance of the  \textsc{colour} domain, i.e. a single colour, as in our model, then the PQC learning such a pure effect for \emph{red or blue} will in fact be simply learning a single colour, intuitively somewhere ``in between" \emph{red} and \emph{blue}. When the domain only comes with a few concepts, such as the 3 concepts used here, this single instance may do well in approximating \emph{red or blue}, as with the 100\% accuracy. However, in the presence of more concepts, we expect that a concept for \emph{red or blue} should involve mixing. And when knowledge of which domains are relevant is not provided to the PQC, \emph{red or blue} can also be successfully learned with the more general PQC in (\ref{eq:concepts-pqc-general}) with 3 layers of rotation and entangling gates, including discarding.



\section{Related Work}
\label{sec:related_work}

The Conceptual VAE is inspired by \shortciteA{beta-vae}, who introduced the $\beta$-VAE for unsupervised concept learning. However, the focus of \shortciteauthor{beta-vae} is on learning the conceptual \emph{domains}, i.e. the underlying factors generating the data \shortcite{bengio:2013}, which they refer to as learning a \emph{disentangled} representation. The main innovation to encourage the VAE to learn a disentangled, or factored, latent space is the introduction of a weighting term $\beta$ on the KL loss. \shortciteauthor{beta-vae} show that setting $\beta$ to a value greater than 1 can result in the dimensions of $\Z$ corresponding to domains such as the lighting or elevation of a face in the celebA images dataset, or the width of a chair in a dataset of chair images.
Our focus is more on the conceptual representations themselves, assuming the domains are already known, and the question of how concept labels can be introduced into the VAE model. 


A paper in NLP that uses a model very similar to the Conceptual VAE is \shortciteA{brazinskas} which introduces the Bayesian skip-gram model for learning word embeddings. One key difference which distinguishes our work from the word embeddings typically used in NLP is that we do not restrict ourselves to the textual domain, meaning that our conceptual representations are \emph{grounded} in some other modality (in our case images) \shortcite{harnad}, bringing them closer to the human conceptual system.
Another relevant paper from the NLP literature, which does consider grounding, is \shortciteA{schlangen-etal-2016-resolving}, where the meanings of words are treated as classifiers of perceptual contexts, similar to how we use classification to induce conceptual representations.

The Conceptual VAE uses Gaussians to represent concepts, since they are the typical distributions used with VAEs and because they are convenient from a mathematical perspective. However, the use of Gaussians is also prevalent in the neuroscience literature, appearing for example  as the \emph{Laplace assumption} in the ``free-energy" or ``predictive processing" framework \shortcite{friston2009predictive,bogacz2017tutorial}. 

In terms of the quantum models, Smolensky has a large body of work arguing for tensor product representations in linguistics and cognitive science more broadly \shortcite{smolensky_book}. Recently these techniques have been integrated into neural models for NLP \shortcite{huang-etal-2018-tensor}. Another line of work which associates tensor-product representations with grammatical structure is the ``DisCoCat" research program attempting to build distributed, compositional representations of language, which began with \shortciteA{coecke2010mathematical}. Recently this work has culminated in the running of quantum NLP models on real quantum hardware \shortcite{qnlp_practice}.

The field of \emph{quantum cognition} \shortcite{pothos2013bbs} has already been mentioned. Some recent work in this area includes \shortciteA{epping:2022} and \shortciteA{epping:2021}, where the latter is concerned with modelling human judgements of colour similarity and uses a Hilbert space representation similar to our models. The learning of concepts containing logical operators has a formal connection to quantum logic \shortcite{birkhoff_1936} and Boolean concept learning in general, for which there is a large literature \shortcite{goodwin:2013}.

\section{Conclusion and Further Work}
\label{sec:further_work}

In this article we have presented a new modelling framework
for structured concepts using a category-theoretic generalisation of G\"ardenfors' conceptual spaces, and shown how the conceptual representations can be learned automatically from data, using two very different instantiations: one classical and one quantum. 
The main contributions of this foundational work are the category-theoretic formalisation, and the two practical demonstrations, especially the quantum implementation which is particularly novel. Substantial further work
is required to demonstrate that the framework can be
applied fruitfully to data from a psychology lab, which would connect our work directly with quantum cognition, and also to agents acting in (virtual) environments, which would connect it to agent-based AI
(Abramson et al., 2020). 


In future more interpretative work on quantum concepts is needed to clarify their advantages, such as those offered by entanglement discussed in Section \ref{sec:entangled-concepts}, and their naturality as a model in cognition.  Another benefit of quantum models over conceptual spaces not explored here is the presence of a ``negation" $C^\bot$ on concepts with $C \leq \discard{}$ \shortcite{rodatz2021conversational,shaikh2021composing}. In contrast, negation is harder to define for concepts in conceptual spaces; for example the complement of a convex region is generally non-convex.

Another interesting question is whether the Conceptual VAE can be applied to data generated from a conceptual hierarchy---for example having shades of colour such as \emph{dark-red}---and whether the learned Gaussian representations for concepts can be partially ordered in an appropriate way \shortcite{clark_concepts}. The quantum concepts as effects have a natural ordering, as discussed in Section~\ref{sec:quantum-models2}, and it would be an interesting comparison to see if hierarchies could be more easily learned with the quantum models.

To make full use of the compositional approach, one should also describe conceptual \emph{processes}, such as reasoning processes and ``metaphorical" mappings between domains, now given by CP maps between quantum models. One could then compare these with the processes in the category $\ConSp$ of fuzzy conceptual processes from \shortciteA{tull2021categorical}. 

Finally, even though all the practical work here has been carried out in simulation on a classical computer, the number of qubits is relatively small, and the circuits relatively shallow, and so the running of these models on real quantum hardware is a distinct possibility. Also left for future work is the search for tasks which could demonstrate advantages for our quantum representations, for example establishing whether non-separable effects in the theory do provide an advantage over classical correlation in modelling conceptual structure.

\section*{Acknowledgements}
Thanks to Lia Yeh, Robin Lorenz and Douglas Brown for extremely detailed and helpful comments on an earlier draft, and also to the rest of the Oxford Quantum Compositional Intelligence team.

\section*{Declarations}

\begin{itemize}
\item Authors' contributions: Sean Tull developed the mathematical formalisation and wrote the theory sections. Razin A. Shaikh wrote the code, ran the experiments, and prepared some of the figures. Sara Sabrina Zemlji\v{c} created the data and helped run the experiments. Stephen Clark oversaw the project, ran some of the experiments, and wrote the remainder of the manuscript. All authors took part equally in setting the general direction of the project.
\item Code availability: The code for generating the data and running the experiments is available at: https://github.com/CQCL/concepts-vae.
\end{itemize}

\bibliographystyle{apacite}
\bibliography{cog,references}

\appendix
\renewcommand{\thesection}{\Alph{section}}

\onecolumn

\section{The Shapes Dataset}
\label{sec:app_shapes}

The parameters used in the Spriteworld software to generate the Shapes dataset:\\

\noindent
\includegraphics[width=15cm]{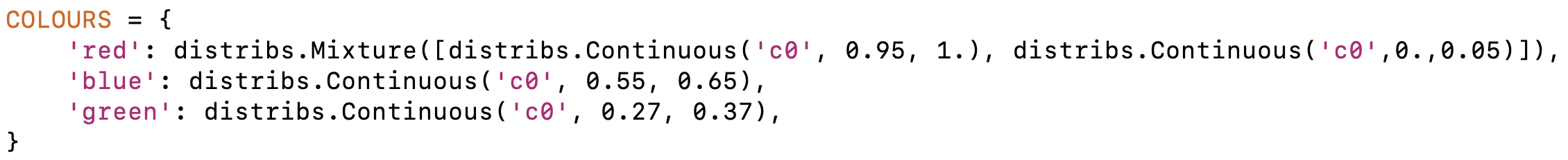}

\noindent
\includegraphics[width=10cm]{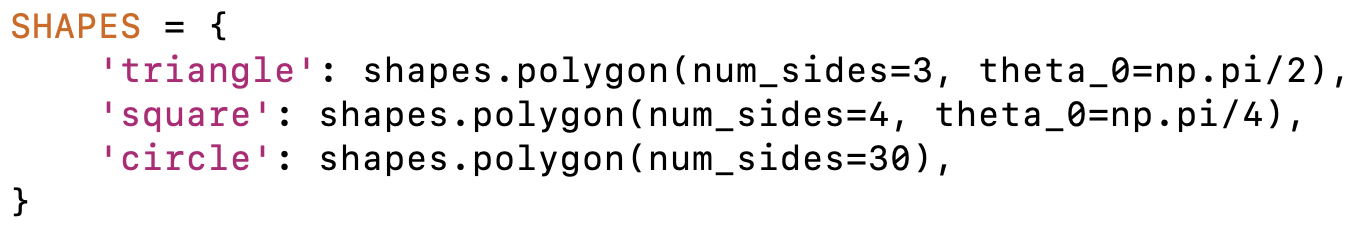}

\noindent
\includegraphics[width=9cm]{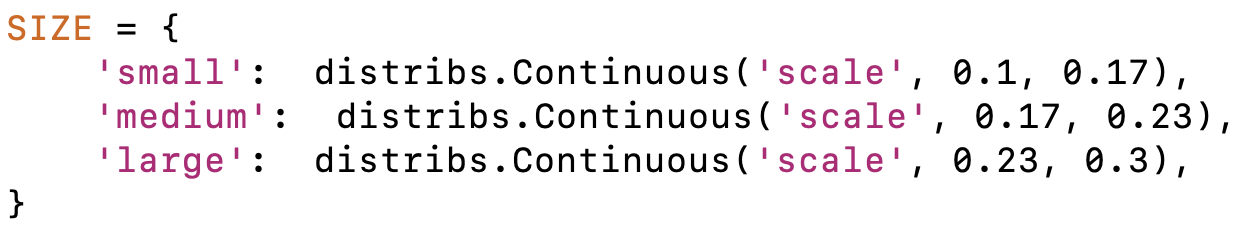}

\noindent
\includegraphics[width=15cm]{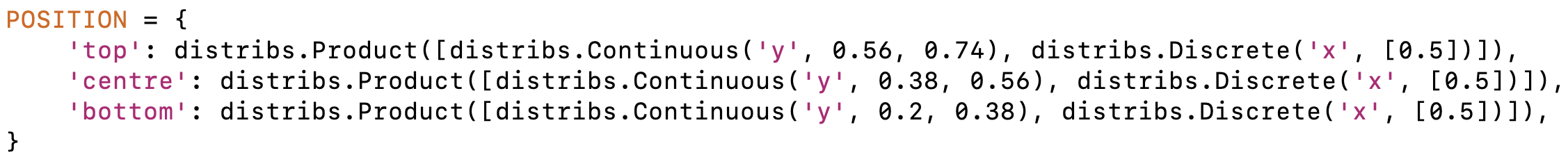}

\noindent
Additional parameters for the \textsc{colour} domain:\\

\noindent
\includegraphics[width=8cm]{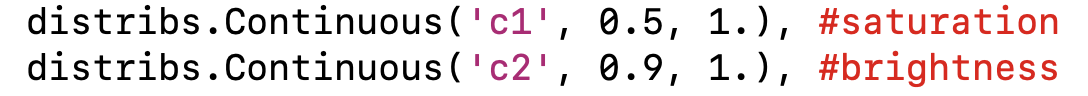}

\subsection{The Extended Colour Dataset}
\label{sec:ext_shapes}

The parameters used in the Spriteworld software to generate the Shapes dataset with more (rainbow) colours:\\

\noindent
\includegraphics[width=15cm]{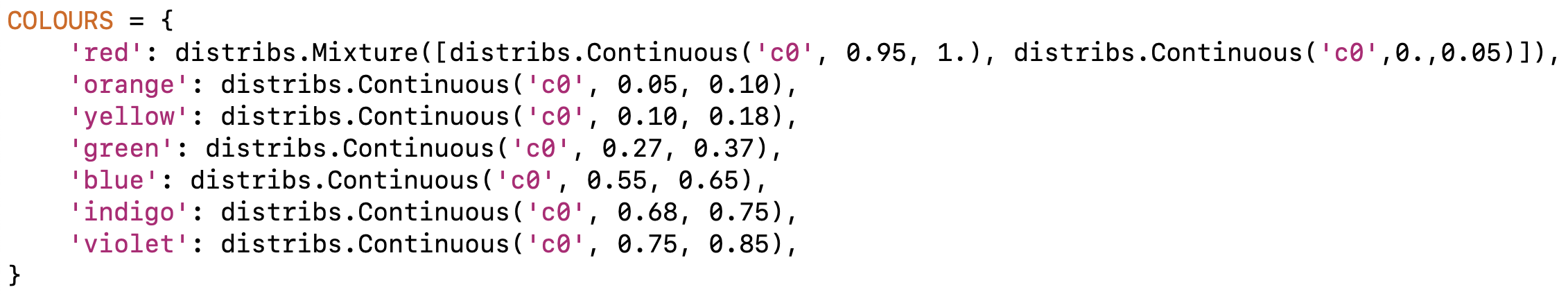}

\newpage

\section{Neural Architectures and Hyper-parameters}
\label{sec:app_neural_nets}

\begin{tabular}{@{}lr@{}}
\toprule
image width & 64\\
image height & 64\\
image channels & 3\\
\midrule
CNN kernel size & $4 \times 4$\\
CNN stride & $2 \times 2$\\
CNN layers & 4\\
CNN filters & 64\\
CNN dense layers & 2\\
CNN dense layer size & 256\\
dimensions of latent space & 6\\
\midrule
initialization interval for means of priors & $[-1.0, 1.0]$\\
initialization interval for log-variances of priors & $[-7.0, 0.0]$\\
\midrule
batch size & 32\\
Adam learning rate & $10^{-3}$\\
Adam $\beta_1$ & 0.9 \\
Adam $\beta_2$ & 0.999\\
Adam $\epsilon$ & $10^{-7}$\\
\bottomrule
\end{tabular}



\end{document}